\documentclass[12pt]{article}

\ifx\pdfoutput\undefined
\usepackage[dvips,bookmarks]{hyperref}
\else
\usepackage{hyperref}
\fi
\hypersetup{colorlinks=false,bookmarksopen,bookmarksnumbered,citecolor=blue,
   pdfstartview=FitH}

\usepackage{latexsym}

\usepackage{amssymb,amsfonts,amsmath}
\usepackage{graphicx} 
\usepackage{indentfirst}

 \usepackage{bbm}

\parskip=\medskipamount

\newcommand {\cA}{{\cal A}}
\newcommand {\cB}{{\cal B}}
\newcommand {\cC}{{\cal C}}
\newcommand {\cD}{{\cal D}}

\newcommand {\cH}{{\cal H}}

\newcommand {\cK}{{\cal K}}
\newcommand {\cL}{{\cal L}}
\newcommand {\cM}{{\cal M}}
\newcommand {\cN}{{\cal N}}

\newcommand {\cT}{{\cal T}}

\def\a{\alpha}

\def\b{\beta}
\def\c{\chi}
\def\d{\delta}
\def\e{\epsilon}
\def\f{\phi}
\def\g{\gamma}
\def\G{\Gamma}

\def\j{\psi}

\def\l{\lambda}
\def\m{\mu}
\def\n{\nu}
\def\o{\omega}
\def\p{\pi}
\def\q{\theta}
\def\r{\rho}
\def\s{\sigma}
\def\t{\tau}
\def\u{\upsilon}
\def\x{\xi}
\def\z{\zeta}

\def\F{\Phi}
\def\J{\Psi}
\def\L{\Lambda}
\def\O{\Omega}

\def\S{\Sigma}
\def\U{\Upsilon}
\def\X{\Xi}

\def\rd{{\rm d}}
\def\ri{{\rm i}}

\newcommand{\ve}{\varepsilon}                            
\newcommand{\DB}{\bar{D}}

\newcommand{\ab}{{\a\b}}

\newcommand{\pa}{\partial}                           
\newcommand{\hf}{\frac12}

\newcommand{\vf}{\varphi}

\newcommand{\be}{\begin{equation}}
\newcommand{\ee}{\end{equation}}
\newcommand{\bea}{\begin{eqnarray}}
\newcommand{\eea}{\end{eqnarray}}
\newcommand{\non}{\nonumber}
\newcommand{\ba}{\begin{array}}
\newcommand{\ea}{\end{array}}

\newcommand{\1}{\underline{1}}

\newcommand{\iu}{\underline{i}}
\newcommand{\ju}{\underline{j}}
\newcommand{\ku}{\underline{k}} 

\newcommand{\Iu}{\underline{I}}
\newcommand{\Ju}{\underline{J}}
\newcommand{\Ku}{\underline{K}}
\newcommand{\mun}{\underline{m}}

\def\dt#1{{\buildrel {\hbox{\LARGE .}} \over {#1}}}    

\newcommand{\bm}[1]{\mbox{\boldmath$#1$}}

\def\double #1{#1{\hbox{\kern-2pt $#1$}}}

\newcommand{\sSp}{\mathsf{Sp}}
\newcommand{\sSU}{\mathsf{SU}}
\newcommand{\sSL}{\mathsf{SL}}
\newcommand{\sGL}{\mathsf{GL}}
\newcommand{\sSO}{\mathsf{SO}}
\newcommand{\sU}{\mathsf{U}}

\newcommand{\sOSp}{\mathsf{OSp}}

\newcommand{\sMat}{\mathsf{Mat}}

\newcommand{\bsubeq}{\begin{subequations}}
\newcommand{\esubeq}{\end{subequations}}

\newcommand{\qb}{{\bar{\theta}}}

\newcommand{\oo}{{\overline{1}}}
\newcommand{\ot}{{\overline{2}}}

\newcommand{\mD}{{\mathbb D}}
\newcommand{\mDB}{{\mathbb \DB}}

\begin{document}
\begin{titlepage}
\begin{flushright}
UUITP-36/10\\
November, 2010\\
\end{flushright}

\begin{center}
{\Large \bf Off-shell superconformal nonlinear sigma-models \\ in three dimensions
}\\ 
\end{center}

\begin{center}

{\bf 
Sergei M. Kuzenko
}

{\footnotesize{
{\it School of Physics M013, The University of Western Australia\\
35 Stirling Highway, Crawley W.A. 6009, Australia}} ~\\
\texttt{kuzenko@cyllene.uwa.edu.au}}\\
\vspace{2mm}

{\bf Jeong-Hyuck Park
}

{\footnotesize{
{\it Department of Physics, Sogang University \\
Shinsu-dong, Mapo-gu, Seoul 121-742, Korea}}~\\
\texttt{park@sogang.ac.kr}}\\
\vspace{2mm}

{\bf Gabriele Tartaglino-Mazzucchelli
} 

{\footnotesize{
{\it Theoretical Physics, Department of Physics and Astronomy,
Uppsala University \\ 
Box 516, SE-751 20 Uppsala, Sweden}}\\
\texttt{gabriele.tartaglino-mazzucchelli@fysast.uu.se}}
\vspace{2mm}

{\bf Rikard von Unge
} 

{\footnotesize{
{\it Institute for Theoretical Physics, Masaryk University \\
Kotl\'a\v{r}sk\'a 2,
61137 Brno, Czech Republic } }~\\
\texttt{unge@physics.muni.cz}}

\end{center}

\begin{abstract}
\baselineskip=14pt
We develop superspace techniques to construct general off-shell 
$\cN \leq 4$ superconformal 
sigma-models in three space-time dimensions. 
The most general $\cN=3$ and $\cN=4$ superconformal sigma-models are constructed 
in terms of $\cN=2$ chiral superfields. 
Several superspace proofs of the folklore statement that 
$\cN=3$ supersymmetry implies $\cN=4$ are presented both in the on-shell and off-shell settings. 
We also elaborate on (super)twistor realisations for (super)manifolds on which 
the three-dimensional  $\cN$-extended superconformal groups act transitively 
and which include Minkowski space as a subspace.
\end{abstract}

\vfill
\end{titlepage}

\newpage
\renewcommand{\thefootnote}{\arabic{footnote}}
\setcounter{footnote}{0}

\tableofcontents{}
\vspace{1cm}
\bigskip\hrule

\section{Introduction}
\setcounter{equation}{0}

Over the last few years  the study of superconformal 
field theories in three space-time dimensions has received a renewed interest. 
The enthusiasm for this topic has been triggered 
by the works of Bagger-Lambert \cite{BL} and Gustavsson \cite{G}
who formulated 
for the first time a maximally supersymmetric $\cN=8$ Chern-Simons 
theory. The 
analysis undertaken in  \cite{BL,G}
was aimed at developing a  description of the world-volume 
theory of multiple M2-branes.
The same quest has later led to the $\cN=6$, $\mathsf{U}(N)\times \mathsf{U}(N)$ 
superconformal Chern-Simons theory introduced by
Aharony, Bergman, Jafferis and Maldacena (ABJM) \cite{ABJM}, which is conjectured to describe 
the low-energy dynamics of a system of $N$ M2-branes.
In the context of the AdS/CFT duality,
 the ABJM theory in the large-$N$ limit should be dual 
 to the dynamics of  M-theory on AdS$_4\times$S$^7/\mathbb{Z}_k$.
At present, extensions of the conjectured duality are a popular
subject of investigation in the  literature.
On the CFT$_3$ side it would be highly desirable to have a formalism in which a maximally
possible amount of supersymmetry is realised  off-shell and the superconformal transformations
of the multiplets 
originate in a simple geometric framework. 
Such properties are useful for investigating various dynamical aspects, including quantum computations,
and may be helpful for  constructing new nontrivial superconformal field theories.
Keeping this in mind, in the present paper we develop superspace formulations for 
general  $\cN \leq 4$ superconformal nonlinear sigma-models in three dimensions.
Our main results concern the extended cases $\cN=3$ and $\cN=4$.

Rigid superconformal sigma-models in three dimensions have recently been constructed in component 
formalism \cite{BCSS} building on earlier works  \cite{A-GF,deWTN,ST,GR,deWRV,deWRV2}.
It was shown, in particular, that for $\cN \leq 4$ the sigma-model target space  $\cM$ is  a  cone 
with an appropriate  Sasakian base manifold. Specifically, $\cM$ is a Riemannian cone for $\cN=1$ \cite{ST}, 
a K\"ahlerian cone for $\cN=2$ \cite{GR},   a hyperk\"ahler cone for $\cN=3$ \cite{deWRV,deWRV2}.
If $\cN=4$, then $\cM$ factorises into a product of two hyperk\"ahler  cones, 
one parametrised by hypermultiplets  and the other  
by twisted hypermultiplets \cite{BCSS,deWTN}. 
In the cases $\cN>4$, the superconformal sigma-models were shown in \cite{BCSS} 
to have necessarily  flat target spaces (symmetric target spaces in the locally supersymmetric case \cite{deWTN}).
Although the approach of \cite{BCSS} is geometric and insightful, it is intrinsically 
on-shell. First of all, the superconformal transformations form a closed algebra only on the mass shell.
Secondly, and most importantly, this formalism does not allow one to generate superconformal 
sigma-models with $\cN=3$ (which in fact  implies $\cN=4$). 
 Given a hyperk\"ahler cone {\cite{GR,deWRV},
one can immediately write down the associated nonlinear sigma-model using the results of \cite{BCSS},
but not vice versa. The approach of \cite{BCSS} does not offer sigma-model techniques to generate 
hyperk\"ahler cones. Such techniques will be developed in the present paper using the power of superspace
model building.

Our approach to construct off-shell superconformal sigma-models in three dimensions 
is an extension of the 5D $\cN=1$ and 4D $\cN=2$ superconformal projective multiplets and their
self-couplings \cite{K-compactified,K-hyper,KLvU,K-duality}. Such multiplets are defined most naturally 
in projective superspace \cite{KLR,LR-projective,LR-projective2}. 
The 4D N=2 projective superspace approach is closely related
to the harmonic  one \cite{GIKOS,GIOS}, for both of them make use of the isotwistor 
superspace ${\mathbb R}^{4|8} \times {\mathbb C}P^1 = {\mathbb R}^{4|8} \times S^2$ 
introduced for the first time  by Rosly \cite{Rosly}. 
The two approaches appear to be  complementary in many respects 
although the latter is more general \cite{K98}. 
As regards
sigma-model applications, 
however, the projective superspace formalism turns out to be more efficient. 
The point is that sigma-models in harmonic superspace (see \cite{GIOS} for a review) do not possess 
a natural decomposition in terms of standard 4D $\cN=1$ superfields, 
a property that is absolutely essential for various applications.  
The existence of such a decomposition is one of the powerful inborn features 
of the 4D $\cN=2$ multiplets in projective superspace, both in the non-superconformal
 \cite{LR-projective,LR-projective2,G-RRWLvU} and superconformal 
 \cite{K-compactified,K-hyper,KLvU,K-duality,KLvU2} cases. 
Similar conclusions apply in three dimensions.

It should be pointed out that hyperk\"ahler cones, which are target spaces for 
4D $\cN=2$  \cite{deWKV1,deWKV2,deWRV}  and 3D $\cN=3$ 
superconformal sigma-models, are intimately related to 
quaternion K\"ahler manifolds which are target spaces for 4D $\cN=2$ \cite{BW}
and 3D $\cN=3$ \cite{deWTN} locally supersymmetric $\s$-models. 
Specifically, there exists a one-to-one correspondence  \cite{Swann}
(see also \cite{Galicki}) between $4n$-dimensional quaternion K\"ahler manifolds
and $4(n+1)$-dimensional hyperk\"ahler cones. 
The superspace techniques presented in this paper provide a formalism for constructing 
$\cN=3$ rigid superconformal sigma-models generated by a Lagrangian of reasonably 
general functional form. Then, for any choice of the Lagrangian, the target space metric must be 
a hyperk\"ahler cone. As a result, the superspace techniques allow us in principle 
to generate new quaternion K\"ahler metrics which in general are difficult to construct.

The 3D $\cN=3,~4$ superconformal sigma-models, which will be constructed
in this paper, can naturally be coupled to conformal supergravity \cite{KLT-M}, 
as a generalisation of the approaches developed for 5D $\cN=1$ \cite{KT-M}
and 4D $\cN=2$ \cite{KLRT-M} supergravity-matter systems. 

The literature on 3D supersymmetric field theories is vast and, unfortunately, we are unable to comment 
upon many interesting developments. However, we wish to mention a few classic  works 
on off-shell superspace formulations. A thorough study of $\cN=1$  supersymmetric theories 
is contained in {\it Superspace} \cite{GGRS}. Important results on $\cN=2$ supersymmetric theories
appeared in \cite{HitchinKLR}. The $\cN=3$ harmonic superspace approach was developed in \cite{Zupnik} 
(it was further reformulated and extended to $\cN=4$ in \cite{Zupnik2}).\footnote{An interesting application of the 3D $\cN=3$ harmonic superspace is the recent
formulation of the ABJM theory given in \cite{ABJMharmonic}.}
The $\cN=4$ projective superspace formalism 
was developed in \cite{HitchinKLR,LR-projective,LR-projective2}.
We should also mention early papers on 3D $\cN\le 4$ Chern-Simons gauge theories in 
superspace  \cite{3DCS,Zupnik}}.

This paper is organised as follows. We start by describing (super)twistor realisations 
for (super)manifolds on which 
the three-dimensional  $\cN$-extended superconformal groups act transitively 
and which include Minkowski space 
as a subspace.\footnote{Such (super)manifolds have been discussed by Howe and Leeming \cite{HL}
in a purely algebraic setting. Our approach elaborates on quite  interesting geometric aspects that did not appear
in \cite{HL}.}  In section 2 we consider the non-supersymmetric case $\cN=0$
and discuss in detail the structure of compactified Minkowski  space.
In section 3 we introduce compactified $\cN$-extended Minkowski superspace.
Its harmonic/projective generalisations are presented in section 4.
Section 5 is devoted to a thorough analysis of $\cN$-extended superconformal Killing vectors, 
and the special cases $\cN\leq 4$ are discussed in much detail.
In section 6 we construct general off-shell $\cN=1$ and $\cN=2$ superconformal sigma-models.
In sections 7 and 8 we derive off-shell $\cN=3$ and $\cN=4$ superconformal sigma-models 
formulated in terms of weight-one polar supermultiplets in projective superspace.
Section 9 provides several proofs  of the statement that $\cN=3$ supersymmetry implies $\cN=4$.
In section 10 we construct the most general $\cN=3$ and $\cN=4$ superconformal sigma-models 
realised in terms of $\cN=2$ chiral superfields. Concluding comments and discussion are given 
in section 11. We have also included three technical appendices. 
Our 3D notation and conventions are collated in Appendix A. Appendix B
provides a supermatrix realisation for the $\cN$-extended super-Poincar\'e group 
and Minkowski superspace. Finally, Appendix C describes the $\cN=2$ reduction for 
$\cN=3$ and $\cN=4$ superconformal Killing vectors.

\section{Compactified  Minkowski space}
\setcounter{equation}{0}

As is  known, the conformal group in $d$ dimensions 
does not act globally 
on Minkowski space ${\mathbb M}{}^d \equiv {\mathbb R}^{d-1,1}$.
However, its action is well-defined on a compactified version of Minkowski space 
$\overline{\mathbb M}{}^d :=(S^{d-1}\times S^1 )/{\mathbb Z}_2$. 
In this section we present a twistor realisation for $\overline{\mathbb M}{}^3 $
which can be naturally generalised to superspace. The realisation given below can be compared 
with the twistor construction of $\overline{\mathbb M}{}^4$ \cite{Uhlmann,twistors,Segal,Tod,WW}
(see \cite{K-compactified} for a recent review).

\subsection{Twistor construction}
Consider a symplectic four-dimensional real vector space. It can be identified with ${\mathbb R}^4$ 
equipped with a skew-symmetric inner product:
\be
\langle \J| \F \rangle_{J}: = \J^{\rm T}J \, \F  \equiv \J_{\hat \a} J^{\hat \a \hat \b} \F_{\hat \b}
=- \langle \F| \J \rangle_{J}
~, \qquad
J =\big(J^{\hat \a \hat \b} \big) 
=\left(
\begin{array}{cc}
0  & {\mathbbm 1}_2\\
 -{\mathbbm 1}_2  &    0
\end{array}
\right) ~,
\label{J}
\ee
for any  vectors 
$\J,\F\in {\mathbb R}^4$. By construction, this inner product is invariant under the group
  $\sSp(2,{\mathbb R})$.  
The vectors $\J,\F\in {\mathbb R}^4$ will be called {\it twistors}.\footnote{In four space-time dimensions, 
twistors are necessarily complex, see e.g. \cite{twistors,WW}.}

The elements of the group\footnote{The group $\sSp(2,{\mathbb R})$ should not be confused 
with its compact sister $\sSp (2) := \sSp(2,{\mathbb C}) \bigcap \sU(4)$.
Sometimes, the group  $\sSp(2,{\mathbb R})$ is denoted  $\sSp(4,{\mathbb R})$, and similarly 
for its complexified  version.} 
$\sSp(2,{\mathbb R})$ will be represented by $4\times 4$ block 
matrices
\bea
g=\big(g_{\hat \a}{}^{\hat \b}\big)= \left(
\begin{array}{cc}
 \cA  & \cB\\
\cC &    \cD 
\end{array}
\right) \in \sSL(4,{\mathbb R}) ~, \qquad 
g^{\rm T} \,J \,g = J~,
\label{SP(2)}
\eea
where  $\cA,\cB,\cC$ and $\cD$ are $2\times 2$ matrices. The symplectic group $\sSp(2,{\mathbb R})$ is the 
2--1 covering of the conformal group in three dimensions, $\sSO_0 (3,2)$. 
The group $\sSp(2,{\mathbb R})$ is generated by its elements of the following three types:
\begin{subequations}
\bea
&&\mbox{Type 1}: \qquad h(\cA):=\left(
\begin{array}{cc}
 \cA  & 0\\
0 &    (\cA^{-1}) ^{\rm T} 
\end{array} \right) ~, 
\qquad \cA \in \sGL (2, {\mathbb R}) ~;\\
&&\mbox{Type 2}: \qquad  s(\cC) :=\left(
\begin{array}{cc}
  {\mathbbm 1}_2  & 0\\
\cC &     {\mathbbm 1}_2
\end{array}
\right) ~, \qquad \cC^{\rm T} =\cC \in \sMat   (2, {\mathbb R})~;\\
&& \mbox{Type 3}: \qquad J~.
\eea
\end{subequations}
The proof of this result is left to the reader as an instructive exercise. 

A {\it Lagrangian subspace} is defined to be a maximal isotropic vector subspace of ${\mathbb R}^4$.
Such a subspace is necessarily two-dimensional. We denote by $\overline{\mathbb M}{}^3$ the space of all 
Lagrangian subspaces of ${\mathbb R}^4$. The group $\sSp(2,{\mathbb R})$ proves to act transitively on 
the compact space  $\overline{\mathbb M}{}^3$
 which can be realised as
\bea
\overline{\mathbb M}{}^3 =\sU (2) / \sf{O} (2)~, 
\eea
 see, e.g., \cite{Berndt} for technical details. 
This homogeneous  space with the transformation group  $\sSp(2,{\mathbb R})$
is called the {\it compactified Minkowski space}.

Let $\cL \in \overline{\mathbb M}{}^3$ be a Lagrangian subspace.
It is generated by two linearly independent twistors $T^\m$, with $\m=1,2$,
such that
\be
\langle T^1| T^2 \rangle_J = 0~. 
\label{nullplane1}
\ee 
Obviously, the basis  chosen, $\{T^\m\}$, is defined only modulo 
the equivalence relation 
\be
\{ T^\m \} ~ \sim ~ \{ \tilde{T}^\m \} ~, \qquad
\tilde{T}^\m = T^\n\,R_\n{}^\m~, 
\qquad R \in \sGL(2,{\mathbb R}) ~.
\label{nullplane2}
\ee
Equivalently, we can think of 
the space $\overline{\mathbb M}{}^3$  as consisting of rank-two
$4\times 2$  real matrices 
\bea
( T^1 ~T^2 )=\left(
\begin{array}{c}
 F\\  G
\end{array}
\right) ~, \qquad
F^{\rm T} \,G =G^{\rm T} \,F~,
\label{two-plane}
\eea
where the $2\times 2$ matrices  $F$ and $G$ are
defined modulo the equivalence relation 
\bea
\left(
\begin{array}{c}
 F\\  G
\end{array}
\right) ~ \sim ~
\left(
\begin{array}{c}
 F\, R\\  G\,R
\end{array}
\right) ~, \qquad R \in \sGL(2,{\mathbb R}) ~.
\label{ER1}
\eea

An open dense subset ${\mathbb M}{}^3$ of  $\overline{\mathbb M}{}^3$ consists of
 those Lagrangian subspaces which are described by $4\times 2$ matrices of the form:
\bea
\left(
\begin{array}{c}
 F\\  G
\end{array}
\right) 
~, \qquad \det F \neq 0~.
\eea
In accordance with the equivalence relation (\ref{ER1}), we then have 
\bea
\left(
\begin{array}{c}
 F\\  G
\end{array}
\right) ~ \sim ~
\left(
\begin{array}{r}
 {\mathbbm 1}_2\\  -x
\end{array}
\right)  ~, \qquad 
x^{\rm T} =x \in \sMat (2, {\mathbb R} )~.
\eea
The subset  ${\mathbb M}{}^3$ can naturally be identified with Minkowski space 
${\mathbb R}^{2,1}$, as demonstrated in Appendix B. In what follows, we will not distinguish 
between  ${\mathbb M}{}^3$ and ${\mathbb R}^{2,1}$.

\subsection{The conformal infinity of Minkowski space}
Let us analyse the structure of the boundary of Minkowski space in $\overline{\mathbb M}{}^3$, that is
\bea
\pa \,{\mathbb M}^3 = \overline{\mathbb M}{}^3 \setminus {\mathbb M}^3~.
\label{2.11}
\eea
For any point from $\pa \,{\mathbb M}^3 $, the $2\times 2$ block $F$ is singular, 
$\det F =0$. Because of the equivalence relation (\ref{ER1}), we can always choose
\bea
F =\left(
\begin{array}{cc}
 f_1  & 0\\
f_2 &    0
\end{array} \right) ~,
\eea
for some two-vector $\vec f \in {\mathbb R}^2$.
There are two different cases to consider: (i) $\vec f =0$; and (ii) $\vec f \neq 0$.
In the case that $\vec f =0$, we have $\det G \neq 0$, and thus
for the corresponding Lagrangian subspace 
\bea
\left(
\begin{array}{c}
 0\\  G
\end{array}
\right) ~ \sim ~
\left(
\begin{array}{c}
0\\   {\mathbbm 1}_2
\end{array}
\right)  ~.
\eea
As a result, the case (i)  leads to a single point in $\pa \,{\mathbb M}^3 $.
The case (ii) is more interesting. Making use of the equivalence relation (\ref{ER1}), we can normalise
$\vec f$ by
\be
\vec f \cdot \vec f = 1~.
\ee 
We still can change the sign of the vector, $\vec f \to -\vec f$, which amounts to residual 
${\mathbb Z}_2$-freedom in the choice of $\vec f$.
For the matrix 
\bea
G =\left(
\begin{array}{cc}
 u_1  & g_1\\
u_2 &    g_2
\end{array} \right) ~, \qquad \vec g \neq 0 ~,
\eea
the isotropy condition (\ref{two-plane}) gives
\be
\vec f \cdot \vec g = 0~.
\ee
Making use of the equivalence relation (\ref{ER1}), we can impose the normalisation condition 
\be \vec g \cdot \vec g =1~.
\ee
Moreover, the same equivalence relation still allows us to choose the direction of $\vec g $ such that 
the set $\{\vec f , \vec g \}$ is an orthonormal basis of standard orientation in ${\mathbb R}^2$.
In other words, the vector $\vec g $ is uniquely determined by the choice of $\vec f$. 
With the vectors  $\{\vec f , \vec g \}$ having been fixed, the freedom  (\ref{ER1}) still allows to perform
transformations 
\bea 
\vec u \to \vec u + c\, \vec g ~, \qquad c \in {\mathbb R}~.
\eea
Thus, we can make $\vec u$ to be a multiple of $\vec f$.
We conclude that any Lagrangian subspace from the boundary (\ref{2.11}) looks like
\bea
\left(
\begin{array}{c}
 F\\  G
\end{array}
\right) ~ \sim ~
\left(
\begin{array}{cc}
f_1 ~&~ 0\\
f_2 &0\\
\hline
 \l f_1  ~& ~g_1\\
\l f_2 &    g_2
\end{array} \right) ~, \qquad \l \in {\mathbb R}~,
\label{2.19}
\eea
where the two-vectors $\vec f$ and $\vec g$ form a standard orthonormal frame in ${\mathbb R}^2$.
The space of two-frames  $\{\vec f , \vec g \}$ is topologically $S^1$. 
On the other hand, in the limit  $\l \to \pm \infty$  the Lagrangian subspace (\ref{2.19}) can be seen to turn into 
that defined in the case (i) above. as a result, we see that the boundary (\ref{2.11}) is topologically 
\be
\overline{\mathbb M}{}^3 \setminus {\mathbb M}^3 = (S^1 \times S^1)/{\mathbb Z}_2~. 
\ee

\subsection{The conformal algebra}
This subsection describes a useful matrix realisation of the conformal algebra
$\mathfrak{sp}(2, {\mathbb R}) \cong  \mathfrak{so}(3,2)$. In what follows we will use
the following notation: 
\bea
\g_a := (\g_a)_\a{}^\b ~, \qquad 
\hat{\g}_a :=  (\g_a)^{\a \b} =\hat{\g}_a^{\rm T}~,
 \qquad \check{\g}_a := (\g_a)_{\a \b}  = \check{\g}_a^{\rm T}~.
\eea
Defining also  $\ve =(\ve^{\a\b})$,  the gamma-matrices are characterised by the property
\bea
\g_a^{\rm T} = -\ve \, \g_a \, \ve^{-1}~.
\eea

Introduce $4 \times 4 $ matrices $\G_A$, with $A=a,3,4$, of the form:
\bea
\G_a =  \left(
\begin{array}{cc}
\g_a ~&~ 0 \\
0 ~& ~(\g_a)^{\rm T} 
\end{array} \right) ~, \qquad 
\G_3 =  \left(
\begin{array}{cc}
0~&~ \ve^{-1} \\
\ve ~& ~ 0 
\end{array} \right) ~, \qquad
\G_4 =  \left(
\begin{array}{cc}
0~&~ -\ve^{-1} \\
\ve ~& ~ 0 
\end{array} \right) ~.
\eea
They obey the anti-commutation relations 
\bea
\{ \G_A , \G_B \} = 2\eta_{AB} {\mathbbm 1}_4~, \qquad 
\eta_{AB}= {\rm diag} \, (-+++-)~.
\label{3.2}
\eea
Clearly, the matrices $\G_A$ constitute  a Majorana representation of the gamma-matrices
for pseudo-Euclidean space ${\mathbb R}^{3,2}$. One easily checks that 
\bea
\G_A^{\rm T} = J \, \G_A \, J^{-1}~,
\label{3.3}
\eea
with $J = \G_0 \G_4$ the symplectic matrix (\ref{J}).

As follows from (\ref{3.2}),  the matrices 
\bea
\S_{AB}:= \frac{1}{4} \big[ \G_A, \G_B\big] 
\eea
form the spinor representation of $\mathfrak{so}(3,2)$. It also follows from 
(\ref{3.3}) that $\S_{AB}$ obey the relations 
\bea
\S_{AB}^{\rm T} \, J +J \, \S_{AB}=0~,
\eea
and therefore they form a basis of  $\mathfrak{sp}(2, {\mathbb R})$. That is, for any 
$\o \in  \mathfrak{sp}(2, {\mathbb R})$, such that $\o^{\rm T} \, J +J \, \o=0$, we have 
\bea
\o = \hf \o^{AB} \S_{AB}~, \qquad \o^{AB}=-\o^{BA}\in {\mathbb R}~.
\eea
The above consideration immediately leads to the famous result 
$\mathfrak{sp}(2, {\mathbb R}) \cong  \mathfrak{so}(3,2)$.

Given an arbitrary element  $\o \in  \mathfrak{sp}(2, {\mathbb R}) $, it can be represented as
\bea
 \o&=& \left(
\begin{array}{c | c}
  \l_\a{}^\b -\hf f \d_\a{}^\b ~& ~ b_{\ab}   \\
\hline
-a^{\a \b}  
\phantom{\Big(}
&  -  \l^\a{}_\b +\hf f \d^\a{}_\b \\
\end{array}
\right) 
\equiv \left(
\begin{array}{c | c}
  \l -\hf  f {\mathbbm 1}_2 ~& ~ \check{b}   \\
\hline
-\hat{a}
\phantom{\Big(}
&  -  \l^{\rm T}  +\hf f {\mathbbm 1}_2 \\
\end{array}
\right)  \in  \mathfrak{sp}(2, {\mathbb R})~,  
\non \\ 
&& \l_\a{}^\a =0~, \qquad {a}^{\a\b} = {a}^{\b \a} ~, \qquad {b}_{\a \b} = {b}_{\b \a}~.
\eea
Associated with $J $ is the following automorphism of the conformal algebra:
\bea
J \left(
\begin{array}{c | c}
  \l -\hf  f {\mathbbm 1}_2 ~& ~ \check{b}   \\
\hline
-\hat{a}
\phantom{\Big(}
&  -  \l^{\rm T}  +\hf f {\mathbbm 1}_2 \\
\end{array}
\right) J^{-1} = 
\left(
\begin{array}{c | c}
-  \l^{\rm T}  +\hf f {\mathbbm 1}_2 
  ~& ~  \hat{a} \\
\hline
- \check{b}
\phantom{\Big(}
& 
  \l -\hf  f {\mathbbm 1}_2 
\\
\end{array}
\right) ~.
\eea

\subsection{Compactified Minkowski space \`a la Dirac}
Let $\cL $ be a Lagrangian subspace of ${\mathbb R}^4$
generated by two linearly independent twistors $T^1$ and $T^2$.
It follows from (\ref{3.3}) that 
\bea
\langle T^\m| \G_A|T^\n \rangle_J =- \langle T^\n | \G_A|T^\m \rangle_J ~.
\eea
Applying the equivalence transformation (\ref{nullplane2}) to $\langle T^\m| \G_A|T^\n \rangle_J$
gives 
\bea
\langle \tilde{T}^\m| \G_A|\tilde{T}^\n \rangle_J = \det R\, \langle T^\m| \G_A|T^\n \rangle_J ~.
\eea
As follows from  the identity 
\bea
(J\, \G^A)^{\hat \a \hat \b} (J\, \G_A)^{\hat \g \hat \d} 
= - J^{\hat \a \hat \b} J^{\hat \g \hat \d} 
+ 2 (J^{\hat \a \hat \g} J^{\hat \b \hat \d} - J^{\hat \a \hat \d} J^{\hat \b \hat \g} )~,
\eea 
the five-vector 
\bea 
X^A := \langle T^1 | \G^A | T^2 \rangle_J 
\label{3.11}
\eea
belongs to the cone $\cC$ in $ {\mathbb R}^{3,2}$ defined by 
\be
\eta_{AB } X^A X^B =0~.
\label{3.12}
\ee
It is an instructive exercise to demonstrate  that any null vector $X^A \in \cC$ can be represented in the form
(\ref{3.11}), for some null twistors $T^1$ and $T^2$ which generate a Lagrangian subspace of  ${\mathbb R}^4$.

It follows from the above consideration that the five-vector (\ref{3.11}) 
is defined modulo the equivalence relation 
\bea
X^A ~\sim ~ \l \, X^A ~, \qquad \l \in {\mathbb R} -\{ 0 \}
\eea
which identifies all points on a straight line in $ {\mathbb R}^{3,2}$.
The space of all straight lines through the origin of the cone $\cC$
 in $ {\mathbb R}^{3,2}$, eq. (\ref{3.12}), is known as Dirac's conformal space\footnote{In \cite{Dirac} Dirac
 credited Oswald Veblen with 
 the general theory of conformal space.} 
 \cite{Dirac}  (although it had been introduced a decade earlier by Weyl \cite{Weyl})
 which is topologically $(S^2 \times S^1)/{\mathbb Z}_2$.
We conclude that $\overline{\mathbb M}{}^3$ can be identified with  Dirac's conformal space, 
\be
\overline{\mathbb M}{}^3 = (S^2 \times S^1)/{\mathbb Z}_2~.
\ee

\section{Compactified  Minkowski superspace}
\setcounter{equation}{0}
In this section we generalise the construction of 3D compactified Minkowski space 
to $\cN$-extended superspace.  Our presentation will be  similar in spirit  to that given
in \cite{K-compactified} for the case of four space-time dimensions. In its turn, 
the work of \cite{K-compactified} built on Manin's approach to the 4D $\cN$-extended superconformal
symmetry \cite{Manin} in conjunction with  Ferber's concept of supertwistors \cite{Ferber}.

\subsection{The superconformal group}
Associated with the symplectic super-metric 
\bea
{\mathbb J} =  \left(
\begin{array}{cc}
J ~&~ 0 \\
0 ~& ~{\rm i} \,{\mathbbm 1}_\cN 
\end{array} \right) 
\label{supermetric}
\eea
is the symmetric purely imaginary quadratic form
\bea
{\bm \S}^{\rm sT}\, {\mathbb J}\, {\bm \S} =\z^{\rm T}  J  \z + {\rm i}\, y^{\rm T} y ~, 
\label{4.2}
\eea
which is defined on the superspace  ${\mathbb R}^{\cN |4}$
parametrised by 4 anticommuting real variables $\z$ and 
$\cN$ commuting real variables $y$,
\bea
{\bm \S} = \left(
\begin{array}{c}
 \z \\
 y 
\end{array}
\right)~, \qquad {\bm \S}^{\rm sT} =\big( \z^{\rm T}~, ~y^{\rm T}\big) = \S^{\rm T}~, 
\qquad \e (\z) =1~, \quad \e(y) =0~.
\label{4.3}
\eea
Here and in what follows $\e (s) $ denotes the Grassmann parity of a supernumber $s$.
Any element $\S \in {\mathbb R}^{\cN |4}$ of the form (\ref{4.3}) will be called an {\it odd real supertwistor}.
Given a linear transformation
\bea
z \to z' = g \, z~, \qquad
g=\left(
\begin{array}{c||c}
 A  & B\\
 \hline \hline
C &    D 
\end{array}
\right) ~,
\label{3.4}
\eea
it leaves 
the quadratic form (\ref{4.2}) invariant if the supermatrix $g$ obeys the equation
\bea
g^{\rm sT} {\mathbb J}\, g = {\mathbb J} ~, \qquad 
g^{\rm sT}=\left(
\begin{array}{c||r}
 A^{\rm T}  & -C^{\rm T}\\
\hline \hline
B^{\rm T}  &    D^{\rm T} 
\end{array}
\right) ~.
\eea
Such supermatrices form the supergroup $\sOSp(\cN|2, {\mathbb R})$ which is the 
$\cN$-extended superconformal group in three space-time dimensions.
The even matrices $A, D$ and the odd matrix $B$ in (\ref{3.4}) have real matrix elements, 
while the odd matrix $C$ has purely imaginary matrix elements. 
Supermatrices $g$ of this type will be called 
{\it real}.

The quadratic form (\ref{4.2}) can naturally be extended to the symmetric inner product 
on ${\mathbb R}^{\cN |4}$ defined by 
\bea
\langle {\bm \S}| {\bm \X} \rangle_{\mathbb J}: = {\bm \S}^{\rm sT}{\mathbb J} \, {\bm \X}
= \langle {\bm \X}| {\bm \S} \rangle_{\mathbb J}~, 
\eea
with $\bm \S$ and $\bm \X$ being arbitrary odd supertwistors.
This inner product is obviously invariant under the action of $\sOSp(\cN|2, {\mathbb R})$.

The superconformal algebra $\mathfrak{osp}(\cN|2, {\mathbb R})$ consists of real supermatrices 
obeying the master equation
\bea
{\bm \o}^{\rm sT} {\mathbb J} + {\mathbb J} \,{\bm \o}=0~.
\eea
The general solution of this equation is
\bea
{\bm \o} 
&=& \left(
\begin{array}{c | c ||c}
  \l -\hf f {\mathbbm 1}_2  ~& ~ \check{b} &~\sqrt{2} \eta^{\rm T}   \\
\hline 
-\hat{a}  ~&   -  \l^{\rm T} +\hf f {\mathbbm 1}_2~&~ -\sqrt{2}\e^{\rm T}
\\
\hline
\hline
{\rm i}\sqrt{2}\,\e ~& ~{\rm i}\sqrt{2}\, \eta &~r
\end{array}
\right) \non \\
&\equiv& \left(
\begin{array}{c | c ||c}
  \l_\a{}^\b -\hf f \d_\a{}^\b  ~& ~ b_{\a \b} &~\sqrt{2} \eta_{\a J}   \\
\hline 
-a^{\a \b}  ~&   -  \l^\a{}_\b +\hf f \d^\a{}_\b~&~ -\sqrt{2}\e^\a{}_J
\\
\hline
\hline
{\rm i}\sqrt{2}\,{\e_I{}^\b} ~& ~{\rm i}\sqrt{2}\, \eta_{I\b}&~r_{IJ}
\end{array}
\right) ~, \qquad I,J =1,\dots, \cN
\label{SP-g} \\
&&  \l_\a{}^\a =0~, \qquad {a}^{\a\b} = {a}^{\b \a} ~, \qquad {b}_{\a \b} = {b}_{\b \a}~, \qquad
r_{IJ}=-r_{JI}~.
\non
\eea
The bosonic  parameters $\l_\a{}^\b$, $f$, $a_{\a\b}$, $b^{\a\b}$ and $r_{IJ}$, as well as  
the fermionic parameters $\e^\a{}_I\equiv \e_I{}^\a $ and $\eta_{\a I} \equiv \eta_{I\a}$
in (\ref{SP-g}) are real.

\subsection{Compactified  Minkowski superspace}
The superconformal group $\sOSp(\cN|2, {\mathbb R})$ naturally acts on ${\mathbb R}^{4|\cN}$
parametrised by elements of the form:
\bea
{\bm S} = \left(
\begin{array}{c}
 f_\a \\
  g^\a \\
  {\rm i} \, \vf_I
  \end{array}
\right)~, \qquad 
\e (f_\a) =\e(g^\a) =0~, 
\quad \e (\vf_I) =1~,  
\eea
with the components $f_\a$ and $g^\a$ being real commuting, and
$\vf_I$ real anticommuting.
 This action preserves the inner product on ${\mathbb R}^{4|\cN}$ defined by 
\bea
\langle {\bm S}| {\bm T} \rangle_{\mathbb J}: = {\bm S}^{\rm sT}{\mathbb J} \, {\bm T}
=- \langle {\bm T}| {\bm S} \rangle_{\mathbb J}
~,  \qquad
{\bm S}^{\rm sT} = \big( f_\a\,,   ~g^\a\,, -   ~{\rm i} \,  \vf_I \big)~, 
\eea
with the super-metric $\mathbb J$ defined in (\ref{supermetric}).
Any element ${\bm S} \in {\mathbb R}^{4|\cN}$ 
is called an {\it even real supertwistor}. 

By analogy with the non-supersymmetric case, we define 
a {\it Lagrangian subspace} of ${\mathbb R}^{4|\cN}$ to be its maximal isotropic subspace.
We denote by $\overline{\mathbb M}{}^{3|2\cN}$ the space of all 
Lagrangian subspaces of ${\mathbb R}^{4|\cN}$. 
Given such a subspace, it is generated by two  supertwistors 
${\bm T}^\m$ such that (i) the bodies of ${\bm T}^1$ and $ {\bm T}^2$ are linearly independent;
(ii) they obey the null condition
\be
\langle {\bm T}^1 | {\bm T}^2  \rangle_{\mathbb J}=0~;
\label{5.3}
\ee
(iii) they are defined only modulo 
the equivalence relation 
\be
\{ {\bm T}^\m \} ~ \sim ~ \{ \tilde{\bm T}^\m \} ~, \qquad
\tilde{\bm T}^\m = {\bm T}^\n\,R_\n{}^\m~, 
\qquad R \in \sGL(2,{\mathbb R}) ~.
\label{super-nullplane2}
\ee
Equivalently, 
the space   $\overline{\mathbb M}{}^{3|2\cN}$ consists of 
rank-two supermatrices of the form 
\bea
\big( {\bm T}^1 ~ {\bm T}^2 \big)=\left(
\begin{array}{c}
{ F}\\ { G} \\  
{\rm i}\, \U
\end{array}
\right) ~, \qquad
G^{\rm T}F =F^{\rm T}G + {\rm i}\,\U^{\rm T} \,\U~,
\label{super-two-plane}
\eea
which are defined modulo the equivalence relation 
\bea
\left(
\begin{array}{c}
 F\\  G \\ {\rm i}\, \U
\end{array}
\right) ~ \sim ~
\left(
\begin{array}{r}
 F\, R\\  G\,R \\ {\rm i}\,\U\,R
\end{array}
\right) ~, \qquad R \in \sGL(2,{\mathbb R}) ~.
\label{5.13}
\eea
Here $F$ and $G$ are $2\times 2$ 
real bosonic matrices, and $\U$ is a $\cN \times 2$ 
real fermionic matrix.

A dense open subset ${\mathbb M}{}^{3|2\cN}$ of  $\overline{\mathbb M}{}^{3|2\cN}$ consists of
those Lagrangian subspaces which are described by supermatrices (\ref{super-two-plane}) 
under the condition
\bea
\det F \neq 0
~.
\label{5.14}
\eea
Using the equivalence relation (\ref{5.13}), such a supermatrix can be brought 
to the following canonical form:
\bea
\big( {\bm T}^1 ~ {\bm T}^2 \big)=
\left(
\begin{array}{c}
 F\\  G \\ {\rm i}\, \U
\end{array}
\right)  \sim 
\left(
\begin{array}{c}
 \d_\a{}^\b \\  -x^{\a\b} +\frac{\rm i}{2}\ve^{\a\b} \q^2  \\ {\rm i}\sqrt{2}\,\q_I{}^\b
\end{array}
\right) ~, \qquad x^{\a\b} =x^{\b\a}~,
\quad \q^2:= \q^\a_I \q_{\a I} ~.~~~
\label{5.8}
\eea
Here the bosonic $x^{\a\b}$ and fermionic $\q_I^\a \equiv \q_I{}^\a$ parameters are real.
Therefore, the subset  ${\mathbb M}{}^{3|2\cN} \subset \overline{\mathbb M}{}^{3|2\cN}$ defined by eq. (\ref{5.14}) 
can be identified with  ${\mathbb R}^{3|2\cN}$.

Given a group element $g \in  \sOSp(\cN|2, {\mathbb R})$, its  (local) action 
on the points  of ${\mathbb M}{}^{3|2\cN}$ can be derived  by the rule
\bea
g \,
\left(
\begin{array}{c}
{\mathbbm 1}_2 \\  -{x}_{(-)} \\ {\rm i}\sqrt{2}\,\q
\end{array}
\right) 
= \left(
\begin{array}{c}
{\mathbbm 1}_2 \\ -{x}'_{(-)} \\ {\rm i}\sqrt{2}\,\q'
\end{array}
\right) R(g;x, \q) ~, \qquad  R(g;x,\q) \in \sGL(2,{\mathbb R}) 
\label{5.9}
\eea
where we have denoted
\bea
x_{(-)}:= \hat{x} -\frac{\rm i}{2}\ve \,\q^2 =\big( x^{\a\b} -\frac{\rm i}{2}\ve^{\a\b} \q^2 \big)~,
\qquad \q :=  \big( \q_I{}^\b \big)~.
\eea
In the case of an infinitesimal superconformal transformation  (\ref{SP-g}), from (\ref{5.9})
we derive 
\begin{subequations} 
\bea
\d x_{(-)} &=& \hat{a} - \l^{\rm T} x_{(-)} - x_{(-)} \l +f \,x_{(-)} 
+x_{(-)}\check{b} \,x_{(-)} + 2{\rm i}\, \e^{\rm T} \q - 2{\rm i} \,x_{(-)} \eta^{\rm T}\q~, \\
\d \q &=& \e -\q\, \l +\hf f \,\q +r \, \q + \q \,\check{b}\,x_{(-)} -2{\rm i}\, \q\, \eta^{\rm T} \q~.
\eea
\end{subequations}
These relations can be rewritten as 
\bea
\d x^m = \x^m +{\rm i}\, \x^\a_I (\g^m)_{\a\b} \q^\b_I~, \qquad
\d \q^\a_I = \x^\a_I~, 
\eea
where $\x^m (x,\q)$ and $\x^\a_I (x,\q)$ are the components of a superconformal 
Killing vector field.  These objects will be studied in section \ref{Superconformal Killing vectors}. 

\section{Compactified  harmonic/projective superspaces}
\setcounter{equation}{0}
\label{CompHPS}

We wish to construct new homogeneous spaces, $\overline{\mathbb M}{}^{3|2\cN} \times {\mathbb X}^\cN_m$, 
of the superconformal group $\sOSp(\cN|2, {\mathbb R})$
that include  $\overline{\mathbb M}{}^{3|2\cN}$ as a submanifold, for any integer
$m\leq [\cN/2]$\footnote{As usual, we denote by $[\cN/2]$
the integer part of $\cN/2$.}.   
Our approach will be analogous to the four-dimensional construction of \cite{K-compactified}
which built on earlier works \cite{Rosly2,LN,HH}. 

\subsection{Projective realisation}
Along with the two linearly independent even real supertwistors ${\bm T}^1 $ and $ {\bm T}^2$ 
under the null condition (\ref{5.3}), we also consider $m$ odd {\it complex} supertwistors
${\bm \S}^{\iu}$, with ${\iu}=1,\dots, m$, such that  (i) the bodies of ${\bm \S}^{\iu}$ are linearly independent; 
(ii) any linear combination of the supertwistors  ${\bm T}^\m $ and ${\bm \S}^{\iu}$ is null, that is
\bea
\langle {\bm T}^\m | {\bm T}^\n  \rangle_{\mathbb J}= \langle {\bm T}^\m | {\bm \S}^{\ju}  \rangle_{\mathbb J}=
\langle {\bm \S}^{\iu} | {\bm \S}^{\ju}  \rangle_{\mathbb J}=0~.
\label{6.1}
\eea
The supertwistors  ${\bm T}^\m $ and ${\bm \S}^{\iu}$ are assumed to be 
defined modulo the equivalence  relation
\bea
({\bm T}^\m ,  {\bm \S}^{\iu}) \sim   ( {\bm T}^\n  , {\bm \S}^{\ju} )\,
\left(
\begin{array}{c|c}
 R_\n{}^\m  &B_\n{}^{\iu}  \\ \hline \hline 
0 &D_{\ju}{}^{\iu}
\end{array}
\right) ~,\quad 
\left(
\begin{array}{c|c}
 R  &B  \\  \hline \hline
 0 & D
\end{array}
\right) \in \sGL(2|m, {\mathbb C})~, \quad  R \in \sGL(2,{\mathbb R})~.~~
\label{6.2}
\eea
We emphasise that both the fermionic $ B_\n{}^{\iu} $ and bosonic $ D_{\ju}{}^{\iu}$ matrix elements  are  complex.
The space $\overline{\mathbb M}{}^{3|2\cN} \times {\mathbb X}^\cN_m$ is defined to consist 
of the equivalence classes associated with all possible  $({\bm T}^\m ,  {\bm \S}^{\iu}) $ 
under the above conditions.

There are several important comments to be made.
Firstly, 
the invariant inner product $\langle ~, ~ \rangle_{\mathbb J}$ 
possesses the following symmetry property 
\bea
\langle {\bm \cT}_1 | {\bm \cT}_2  \rangle_{\mathbb J} 
= -(-1)^{\e_1 \e_2} \langle {\bm \cT}_2 | {\bm \cT}_1  \rangle_{\mathbb J} ~,
\eea
where $\e_i$ denotes the Grassmann parity of ${\bm \cT}_i$.
Secondly, associated with the odd supertwistors  ${\bm \S}^{\iu}$ are their complex conjugates
 $\bar{\bm \S}^{\iu}$ which possess analogous properties 
 \bea
\langle {\bm T}^\m | \bar{\bm \S}^{\ju}  \rangle_{\mathbb J}=
\langle \bar{\bm \S}^{\iu} | \bar{\bm \S}^{\ju}  \rangle_{\mathbb J}=0~.
\eea
One can also see that the $2m$ supertwistors  ${\bm \S}^{\iu}$ and $\bar{\bm \S}^{\ju}$
are linearly independent,
\bea
\det \,\langle {\bm \S}^{\iu} | \bar{\bm \S}^{\ju}  \rangle_{\mathbb J}\neq 0~.
\eea

We are mostly interested in the dense open subset ${\mathbb M}{}^{3|2\cN} \times {\mathbb X}^\cN_m$ 
of $\overline{\mathbb M}{}^{3|2\cN} \times {\mathbb X}^\cN_m$. For its points, 
the even supertwistors ${\bm T}^\m$ can be chosen as in (\ref{5.8}). 
Making use of the null conditions (\ref{6.1}) and the equivalence relation (\ref{6.2}), 
it is not difficult to show that the points of  ${\mathbb M}{}^{3|2\cN} \times {\mathbb X}^\cN_m$ can 
be parametrised by supermatrices of the form:
\bea
\big( {\bm T}^1~ {\bm T}^2 \big) \sim 
\left(
\begin{array}{c}
 \d_\a{}^\b \\  -x^{\a\b} +\frac{\rm i}{2}\ve^{\a\b} \q^2  \\ {\rm i}\sqrt{2}\,\q_I{}^\b
\end{array}
\right) ~, \qquad 
\big( {\bm \S}^1\cdots  {\bm \S}^m \big) \sim 
\left(
\begin{array}{c}
 0 \\  - \sqrt{2}\q_K^\a Z_K{}^{\ju} \\ 
 Z_I{}^{\ju}
\end{array}
\right) ~.
\label{Projective-MP}
\eea
Here the $m$ complex vectors $Z^{\ju} =(Z_I{}^{\ju}) \in {\mathbb C}^\cN -\{ 0\} $ are required to be linearly 
independent and subject to the null conditions
\bea
Z^{\ju} \cdot Z^{\ku} :=Z_I{}^{\ju} Z_I{}^{\ku} =0~, \qquad \forall {\ju}, {\ku} = 1, \dots, m
\label{6.4}
\eea
and are defined modulo the equivalence relation 
\bea
Z_I{}^{\ju} ~\sim ~Z_I{}^{\ku} \, D_{\ku}{}^{\ju} ~, \qquad D =(D_{\ku}{}^{\ju} ) \in \sGL (m, {\mathbb C})~.
\label{6.5}
\eea
The complex $\cN$-vectors $Z^{\ju}$ can be 
represented as a superposition of 
their real and imaginary parts, 
$Z^{\ju}=  \frac{1}{\sqrt 2}(X^{\ju}+{\rm i}\, Y^{\ju})$.
Then, the null conditions (\ref{6.4}) take the form:
\bea
X^{\ju} \cdot X^{\ku} = Y^{\ju} \cdot Y^{\ku} ~, \qquad
X^{\ju} \cdot Y^{\ku} = -X^{\ku} \cdot Y^{\ju}~, 
\qquad  \forall {\ju}, {\ku} = 1, \dots, m~.
\eea

\subsection{Harmonic realisation}
We would like to  describe an alternative 
realisation of the space $\overline{\mathbb M}{}^{3|2\cN} \times {\mathbb X}^\cN_m$. 
It differs from  the projective realisation considered above, but is equivalent to it.
Along with the two linearly independent even real supertwistors ${\bm T}^1 $ and $ {\bm T}^2$ 
and the $m$  odd {\it complex} supertwistors ${\bm \S}^{\iu}$ under the null conditions (\ref{6.1}), 
we also consider $\cN- m \geq m$ odd {\it complex} supertwistors ${\bm \U}^{\Iu}$, 
with ${\Iu}=1,\dots, \cN-m $, such that  
(i) the bodies of ${\bm \S}^{\iu}$ and ${\bm \U}^{\Iu}$ form a basis for ${\mathbb C}^\cN$;
(ii) the odd supertwistors ${\bm \U}^{\Iu}$ are orthogonal to the even supertwisors,
\bea
\langle {\bm T}^\m | {\bm \U}^{\Ju}  \rangle_{\mathbb J}=0~.
\eea
The set of  $2+\cN$ supertwistors  ${\bm T}^\m $,
${\bm \S}^{\iu}$ and ${\bm \U}^{\Iu}$
is assumed to be 
defined modulo the equivalence  relation:
\begin{subequations}
\bea
({\bm T}^\m ,  {\bm \S}^{\iu} ,  {\bm \U}^{\Iu}) \sim   ( {\bm T}^\n  , {\bm \S}^{\ju},  {\bm \U}^{\Ju} )\,
\left(
\begin{array}{c|cc}
 R_\n{}^\m  &B_\n{}^{\iu} & C_\n{}^{\Iu}  \\  \hline \hline
0 &D_{\ju}{}^{\iu} &E_{\ju}{}^{\Iu} \\
0 & 0& F_{\Ju}{}^{\Iu}
\end{array}
\right) ~.
\label{Her1}
\eea
Here the $(2+\cN)\times (2+\cN)$ supermatrix on the right is chosen such that 
\bea
\left(
\begin{array}{c|cc}
 R  &B  &C\\  \hline \hline
 0 & D&E\\
 0&0& F
\end{array}
\right) \in \sGL(2|\cN, {\mathbb C})~, 
\qquad
 R \in \sGL(2,{\mathbb R})
 \label{Her2}
\eea
\end{subequations}
but otherwise it is arbitrary.
It is not difficult to show that the space $\overline{\mathbb M}{}^{3|2\cN} \times {\mathbb X}^\cN_m$
can be identified with the space of   
equivalence classes associated with all possible triplets $({\bm T}^\m ,  {\bm \S}^{\iu},  {\bm \U}^{\Iu})$ 
under the above conditions.

We consider the Minkowski patch and choose the isotwistors ${\bm T}^\m  $ and   ${\bm \S}^{\iu}$ 
as in (\ref{Projective-MP}). The equivalence relation (\ref{Her1}) allows us to choose
the $\cN-m$ odd supertwistors ${\bm \U}^{\Iu}$ in the form:
\bea
\big( {\bm \U}^1\cdots  {\bm \U}^{\cN-m} \big) \sim 
\left(
\begin{array}{c}
 0 \\  - \sqrt{2}\q_K^\a W_K{}^{\Ju} \\ 
 W_I{}^{\Ju}
\end{array}
\right) ~,
\label{Harmonic-MP}
\eea
for some complex $\cN$-vectors $W^{\Ju}=(W_I{}^{\Ju})$. By construction, 
the set of $m$ vectors $Z^{\ju}$ in  (\ref{Projective-MP}) and the set of $\cN-m$ vectors $W^{\Ju}$
must form a basis for ${\mathbb C}^\cN$. The former is defined modulo the equivalence 
relation (\ref{6.5}). Similarly, the latter is defined modulo the following equivalence relation
\bea
W_I{}^{\Ju} ~\sim ~ Z_I{}^{\ku} \,E_{\ku}{}^{\Ju}
+W_I{}^{\Ku} \, F_{\Ku}{}^{\Ju} ~, \quad E_{\ku}{}^{\Ju} \in {\mathbb C}~, \quad 
\quad (F_{\Ku}{}^{\Ju} ) \in \sGL (\cN-m, {\mathbb C})~.
 \label{Her3}
\eea
On an 
open subset of the space (\ref{6.4}),  
the equivalence relation (\ref{6.5}) can be used to choose 
\bea
( Z_I{}^{\ju}) = \left( \begin{array}{c}
 {\mathbbm 1}_m  \\  \z 
\end{array} 
\right)~,
\eea
where $\z$ is a complex $(\cN-m) \times m$ matrix obeying the equation 
\be
\z^{\rm T} \z = - {\mathbbm 1}_{\cN -m}~.
\ee
After that, the $E$-gauge freedom (\ref{Her3}) allows us to choose
\bea
( W_I{}^{\Ju}) = \left( \begin{array}{c}
0  \\  \o 
\end{array} 
\right)~, \qquad   \o \in \sGL (\cN-m, {\mathbb C})~.
\eea
Finally, the $F$-gauge freedom (\ref{Her3}) allows us to choose
\bea
( W_I{}^{\Ju}) = \left( \begin{array}{c}
0  \\  {\mathbbm 1}_{\cN-m} 
\end{array} 
\right)~.
\eea
We see that the $\cN-m$ odd supertwistors ${\bm \U}^{\Iu}$  do not add any new degrees of freedom.

Up to an equivalence transformation, we can always choose $W^{\Iu}$ to consist of two subsets
$W^{\Iu} =\{ {\bar Z}^{\iu}, R^{\underline \a} \}$, where ${\bar Z}^{\iu}$ is the complex conjugate of 
${ Z}^{\iu}$, while $2m$ vectors $ R^{\underline \a}$ are real and orthogonal to ${ Z}^{\iu}$
and $\bar{ Z}^{\iu}$,
\bea
{ Z}^{\iu} \cdot R^{\underline \a}= \bar{ Z}^{\iu} \cdot R^{\underline \a}=0~, \qquad \iu= 1, \dots, m~, 
\quad \underline{\a}=1,\dots \cN-2m~.
\eea
Using the freedom to perform equivalence transformations  (\ref{6.5}), 
we can bring the Hermitian nonsingular matrix 
$   \bar{ Z}^{\iu} \cdot { Z}^{\ju}$ to the form 
\bea
 \bar{ Z}^{\iu} \cdot { Z}^{\ju} =  \d^{\iu \ju}~.
 \label{ZbarZ}
\eea
If  the null $\cN$-vectors  $Z^{\iu}$ are represented as a sum 
of  their real and imaginary parts,
$Z^{\iu}= \frac{1}{\sqrt 2}( X^{\iu}+{\rm i}\, Y^{\iu})$, then eq. (\ref{ZbarZ})  is equivalent to 
\bea
X^{\iu} \cdot X^{\ju} = Y^{\iu} \cdot Y^{\ju} =  \d^{\iu \ju}~, \qquad
X^{\iu} \cdot Y^{\ju} =0
\qquad  \forall {\iu}, {\ju} = 1, \dots, m~.
\eea
Using the freedom to perform equivalence transformations  (\ref{6.5}), 
we can also choose the real vector $R^{\underline \a}$ to form an orthonormal set, 
\bea
R^{\underline \a} \cdot R^{\underline \b} = \d^{ \underline{\a} \underline \b}~.
\eea
Now, the $\cN$ vectors $\{ X^{\iu} , Y^{\ju} , R^{\underline \a} \} $ generate a group element 
$g:= (X_I{}^{\iu} , Y_I{}^{\ju} , R_I{}^{\underline \a} ) \in \sSO(\cN)$.

\subsection{Some special cases}
${}$For $\cN>2$ and $m=1$ the internal manifold  ${\mathbb X}^\cN_1$ proves to be a symmetric 
space, 
\bea
{\mathbb X}^\cN_1= \sSO (\cN) / \sSO (\cN-2) \times \sSO(2) ~, \qquad \cN>2~.
\eea
In the $\cN=2$ case, the  space ${\mathbb X}^2_1$ reduces to  just two points, 
$Z^{(+)}_I = (1/\sqrt{2})  (1, +{\rm i} )$ and $Z^{(-)}_I = (1/\sqrt{2})  (1, -{\rm i} )$, 
which can be seen to correspond to the chiral and antichiral subspaces of ${\mathbb M}{}^{3|4}$.

In the case $\cN=3$ and $m=1$, to solve the null condition $Z_I Z_I =0$, it is useful to replace 
the $\sSO(3)$  index of $Z_I$ by a pair  of isospinor ones, 
\bea
Z_I ~\to ~ Z_i{}^j :=\frac{\rm i}{\sqrt 2} (\vec{Z} \cdot \vec{\s})_i{}^j 
\equiv Z_I (\t_I)_i{}^j~, \qquad Z_i{}^i=0~, 
\label{N=3conversion}
\eea
with $\vec \s$ the Pauli matrices. Then, the null condition $Z_I Z_I =0$ is solved by
\bea
Z^{ij} = v^i\, v^j ~, \qquad v^i \in {\mathbb C}^2 \setminus\{0\}~.
\eea
The equivalence relation (\ref{6.5}) turns into 
\be
v^i ~ \to ~\l \,v^i ~, \qquad \l \in {\mathbb C}-\{ 0 \}~
\ee
and thus the internal space ${\mathbb X}^3_1$ is ${\mathbb C}P^1$.

In the case $\cN=4$ and $m=1$, to solve the null condition $Z_I Z_I =0$, it is useful to replace 
the $\sSO(4) \cong \big( \sSU_{\rm L}(2)\times \sSU_{\rm R} (2) \big) /{\mathbb Z}_2$  index of $Z_I$ by 
a pair  of different isospinor ones corresponding to the groups $\sSU_{\rm L}(2)$ and  $\sSU_{\rm R} (2)$, 
\bea
Z_I ~\to ~ Z_{ i}{}^{\bar k} :=\frac{\rm i}{\sqrt 2} (\vec{Z} \cdot \vec{\s})_{ i}{}^{\bar k}
+ \frac{1}{\sqrt 2}Z_4 \d_{ i}{}^{\bar k} \equiv Z_I (\t_I)_i{}^{\bar k}~.  
\label{N=4conversion}
\eea
Then, the null condition $Z_I Z_I =0$ is solved by 
\bea
 Z^{ i \bar k} =v_{\rm L}^{ i }  \, v_{\rm R}^{ \bar k} 
\eea
for two non-zero isospinors $v_{\rm L}^{ i } $ and $ v_{\rm R}^{ \bar k}$
transforming under  the groups   $\sSU_{\rm L}(2)$ and  $\sSU_{\rm R} (2)$ respectively.
The equivalence relation (\ref{6.5}) turns into 
\bea
v_{\rm L}^{ i }  ~\to~ \l \m \,v_{\rm L}^{ i }  ~, 
\qquad v_{\rm R}^{ \bar k} ~\to~ \frac{\l}{\m}\,v_{\rm R}^{\bar k}  ~, 
\qquad  \l, \m \in {\mathbb C}\setminus\{ 0 \}~.
\eea
We conclude that ${\mathbb X}^4_1 ={\mathbb C}P^1 \times {\mathbb C}P^1$.

In conclusion, consider  the case $\cN=4$ and $m=2$. We have to deal with two linearly 
independent four-vector
$Z_\pm \in {\mathbb C}^4$ obeying the null conditions
\bea
Z_+ \cdot Z_+ = Z_-\cdot Z_- =Z_+\cdot Z_- =0~.
\eea
A general solution to these conditions proves to be a sum of two partial solutions:
\begin{subequations}
\bea
Z_\pm^{ i \bar k} &=& w_{\pm}^{ i }  \, v_{\rm R}^{ \bar k} ~, \qquad  
\big( w_+^i ~w_-^i \big) \in \sGL(2,{\mathbb C})~, 
\qquad  v_{\rm R}^{ \bar k} \in {\mathbb C}^2 \setminus\{0\}~; 
\label{solutionA}\\
Z_\pm^{ i \bar k} &=&v_{\rm L}^{ i }  \, w_{\pm}^{ \bar k}~,
 \qquad  \big( w_+^ { \bar k}~w_-^{ \bar k}\big) \in \sGL(2,{\mathbb C})~, 
\qquad  v_{\rm L}^i \in {\mathbb C}^2 \setminus\{0\}~.
\label{solutionB}
\eea
\end{subequations}
For the family of first solutions, eq. (\ref{solutionA}), the gauge freedom (\ref{6.5}) can be used to choose
\be
\big( w_+^i ~w_-^i \big) ={\mathbbm 1}_2~.
\ee
After that we are still left with a residual gauge freedom of the form:
\be
v_{\rm R}^{ \bar k} \sim \l\, v_{\rm R}^{ \bar k} ~, 
\qquad  \l  \in {\mathbb C} \setminus\{0\}~. 
\ee
Therefore, the internal space in this case is ${\mathbb C}P^1$.
The same consideration applies to the family of solutions  (\ref{solutionB}).
We conclude that ${\mathbb X}^4_2 ={\mathbb C}P^1 \bigcup {\mathbb C}P^1$.

\section{Superconformal Killing vector fields} 
\label{Superconformal Killing vectors} 
\setcounter{equation}{0}
${}$For various studies  of four-dimensional $\cN=1$ superconformal field theories in superspace, 
the concept of superconformal Killing vectors \cite{Sohnius,Lang,Shizuya,BK}
has been shown to be indispensable, see e.g. \cite{Park:1997bq,Osborn}. 
This concept has been generalised to $\cN$-extended superconformal symmetry in 
three \cite{Park3}, four \cite{HH,Park4} and six \cite{Park6} dimensions. 
The five-dimensional case has been worked out in \cite{K-compactified}.
Here we elaborate on the salient properties of the 3D $\cN$-extended superconformal Killing 
vectors in a form that is more close in spirit to the presentation given in \cite{K-compactified}.

\subsection{N-extended superconformal Killing vector fields} 
Consider the 3D $\cN$-extended Minkowski superspace ${\mathbb M}^{3|2\cN}$ parametrised by 
real bosonic and fermionic coordinates $z^A =(x^a, \q^\a_I)$, where $I=1,\dots, \cN$.
We recall that the spinor covariant derivatives $D_\a^I$ obey  the anticommutation relations 
\bea
\big\{ D^I_\a , D^J_\b \big\} = 2{\rm i}\, \d^{IJ}  (\g^m)_{\a\b}\,\pa_m~.
\eea
Superconformal transformations, $z^A \to z^A +\d z^A = z^A +\x^A(z)$, 
are generated by superconformal Killing vectors. By definition, a superconformal Killing vector 
\be
\x  = \x^{a} (z) \, \pa_{ a} 
+  \x^{\a}_I (z) \, D_{\a}^I 
\ee
is a real vector field obeying the condition
\be
[\x, D_\a^I ] \propto D_\b^J~
\ee
which is equivalent to
\bea
D^I_\a \x^b = 2{\rm i} \, \d^{IJ} (\g^b)_{\a \b} \x^\b_J~.
\label{sck-5.2}
\eea
Using this condition, it is a short calculation to show that $\x^a (x,\q)$ is an ordinary conformal Killing vector
(parametrically depending on $\q$)
\be
\pa_a \x_b + \pa_b \x_a = \frac{2}{3} \eta_{ab} \pa_c \x^c~.
\ee
An important implication of (\ref{sck-5.2}) is 
\bea
\pa_{(\a\b}\, \x^I_{\g)} =0~.
\eea

Direct calculations give 
\bea
[\x, D_\a^I ] = -(D^I_\a \x^\b_J) D^J_\b = \o_\a{}^\b (z)D_\b^I + \L^{IJ}(z) D_\a^J -\hf \s (z) D^I_\a~,
\label{master1}
\eea
where we have defined 
\begin{subequations}
\bea
\o_{\a\b} &:=& -\frac{1}{\cN} D^J_{(\a} \x_{\b)}^J =-\frac{1}{4} \pa^\g{}_{(\a} \x_{\b )\g} ~,
\label{Lorentz}\\
\L^{IJ} &:=& -2 D_\a^{ [I} \x^{J]\a}~, \label{rotation}\\
\s&:=& \frac{1}{\cN} D^I_\a \x^\a_I =\frac{1}{3} \pa_a \x^a~.
\label{scale}
\eea
\end{subequations}
Here the parameters $\o_{\a \b} =\o_{\b \a}$, $\L^{IJ}=- \L^{JI}$ and $\s$ correspond
to  $z$-dependent Lorentz, $\sSO (\cN )$ and scale transformations.
These transformation parameters are related to each other as follows:
\begin{subequations}
\bea
D^I_\a \o_{\b \g} &=&  \ve_{ \a ( \b} D^I_{ \g)} \s~,  \label{Lorentz-scale}\\
D^I_\a \L^{JK} &=& -2 \d^{ I [ J } D_\a^{ K] } \s~. \label{rotation-scale}
\eea
\end{subequations}
Making use of (\ref{Lorentz}) gives
\bea
\pa^{\a\b} \o_{\a\b}=0~.
\eea 
In conjunction with (\ref{Lorentz-scale}), this identity implies that 
\be
D^{\a I} D_\a^I \s =0 ~. \qquad \qquad \mbox{(no sum in $I$)}~
\label{D2sigma}
\ee

\subsection{N = 1 superconformal Killing vector fields} 
In this subsection, we specialise the above results to the case $\cN=1$.

The $\cN=1$ superconformal Killing vectors 
\be
\x  = \x^{a}   \pa_{ a} 
+  \x^{\a}   D_{\a} 
\ee
are characterised by the property
\bea
[\x, D_\a ] = 
\o_\a{}^\b  D_\b -\hf \s  D_\a~,
\eea
where the $z$-dependent parameters of Lorentz ($\o_{\a \b}$) and
scale ($\s$) transformations are
\begin{subequations}
\bea
\o_{\a\b} &:=& - D_{(\a} \x_{\b)} =-\frac{1}{4} \pa^\g{}_{(\a} \x_{\b )\g} ~,
\label{Lorentz N=1}\\
\s&:=& D_\a \x^\a =\frac{1}{3} \pa_a \x^a~.
\eea
\end{subequations}
These parameters are related to each other by the relation
\bea
D_\a \o_{\b \g} &=&  \ve_{ \a ( \b} D_{ \g)} \s~,  
\eea
which implies 
\be
D^2 \s=0~.
\ee

\subsection{N = 2 superconformal Killing vector fields} 
\label{N2superKilling}
In the $\cN=2$ case, it is useful to introduce a new basis for the spinor covariant derivatives.
Instead of the covariant derivatives
$D^I_\a$, with $I = {\bf 1}, {\bf 2}$,  we introduce new operators ${\mathbb D}_\a$
and ${\mathbb \DB}_\a$ defined as
\bea
&&
{\mathbb D}_\a=\frac{1}{ \sqrt{2}}(D_\a^{\bf{1}}-\ri D_\a^{\bf{2}})~,\qquad
{\mathbb \DB}_\a=-\frac{1}{ \sqrt{2}}(D_\a^{\bf{1}}+\ri D_\a^{\bf{2}})~.~~~
\eea
They have definite $\sU (1)$ charges 
with respect to 
the group $\sSO(2) \in \sOSp(2|2, {\mathbb R})$
and satisfy the anticommutation relations
\bea
\{{\mathbb D}_\a,{\mathbb \DB}_\b\}=-2\ri\, \pa_{\a\b}~,\qquad
\{{\mathbb D}_\a,{\mathbb D}_\b\}=\{ {\mathbb \DB}_\a, {\mathbb \DB}_\b\}=0~.
\label{N=2acd}
\eea
The superconformal Killing vector, 
$\x  = \x^{a}   \pa_{ a} +  \x^{\a}_I   D_{\a}^I $,
takes the form 
\bea
\x=
\x^{a} \pa_{ a} 
+\x^\a{\mathbb D}_\a
+{\bar{\x}}_\a{\mathbb \DB}^\a~,
\eea
where the spinor  $\x^\a$ and its complex conjugate 
$\bar{\x}^\a$ are defined by 
\bea
\x^\a &=&\frac{1}{ \sqrt{2}}(  \x^{\a}_{\bf{1}}+ \ri \x^{\a}_{\bf{2}})~,\qquad 
\bar{\x}^\a=\frac{1}{ \sqrt{2}}( \x^{\a}_{\bf{1}} -\ri\x^{\a}_{\bf{2}})~.
\eea
It is easy to see that $\x^\a$ is chiral, 
\be
\bar{\mathbb D}_\b \x^\a=0~.
\ee

Our next task is  to express the parameters (\ref{Lorentz}), (\ref{rotation}) and (\ref{scale})
in terms of $\x^\a$ and its conjugate. Direct calculations give\footnote{Here the antisymmetric tensor $\ve^{IJ}$ is 
defined by $\ve^{{\bf 1}{\bf 2}}=1$.}
\begin{subequations}
\bea
\o_{\a\b}&=& -{\mathbb D}_{(\a}\x_{\b)}={\mathbb \DB}_{(\a}\bar{\x}_{\b)}
=-\frac{1}{4}\pa^\g{}_{(\a}\x_{\b)\g}~, \\
\L^{IJ}&=& \ve^{IJ} \L ~, \qquad 
\L=\frac{\rm i}{4} \Big({\mathbb D}_\g\x^\g+{\mathbb \DB}_\g\bar{\x}^\g\Big)  ~, \\
\s & = & \hf \Big({\mathbb D}_\g\x^\g-{\mathbb \DB}_\g\bar{\x}^\g\Big) =\frac{1}{3}\pa_a\x^a~.
\eea
\end{subequations}
If we define 
\bea
{\bm \s} := \s +{\rm i} \, \L ~, \qquad \s =\hf ( {\bm \s} + {\bar {\bm \s}})~,
\eea
then eq. (\ref{rotation-scale}) is equivalent to the fact that $\bm \s$ is chiral, 
\be
\bar{\mathbb D}_\a {\bm \s} =0~.
\ee
It follows from eq. (\ref{D2sigma}) that 
\bea
{\mathbb D}^2 {\bm \s} =0~.
\eea

${}$Finally, the master relation (\ref{master1}) turns into
\bea
{[}\x,{\mathbb D}_\a{]}= \o_{\a}{}^{\b}{\mathbb D}_\b
+\frac{1}{4}\Big({\bm \s} -3 \bar{\bm \s} \Big){\mathbb D}_\a~.
\eea

\subsection{N = 3 superconformal Killing vector fields} 
\label{N=3superKillings}

As follows from eq. (\ref{scale}), the case $\cN=3$ is special
in the sense that the superconformal transformations preserve the volume,
${\rm d}^3x \,{\rm d}^6\q$, of Minkowski superspace ${\mathbb M}^{3|6}$
\bea
(-1)^A D_A \x^A=0~, \qquad D_A :=(\pa_a, D^I_\a )~.
\label{N3SupVol}
\eea
This is similar to the four-dimensional  $\cN=2$ case, see e.g. \cite{K-hyper}.

It is useful to convert each $\sSO(3)$ vector index into a pair of two isospinor ones by the rule
(\ref{N=3conversion}). The isospinor indices will be raised and lowered using the antisymmetric 
$\sSU(2) $ invariant tensor $\ve_{ij}$ and $\ve^{ij}$ (normalised as $\ve^{12}=\ve_{21}=1$). The rules for raising and lowering 
the isospinor indices are
\bea
\psi^i=\ve^{ij}\psi_j~, \qquad \psi_i=\ve_{ij}\psi^j~.
\eea 
In particular, associated with the matrices $(\t)_i{}^j$, eq. (\ref{N=3conversion}), 
are the symmetric matrices $(\t_I)_{ij} =(\t_I)_{ji} $ and $(\t_I)^{ij} =(\t_I)^{ji} $ 
which are related to each other by complex conjugation:
\bea
\overline{ (\t_I)_{ij} } =(\t_I)^{ij}  ~.
\eea
If $A_I$ and $B_I$ are $\sSO(3)$ vectors and $A_{ij}$ and $B_{ij}$ the associated symmetric isotensors, 
then the dot-product reads as 
\bea
A\cdot B :=A_I B_I = A_{ij} B^{ij}~.
\eea
Given an antisymmetric second-rank $\sSO(3)$ tensor,  $\L^{IJ} =- \L^{JI}$, its  counterpart 
with isospinor indices
$\L^{ijkl}=-\L^{klij}= \L^{IJ}(\t_I)^{ij}(\t_J)^{kl}$ can be decomposed as
\bea
\L^{ijkl}= \ve^{jl}\L^{ik}+ \ve^{ik}\L^{jl}~,\qquad
\L^{ij}=\L^{ji}~.
\eea
Note that the spinor covariant derivatives $D_\a^{ij}=(\t_I)^{ij}D_\a^I$ satisfy the algebra
\bea
\{D_\a^{ij},D_\b^{kl}\}=-2\ri\ve^{i(k}\ve^{l)j}\pa_{\a\b}~.
\eea
In terms of the superspace coordinates $z^A=(x^a,\q^\a_{ij})$, with $\q^\a_{ij}:=(\t_I)_{ij}\q^\a_I$, the explicit 
realisation of the covariant derivatives is
\bea
D_\a^{ij}=\frac{\pa}{\pa\q^\a_{ij}}+\ri\,\q^{\b}_{ij}\pa_{\a\b}~.
\eea

The master relation (\ref{master1}) in the $\sSU(2)$ notation takes the form
\bea
[\x, D_\a^{ij} ] = \o_\a{}^\b D_\b^{ij}- \L^i{}_k D_\a^{k j} -\L^j{}_k D_\a^{ik } 
- \hf \s D_\a^{ij}~,
\label{master-5.33}
\eea
where we now have ($\x^\a_{ij}=(\t^I)_{ij}\x^\a_I$)
\bsubeq
\bea
\o_{\a\b}&=&- \frac{1}{ 3}D_{(\a}^{kl}\x_{\b)kl}=- \frac{1}{ 4}\pa_{(\a}{}^\g\x_{\b)\g} ~, \\
\L_{ij}&=&
\frac{1}{ 4}D_{\a(i}^{k} \x_{ j) k}^{ \a}
~,
\label{N=3sckv-Lambda}\\
\s&=& \frac{1}{3}D_\a^{ij}\x^\a_{ij}=\frac{1}{ 3}\pa^a\x_a~.
\label{N=3sckv-sigma}
\eea
\esubeq
The above parameters are related to each other by the following relations:
\bsubeq
\bea
D_{\a}^{ij}\o_{\b\g}&=&\ve_{\a(\b}D_{\g)}^{ij}\s
~,
\\
D_{\a}^{ij} \L^{kl}
&=&
\hf  \ve^{i(k}D_\a^{l)j}\s
+\hf\ve^{j(k}D_\a^{l)i}\s
~.
\eea
\esubeq

According to the analysis of section \ref{CompHPS}, in the $\cN=3$ case it natural to introduce a 
$\sSU(2)$ complex isotwistor $v^i$ which defines a null $\sSO(3)$ vector $Z_I=v^iv^j(\t_I)_{ij}$.
It is useful to introduce two complex isospinor $v^i$ and $u^i$ which can be used to 
change basis for the isospinor indices e.~g. by using the completeness relation
\bea
\d_j^i=\frac{1}{ (v,u)}\big(v^{i}u_j- v_j u^i \big)~,\qquad
(v, u):=v^{i}u_i~.
\label{completeness}
\eea
The choice of $u_i$ is restricted only by the condition $(v,u)\ne0$.

In the $\{v,u\}$-basis,  the spinor covariant derivatives $D_\a^{ij}$ turn into
$D^{(2)}_\a$, $D^{(0)}_\a$ and $D^{(-2)}_\a$
defined as follows
\bea
D^{(2)}_\a:=v_i v_j D_\a^{ij}~,\qquad 
D^{(0) }_\a:=   \frac{1}{(v,u)}v_i u_jD_\a^{ij}~,\qquad
D^{(-2)}_\a:= \frac{1}{(v,u)^2 }u_i u_jD_\a^{ij}~,
\label{20-2}
\eea
where the superscript on $D^{(2)}_\a$, $D^{(0)}_\a$ and $D^{(-2)}_\a$ indicates
the degree of homogeneity in $v$s.
The spinor covariant derivatives  satisfy the algebra
\bsubeq
\bea
&\{D_\a^{(2) },D_\b^{(2)}\}=\{D_\a^{(-2)},D_\b^{(-2)}\}=0~,\\
&\{D_\a^{(2)},D_\b^{(-2)}\}=
2\ri  \pa_{\a\b}~,\qquad
\{D_\a^{(0)},D_\b^{(0)}\}=-\ri \pa_{\a\b}~,
\\
&\{D_\a^{(2)},D_\b^{(0)}\}=\{D_\a^{(-2)},D_\b^{(0)}\}=0
~.
\eea
\esubeq
The derivatives $D_\a^{(2)}$ can be used to define analyticity constraints on 
superfields.

In the basis for  the covariant derivatives introduced, the superconformal Killing vector takes the form
\bea
\x=
\x^a\pa_a
+ \x^{ (2) \a}D_\a^{(-2)}
- 2\x^{(0) \a}D_\a^{(0)}
+ \x^{(-2)\a}D_\a^{(2)}
~,
\eea
where we have introduced the spinor components
\bea
\x^{(2) \a}:=v^i v^j \x^\a_{ ij}~,\qquad
\x^{(0) \a}:= \frac{1}{(v,u)} v^i u^j\x^{\a}_{ ij}~,\qquad
\x^{(-2) \a}:= \frac{1}{(v,u)^2} u^i u^j \x^{\a}_{ ij}~.
\eea
Note that the  $\x^{(2) \a}$ satisfies the analyticity condition
\bea
D_\b^{(2)}\x^{(2) \a}=0~.
\eea

The covariant derivatives $D_\a^{(2)}$ satisfy the master relation
\bea
\Big{[}\x- \L^{(2)} {\bm \pa}^{(-2)},D_\a^{(2)}\Big{]}=
\o_{\a}{}^{\b}D_\b^{(2)}
- \big(2\S - \hf\s)D_{\a}^{(2)}~.
\label{N=3master}
\eea
Here we have introduced the scalar superfields
\bea
\L^{(2)}:=v_i v_j\L^{ij}~,\qquad
\L^{(0)}:= \frac{1}{(v,u)} v_i u_j \L^{ij}~,\qquad
\S:=  \hf\s+  \L^{(0)}  ~,
\eea
and used the isotwistor derivatives
\bea
{\bm \pa}^{(-2)}= \frac{1}{(v,u)}u^{i} \frac{\pa}{\pa v^{i}}~,\qquad
{\bm \pa}^{(2)}= (v,u)  v^{i}\frac{\pa}{ \pa u^{i}}
~.
\eea
One can check that 
\bea
&&~~~
D_\g^{(0)}\x^{(2) \g}
=-D_\g^{(2)}\x^{(0) \g}
=- 2\L^{(2)}~.
\eea
The following important relations can be easily derived
\bea
D_\a^{(2)}\L^{(2)}=D_{\a}^{(2)} \S=0~,\qquad
{\bm \pa}^{(2)}\S= \L^{(2)}~.
\label{N=3Sigma}
\eea

For later use, it is important to note that the volume-preservation identity eq. (\ref{N3SupVol})
can be rewritten in the form
\bea
\pa_a\x^a
-D_\a^{(-2)}\x^{\a(2)}
+2D_\a^{(0)}\x^{\a(0)}- {\bm \pa}^{(-2)}\L^{(2)} =2\S~.
\label{N=3volume}
\eea

\subsection{N = 4 superconformal Killing vector fields} 
\label{N=4superKillings}

In complete analogy with $\cN=3$ supersymmetry, 
in the $\cN=4$ case it is also useful to convert the $\sSO(4)$ vector 
indices into pairs of isospinor ones.
This is achieved by making use of the matrices $(\t_I)_i{}^{\bar k}$ through the conversion rule 
(\ref{N=4conversion}).
 The isospinor indices $i$ and ${\bar k}$ respectively transform under 
 $\sSU_{\rm L}(2)$ and $\sSU_{\rm R} (2) $
of 
 $\sSO(4) \cong \big( \sSU_{\rm L}(2)\times \sSU_{\rm R} (2) \big) /{\mathbb Z}_2$.
 By using the antisymmetric 
 invariant tensors $\ve_{ij},\ve^{ij}$ and
  $\ve^{{\bar k}{\bar l}},\ve_{{\bar k}{\bar l}}$,
 we will raise and lower the $\sSU(2)$s indices according to the rule\bsubeq
\bea
&\psi^i=\ve^{ij}\psi_j~, \qquad \psi_i=\ve_{ij}\psi^j~;
\\
&\chi_{\bar k}=\ve_{{\bar k}{\bar l}}\chi^{\bar l}~, \qquad \chi^{\bar k}=\ve^{{\bar k}{\bar l}}\chi_{\bar l}~.
\eea 
\esubeq
Then, 
associated with the matrices $(\t_I)_i{}^{\bar k}$
there are the matrices $(\t_I)_{i{\bar k}} =\ve_{{\bar k}{\bar l}}(\t_I)_{i}{}^{\bar l} $ 
and $(\t_I)^{i{\bar k}} =\ve^{ij}(\t_I)_j{}^{{\bar k}} $ 
which are related to each other by complex conjugation:
\bea
\overline{ (\t_I)_{i{\bar k}} } =(\t_I)^{i{\bar k}}  ~.
\eea
Given two $\sSO(4)$ vectors $A_I$ and $B_I$ and $A_{i{\bar k}}$ and $B_{i{\bar k}}$ their 
associated isotensors, the dot-product reads as 
\bea
A\cdot B :=A_I B_I = A_{i{\bar k}} B^{i{\bar k}}~.
\eea
It follows that in the isospinor notation the covariant derivatives 
$D_\a^{i{\bar k}}=(\t_I)^{i{\bar k}}D_\a^I$ 
satisfies the algebra
\bea
\{D_\a^{i{\bar k}},D_\b^{j{\bar l}}\}=2\ri\ve^{ij}\ve^{{\bar k}{\bar l}}\pa_{\a\b}~.
\eea
In terms of the superspace coordinates $z^A=(x^a,\q^\a_{k{\bar l}})$, with 
$\q^\a_{k{\bar l}}:=(\t_I)_{k{\bar l}}\, \q^\a_I$, the explicit realisation
of the covariant derivatives is
\bea
D_\a^{k{\bar l}}=\frac{\pa}{\pa\q^\a_{k{\bar l}}}+\ri\,\q^{\b}_{k{\bar l}}\pa_{\a\b}~.
\eea

An antisymmetric second-rank $\sSO(4)$ tensor, like the superfield $\L^{IJ} =- \L^{JI}$, 
is transformed to the isotensor 
$\L^{i{\bar k}j{\bar l}}=-\L^{j{\bar l}i{\bar k}}= \L^{IJ}(\t_I)^{i{\bar k}}(\t_J)^{j{\bar l}}$.
This can be expressed in terms of  its $\sSO(4)$ self-dual and antiself-dual parts
according to the following decomposition
\bea
\L^{i{\bar k}j{\bar l}}= \ve^{{\bar k}{\bar l}}\L_{\rm L}^{ij}+ \ve^{ij}\L_{\rm R}^{{\bar k}{\bar l}}~,
\qquad
\L_{\rm L}^{ij}=\L_{\rm L}^{ji}~,~~
\L_{\rm R}^{{\bar k}{\bar l}}=\L_{\rm R}^{{\bar l}{\bar k}}~.
\eea
The symmetric $\L_{\rm L}^{ij}$ and $\L_{\rm R}^{{\bar k}{\bar l}}$ 
parameters together with $\o_{\a\b}$ and
$\s$ of (\ref{Lorentz}), (\ref{scale}) are expressed in terms of 
the spinor $\x^\a_{i{\bar k}}=(\t_I)_{i{\bar k}}\x^\a_I$ as
\bsubeq
\bea
\L_{\rm L}^{ij}&=&\frac{1}{4}D_{{\bar k}}^{\a(i}\x_\a^{j){\bar k}}~,~~~
\L_{\rm R}^{{\bar k}{\bar l}}=\frac{1}{4}D_{\a}^{i({\bar k}}\x_i^{\a{\bar l})}~,
\\
\o_{\a\b}&=&-\frac{1}{4}D_{(\a}^{i{\bar k}}\x_{\b)i{\bar k}}=-\frac{1}{ 4}\pa_{(\a}{}^\g\x_{\b)\g}~,~~~
\\
\s&=&\frac{1}{4}D_\a^{i{\bar k}}\x^\a_{i{\bar k}}=\frac{1}{ 3}\pa^a\x_a~.
\eea
\esubeq
The above superfields turn out to be related to each other by the following equations
\bsubeq
\bea
&
D_{\a}^{i{\bar k}}\o_{\b\g}=\ve_{\a(\b}D_{\g)}^{i{\bar k}}\s~,
\\
&
D_{\a}^{i{\bar l}} \L_{\rm L}^{jk}=
\ve^{i(j}D_\a^{k){\bar l}}\s
~,~~~
D_{\a}^{i{\bar k}} \L_{\rm R}^{{\bar l}{\bar p}}=
\ve^{{\bar k}({\bar l}}D_\a^{i{\bar p})}\s ~.
\eea
\esubeq
Note also that the equation (\ref{master1}) in isospinor notation takes the form
\bea
{[}\x,D_\a^{i{\bar k}}{]}=
\o_{\a}{}^{\b}D_\b^{i{\bar k}}
+(\L_{\rm L})^{ij}D_{\a j}^{{\bar k}}
+(\L_{\rm R})^{{\bar k}{\bar l}}D_{\a {\bar l}}^i
-\frac{1}{2}\s D_\a^{i{\bar k}}
~.
\eea

By using two complex isotwistors, similarly to the analysis of subsection \ref{N=3superKillings},
we can change basis for the isospinor indices.
For example, we introduce two left-isospinors $v_{\rm L}:=(v^i)$ and $u_{\rm L}:=(u^i)$, 
which satisfy the very same
relations as (\ref{completeness}), and define new spinor covariant derivatives
\bea
D^{(1){\bar k}}_\a:=v_{i}D_\a^{i{\bar k}}~,\qquad 
D^{(-1){\bar k}}_\a:=\frac{1}{(v_{\rm L},u_{\rm L})}u_{i}D_\a^{i{\bar k}}
~.
\eea
Here the superscript on $D^{(1){\bar k}}_\a$ and $D^{(-1){\bar k}}_\a$ indicates
the degree of homogeneity in $v$s.
The spinor covariant derivatives  satisfy the algebra
\bsubeq
\bea
&\{D_\a^{(1){\bar k}},D_\b^{(1){\bar l}}\}=\{D_\a^{(-1){\bar k}},D_\b^{(-1){\bar l}}\}=0~, \label{5.57a}\\
&\{D_\a^{(1){\bar k}},D_\b^{(-1){\bar l}}\}=
-2\ri \ve^{{\bar k}{\bar l}} \pa_{\a\b}~.
\label{5.57b}
\eea
\esubeq
In particular the $D_\a^{(1){\bar{k}}}$ derivatives represent a maximal anti-commuting subset of the 
$D_\a^{i{\bar k}}$ derivatives 
and can be used to define consistent analyticity constraints.

In the  covariant derivative basis just introduced, the superconformal Killing vector takes the 
form
\bea
\x=
\x^a\pa_a
-\x^{ (1)}{}^{ \a}_{\bar k}D_\a^{(-1){\bar k}}
+ \x^{(-1)}{}^{\a}_{\bar k}D_\a^{(1){\bar k}}
\label{5.58-add}
~,
\eea
where we have introduced the spinor components
\bea
\x^{(1)}{}_{\bar k}^{ \a}:= -v^{i} \x^\a_{ i{\bar k}}~,\qquad
\x^{(-1)}{}_{\bar k}^{ \a}:= -\frac{1}{(v_{\rm L},u_{\rm L})} u^{i} \x^\a_{ i{\bar k}}
~.
\eea
Note that the  $\x^{(1)}{}^{ \a}_{\bar k}$ superfield is constrained by the condition
\bea
D_{\b({\bar k}}^{(1)}\x^{(1)}{}^{\a}_{{\bar l})}=0~.
\eea

The covariant derivatives $D_\a^{(1){\bar k}}$ satisfy the master relation
\bea
\Big{[}\x- \L_{\rm L}^{(2)} {\bm \pa}_{\rm L}^{(-2)},D_\a^{(1){\bar k}}\Big{]}=
\o_{\a}{}^{\b}D_\b^{(1){\bar k}}
-\big(\ve^{{\bar k}{\bar l}}\S_{\rm L} -\L_{\rm R}^{{\bar k}{\bar l}}\big)D_{\a {\bar l}}^{(1)}
~.
\eea
Here we have introduced the scalar superfields
\bea
\L_{\rm L}^{(2)}:=v_i v_j\L_{\rm L}^{ij}~,\qquad
\L_{\rm L}^{(0)}:= \frac{1}{(v_{\rm L},u_{\rm L})} v_i u_j \L_{\rm L}^{ij}~,\qquad
\S_{\rm L}:=\hf\s+\L_{\rm L}^{(0)}
~.
\eea
One can derive the following relations between the previous parameters
\bsubeq
\bea
&
D_\a^{(1) {\bar k}}\x^{(1)}{}_{{\bar k}}^{\a}=4\L_{\rm L}^{(2)}~,\\
&
D_{\a}^{(1){\bar k}} \L_{\rm L}^{(2)}=D_\a^{(1){\bar k}}\S_{\rm L}=0~,
\qquad
{\bm \pa}_{\rm L}^{(2)}\S_{\rm L}=\L_{\rm L}^{(2)}~.
\eea
\esubeq
We conclude by giving the following equation
\bea
\pa_a\x^a 
+D_\a^{(-1){\bar k}}\x^{(1)}{}^{ \a}_{{\bar k}}
-{\bm \pa}_{\rm L}^{(-2)}\L_{\rm L}^{(2)}
=2\S_{\rm L}
\label{5.64-add}
\eea
which will be crucial in the formulation of a manifestly superconformal $\cN=4$ action principle.

In complete analogy with the previous discussion, we also introduce right isotwistors
$v_{\rm R}:=(v^{\bar k})$ and
$u_{\rm R}:=(u_{\bar k})$ such that $(v_{\rm R},u_{\rm R}):=v^{\bar k}u_{\bar k}\ne0$,
and use them to change basis for the right-isospinor indices.
We then define new covariant derivatives
\bea
D^{(1)i}_\a:=v_{\bar k}D_\a^{i{\bar k}}~,\qquad
D^{(-1)i}_\a:=\frac{1}{(v_{\rm R},u_{\rm R})}u_{\bar k}D_\a^{i{\bar k}}
~.
\eea
They satisfy anti-commutation relations analogous to eq. (\ref{5.57a}) and (\ref{5.57b}).
Note that the spinor derivatives $D^{(1)i}_\a$ represent a second maximal anti-commuting subset of
the operators $D_\a^{i{\bar k}}$  and can be used to define a new type of constrained superfields.
One can then rewrite the superconformal Killing vector in the  $D^{(1)i}_\a$  
basis, and all the results in (\ref{5.58-add})--(\ref{5.64-add}) 
carry over 
with the only modification that the left
sector is changed everywhere with the right one.

The existence of the two independent sets of anti-commuting covariant derivatives
$D^{(1){\bar k}}_\a$ and $D^{(1)i}_\a$ is crucial to build the two type of
hypermultiplets which describe matter in the $\cN=4$ case. We will come back to this points in section 8.

\section{N = 1 and N = 2 superconformal sigma-models} 
\setcounter{equation}{0}

As a warm-up exercise in using the formalism of superconformal Killing vectors,
in this section we construct the most general $\cN=1,~2$ superconformal sigma-models.

\subsection{N = 1 superconformal sigma-models} 
To generate $\cN=1$ superconformal actions, we need real scalar densities with the 
following transformation law 
\bea
\d \cL = -\x \cL - 2\s\, \cL~,
\eea
with respect to the superconformal group. Given such a Lagrangian $\cL$, the action
\bea
\int {\rm d}^3x {\rm d}^2 \q  \, \cL
\eea
is $\cN=1$ superconformal. 

Consider a general massless
$\cN=1$ sigma-model action 
\bea
S= -\hf \int {\rm d}^3x {\rm d}^2 \q \, g_{\m \n }(\vf) (D^\a \vf^\m ) D_\a \vf^\n~,
\label{N=1s-m}
\eea
where $g_{\m \n }(\vf) $ is the metric on the target space. 
We wish to determine those restrictions on the target space geometry which guarantee the sigma-model 
to be superconformal. Without loss of generality, 
a superconformal transformation of $\vf^\m$ can be chosen to be 
\bea
\d \vf^\m = -\x \vf^\m - \hf \s\, \c^\m (\vf)~,
\eea
where $ \c^m (\vf)$ is a vector field on the target space.
Using the properties of the $\cN=1$ superconformal Killing vector fields, 
the variation of the action can be brought to the following form:
\bea
\d S &=&\phantom{+} \hf  \int {\rm d}^3x {\rm d}^2 \q \, g_{\m \n }(\vf) (D^\a\vf^\m ) (D_\a \vf^\l)
\Big( \nabla_\l \c^\n (\vf) -\d_\l^\n \Big)\s \non \\
&&+ \hf \int {\rm d}^3x {\rm d}^2 \q \,g_{\m \n }(\vf) (D^\a \vf^\m ) \c^\n D_\a \s~.
\label{7.3}
\eea
In the case that $D_\a \s =0 $ and $\s$ is  non-zero, the expression in the second line  of (\ref{7.3}) vanishes. 
Then, the remaining variation is equal to zero  only  if 
\bea
\nabla_\m \c^\n =\d_\m^\n \quad \longleftrightarrow \quad
\nabla_\m \c_\n =g_{\m \n}~.
\label{7.4}
\eea
We observe that $\c= \c^\m (\vf ) {\pa}_\m $ should be 
a homothetic conformal Killing vector field such that $\c_\m (\vf)$ is the gradient of 
a function over the target space,
\bea
\c_\m (\vf) =\pa_\m f(\vf)~, \qquad f (\vf) =\hf g_{\m \n} (\vf) \c^\m (\vf ) \c^\n (\vf)~.
\label{7.5}
\eea
The sigma-model target space is a Riemannian cone \cite{GR}.

If the equations (\ref{7.4}) and (\ref{7.5}) hold, the variation (\ref{7.3}) turns into
\bea
\d S &=& \hf  \int {\rm d}^3x {\rm d}^2 \q \, (D^\a f) D_\a \s 
= - \hf  \int {\rm d}^3x {\rm d}^2 \q \, f D^2 \s =0~,
\eea
since $D^2 \s =0$.

The action (\ref{N=1s-m}) can be generalised by adding a potential term
\bea
S= -\hf \int {\rm d}^3x {\rm d}^2 \q \, g_{\m \n }(\vf) (D^\a \vf^\m ) D_\a \vf^\n
+ \ri\int {\rm d}^3x {\rm d}^2 \q \, V(\vf^\m) ~,
\label{N=1s-m+pot}
\eea
for some  scalar field $V(\vf)$ on the target space. If the target space is a cone, 
and $V(\vf)$ obeys the homogeneity condition
\bea
\c^\m (\vf) V_\m (\vf  ) = 4 V(\vf )~,
\eea
then the action (\ref{N=1s-m+pot}) is $\cN=1$ superconformal.

\subsection{N = 2 superconformal sigma-models} 
\label{N2 superconformal sigma-models} 
There are two simple constructions to generate $\cN=2$ superconformal actions. 
First, given a real superfield $\cL$ transforming by the rule 
\bea
\d \cL =- \x \cL - \hf \big( {\bm \s} +\bar{\bm \s}\big) \cL
=-\pa_a(\x^a \cL) +{\mathbb D}_\a (\x^\a \cL)  +\bar{\mathbb D}^\a ({\bar \x}_\a \cL)~, 
\eea
the functional 
\bea
\int {\rm d}^3x {\rm d}^2 \q {\rm d}^2 {\bar \q}\, \cL
\eea
is $\cN=2$ superconformal. Secondly, given a chiral superfield $\cL_{\rm c} $, 
$\bar{ {\mathbb D}}_\a \cL_{\rm c} =0$, with the superconformal transformation
\bea
\d \cL_{\rm c} =- \x \cL_{\rm c} - 2  {\bm \s}  \cL_{\rm c}
=-\pa_a(\x^a \cL_{\rm c}) +{\mathbb D}_\a (\x^\a \cL_{\rm c})  ~, 
\eea
the functional 
\bea
\int {\rm d}^3x {\rm d}^2 \q \, \cL_{\rm c}
\eea
is $\cN=2$ superconformal.

Consider the general $\cN=2$  supersymmetric sigma-model  \cite{Zumino}
\bea
S= \int {\rm d}^3x {\rm d}^2 \q {\rm d}^2 {\bar \q}\, K(\F^I, {\bar \F}^{\bar J})~, \qquad {\bar {\mathbb D}}_\a \F^I=0~,
\label{7.7}
\eea
where  $K$  is the K\"ahler potential of a K\"ahler manifold $\cM$. As usual, we denote by
 $g_{I \bar J} (\F , \bar \F) $ the K\"ahler metric on $\cM$.
 Our goal is to determine those restrictions on the target space geometry which guarantee the sigma-model 
to be superconformal. Since (\ref{7.7}) is a special $\cN=1$ supersymmetric nonlinear sigma-model, 
it is $\cN=1$ superconformal if the target space possesses a homothetic conformal Killing vector field
$\c^\m = (\c^I , {\bar \c}^{\bar J})$
such that $\c_\m (\F)$ is the gradient of a function over the target space.
The relations (\ref{7.4}) and (\ref{7.5}) turn into\begin{subequations}
\bea
\nabla_I \c^J &=& \d_I^J~, \qquad {\bar \nabla}_{\bar I} \c^J ={\bar \pa}_{\bar I} \c^J =0
 \label{7.8a} \\
 \c_I := g_{I\bar J}  {\bar \c}^{\bar J}&=& \pa_I K~, \qquad g_{I\bar J} 
=\pa_I {\bar \pa}_{\bar J} K
~, \label{7.8b}
\eea
\end{subequations}
where $K$ can be chosen to be 
\bea
 K =g_{I\bar J} \c^I  {\bar \c}^{\bar J}~.
 \label{7.9}
 \eea
 In accordance with (\ref{7.8a}), the homothetic conformal Killing vector field is holomorphic, 
 $\c^I =\c^I(\F)$.
The target space $\cM$ is called a K\"ahlerian cone \cite{GR}.

The action (\ref{7.7}) is invariant under the $\cN=2$ superconformal transformations
(compare with the 4D $\cN=1$ case \cite{K-duality})
\bea
\d \F^I =- \x \F^I - \hf {\bm \s} \c^I (\F)~, 
\label{genN=2sctr}
\eea
with $\x$ an arbitrary $\cN=2$ superconformal Killing vector.
This follows from the identity 
\bea
\c^I (\F) K_I (\F , \bar \F ) = K(\F , \bar \F)~.
\eea
 
The sigma-model (\ref{7.7}) can be generalised  by including a superpotential.
\bea
S= \int {\rm d}^3x {\rm d}^2 \q {\rm d}^2 {\bar \q}\, K(\F^I, {\bar \F}^{\bar J})
+\Big\{  \int {\rm d}^3x {\rm d}^2 \q  \, W(\F^I)  +{\rm c.c.}\Big\}~,
\eea
with $W(\F)$  a holomorphic field on the target space. If $\cM$ is a K\"ahlerian cone
and $W(\F)$ obeys the homogeneity condition
\bea
\c^I (\F) W_I (\F  ) = 4 W(\F )~,
\eea
the sigma-model under consideration is $\cN=2$ superconformal.

Local complex coordinates, $\F^I$, on $\cM$ can be chosen in such a way that $\c^I =\F^I$.
Then $K(\F^I, {\bar \F}^{\bar J}) $ obeys the following homogeneity condition:
\bea
\F^I \frac{\pa}{\pa \F^I} K(\F, \bar \F) =  K( \F,   \bar \F)~.
\label{Kkahler2}
\eea

Supersymmetric nonlinear sigma-models can also be generated from self-couplings
of vector multiplets. Consider a dynamical system of several Abelian vector multiplets 
realised in terms of gauge-invariant  real field strengths $G^i $, with $i=1,\dots, n$,
constrained as follows \cite{HitchinKLR}:
\be
{\mathbb D}^2 G^i = \bar{\mathbb D}^2 G^i =0~, \qquad i=1, \dots, n~.
\label{vmc}
\ee
Dynamics of the vector multiplets can be described by an action
\bea
\int {\rm d}^3x {\rm d}^2 \q {\rm d}^2 {\bar \q}\, L(G^i)~.
\label{vm-action}
\eea
The constraints (\ref{vmc}) uniquely fix the superconformal transformation of $G^i$:
\bea
\d G^i =- \x G^i - \hf \big( {\bm \s} +\bar{\bm \s}\big) G^i~.
\eea
Therefore, the action (\ref{vm-action}) is superconformal if the Lagrangian $L(G^i)$
is a homogeneous function of the $n$ variables $G^i$ of degree one, 
\be
G^i L_i (G) =L (G)~.
\ee
In the case of a single vector multiplet, $n=1$, there is a unique superconformal model 
generated by a Lagrangian $L(G) \propto - G \ln G$. It describes an improved vector multiplet 
\cite{HitchinKLR}. Such a model is generated as a low-energy effective action in quantum 3D $\cN=2$ supersymmetric Yang-Mills theories \cite{BPS}.

\section{Off-shell N = 3 superconformal sigma-models} 
\setcounter{equation}{0}
\label{N=3conf}
In this section we develop an efficient formalism to generate off-shell $\cN=3$
superconformal sigma-models. It should be remarked that 3D $\cN=3$ supersymmetry is quite interesting
in its own right, since it can not be obtained
by naive dimensional reduction from higher dimensions.

\subsection{N = 3 superconformal projective multiplets}
We start with defining a family of off-shell $\cN=3$ superconformal multiplets 
living in $\cN=3$ projective superspace
\bea
{\mathbb M}^{3|6} \times {\mathbb C}P^1~.
\label{N=3ps}
\eea
The definition of such supermultiplets as well as the superconformal action principle (to be introduced 
in the next subsection) make use of the spinor covariant derivatives
$D^{(2)}_\a$, $D^{(0)}_\a$ and $D^{(-2)}_\a$ introduced earlier,  eq. (\ref{20-2}).

A superconformal projective multiplet of {\it integer} weight $n$,
$Q^{(n)}(z,v)$, is a superfield that 
lives on  ${\mathbb M}^{3|6}$, 
is holomorphic with respect to the isotwistor variables $v^i $ on an open domain of 
${\mathbb C}^2 \setminus  \{0\}$ 
 such that the following conditions hold:
\\
(i) it obeys the analyticity constraints 
\be
D^{(2)}_{\a} Q^{(n)} 
=0~;
\label{ana}
\ee  
(ii) it is  a homogeneous function of $v^i$ 
of degree $n$, that is  
\be
Q^{(n)}(z,c\,v)\,=\,c^n\,Q^{(n)}(z,v)~, \qquad c\in \mathbb{C}^*:={\mathbb C} \setminus  \{0\}~;
\label{weight}
\ee
(iii) it possesses the  superconformal transformation law:
\be
\d Q^{(n)} = - \Big(  \x -  \L^{(2)}  {\bm \pa}^{(-2)} \Big) \, Q^{(n)} 
-n \,\S \, Q^{(n)} ~.
\label{harmult1}
\ee
As a consequence of eqs. (\ref{N=3master}) and (\ref{N=3Sigma}), 
the variation $\d Q^{(n)} $ is analytic.
By construction, $Q^{(n)}$ is independent of the auxiliary isotwistor $u_i$,
\be
\frac{\pa}{\pa u^{i}}\, Q^{(n)}=0 \quad \Longleftrightarrow \quad
{\bm \pa}^{(2)}Q^{(n)}=0~.
\ee 
Eq. (\ref{weight}) implies that $\d Q^{(n)}$ 
is also independent of $u_i$,
\bea
{\bm \pa}^{(2)}
\d Q^{(n)} =0~, 
\eea
although separate contributions to the right-hand side  of (\ref{harmult1}) involve $u_i$. 

As is clear from the above consideration, the isotwistor
$ v^i \in {\mathbb C}^2 \setminus\{0\}$ is   defined modulo the equivalence relation
$ v^i \sim c\,v^i$,  with $c\in {\mathbb C}^*$, {hence it parametrises ${\mathbb C}P^1$}.
Therefore, the projective multiplets live in ${\mathbb M}^{3|6} \times {\mathbb C}P^1$.

The definition given above is similar to that of 4D $\cN=2$ superconformal projective multiplets 
\cite{K-hyper}. The main difference is that the operators $D_\a^{(2)} $ in the analyticity constraints 
(\ref{ana}) are quadratic in $v^i$, while the corresponding 4D operators are linear in $v^i$.

Given a  superconformal  weight-$n$ multiplet $ Q^{(n)} (v^i)$, 
its {\it smile conjugate},\footnote{The   smile conjugation is the real structure
introduced in  \cite{Rosly,GIKOS,KLR,HitchinKLR}.}
$ \breve{Q}^{(n)} (v^i)$, is defined by 
\bea
 Q^{(n)}(v^i) \longrightarrow  {\bar Q}^{(n)} ({\bar v}_i) 
  \longrightarrow  {\bar Q}^{(n)} \big({\bar v}_i \to -v_i  \big) =:\breve{Q}^{(n)}(v^i)~,
\label{smile-iso}
\eea
with ${\bar Q}^{(n)} ({\bar v}_i)  :=\overline{ Q^{(n)}(v^i )}$
the complex conjugate of  $ Q^{(n)} (v^i)$, and ${\bar v}_i$ the complex conjugate of 
$v^i$. One can show that $ \breve{Q}^{(n)} (v)$ is a superconformal  weight-$n$ multiplet,
unlike the complex conjugate of $Q^{(n)}(v) $.
One can also check that 
\bea
\breve{ \breve{Q}}^{(n)}(v) =(-1)^n {Q}^{(n)}(v)~.
\label{smile-iso2}
\eea
Therefore, if  $n$ is even, one can define real isotwistor superfields, 
 $\breve{Q}^{(2m)}(v) = {Q}^{(2m)}(v )$.

Consider a  superconformal Killing vector  $\x_{\rm K}$ that obeys the conditions
\bea
\L_{ij} (z) = \s (z) =0~,
\eea
with $\L_{ij} (z) $ and $  \s (z)$  defined in eqs. 
(\ref{N=3sckv-Lambda}) and (\ref{N=3sckv-Lambda}), respectively.
It is called a $\cN=3$ Killing vector, for 
the set of all such vectors can be seen to form a superalgebra isomorphic to the $\cN=3$ super-Poincar\'e algebra. 
In the super-Poincar\'e case, the transformation law (\ref{harmult1}) reduces to 
the universal (weight-independent) form:
\bea
\d Q^{(n)} = -  \x_{\rm K} \, Q^{(n)} ~.
\label{promult2}
\eea
If we are interested in general $\cN=2$ supersymmetric (i.e. super-Poincar\'e invariant)  theories, 
not necessarily superconformal ones, 
projective multiplets should be defined by the relations  (\ref{ana}), (\ref{weight}) and (\ref{promult2}).

\subsection{N = 3 superconformal action}
Consider a real weight-2 projective multiplet $\cL^{(2)}(z,v)$. We can 
associate with  $\cL^{(2)}$  the following functional 
\bea
S=\frac{1}{8\p} \oint_{\g}  { v_i {\rm d} v^i }
\int {\rm d}^3x \, \big(D^{(-2)}\big)^2 \big(D^{(0)}\big)^2 \cL^{(2)} (z,v) \Big|_{\q =0}~.
\label{N=3action}
\eea
Here the line integral  is carried out over a closed contour,
$\g =\{v_i(t)\}$, in ${\mathbb C}P^1$.
The covariant derivatives $D^{(-2)}_\a$ and $D^{(0)}_\a$ in (\ref{N=3action}) depend on a constant ($t$-independent)
isotwistor $u_i$, in accordance with eq. (\ref{20-2}), which is 
subject to the only condition that $v(t)$ and $u$ form 
a linearly independent basis at each point of the contour $\g$,
that is $\big(v(t),u\big) \neq 0$. The functional is actually independent of $u$, since it is
 invariant under arbitrary
projective transformations of the form
\be
(u_i{}\,,\,v_i{})~\to~(u_i{}\,,\, v_i{} )\,R~,~~~~~~R\,=\,
\left(\begin{array}{cc}a~&0\\ b~&c~\end{array}\right)\,\in\,{\rm \sGL(2,\mathbb{C})}~.
\label{projectiveGaugeVar}
\ee
This invariance follows from the following three observations. First of all, an infinitesimal  transformation 
$\d u_i = b v_i$ acts  on the covariant derivatives in (\ref{N=3action}) 
as 
\bea
\d D^{(-2)}_\a = \frac{2b}{(v,u)} D^{(0)}_\a~, \qquad \d D^{(0)}_\a =  \frac{b}{(v,u)}  D^{(2)}_\a~.
\eea
The second observation is that $\big(D^{(0)}\big)^3 \propto \pa\, D^{(0)}$. The third observation is 
the analyticity of $\cL^{(2)}$, that is $D_\a^{(2)} \cL^{(2)} =0$.

The action (\ref{N=3action}) proves to be $\cN=3$ superconformal.
Indeed, making use of 
eq. (\ref{N=3volume}), the superconformal transformation of $\cL^{(2)} $ can be rewritten in the form
\bea
\d \cL^{(2)} = -\pa_a \Big( \x^a \cL^{(2)} \Big) +D^{(-2)}_\a \Big( \x^{(2) \a} \cL^{(2)} \Big) 
-2 D^{(0)}_\a \Big( \x^{(0) \a} \cL^{(2)} \Big) + {\bm \pa}^{(-2)} \Big( \L^{(2)} \cL^{(2)} \Big) ~.~~~~
\eea
Here the first three terms on the right do not contribute  to the variation of the action (\ref{N=3action}). 
It remains to
show that the last term also  produces no contribution to the variation of the action. 
To achieve this, we first point out the identity
\bea
 \big(D^{(-2)}\big)^2 \Big[ {\bm \pa}^{(-2)} , \big(D^{(0)}\big)^2 \Big] =0~.
 \eea
Our second observation is 
that (compare with eq. (2.50) in \cite{K-hyper})
\bea
(\dt{v}, v)  \big(D^{(-2)}\big)^2 \big(D^{(0)}\big)^2 {\bm \pa}^{(-2)} \Big( \L^{(2)} \cL^{(2)} \Big) 
&=& \big(u_iu_j D^{ij}\big)^2
\frac{ (\dt{v}, v) }{ (v,u)^4 } {\bm \pa}^{(-2)} \Big( 
 \big(D^{(0)}\big)^2 \L^{(2)} \cL^{(2)} \Big) \non \\
&=& -\frac{\rm d}{{\rm d} t } \Big(  \big(D^{(-2)}\big)^2 \big(D^{(0)}\big)^2  \L^{(2)} \cL^{(2)} \Big) ~,
\eea
where $(\dt{v},v)\, {\rm d}t = v_i \rd v^{i}$ is part of the line integral measure 
in (\ref{N=3action}). The result obtained is a special case of the following general property:
if $f^{(n)}(v) $ is a homogeneous function of $v^i$ of degree $n$, 
$f^{(n)}(c\, v) = c^n f^{(n)}(v) $,
then 
\bea
\frac{ (\dt{v}, v) }{ (v,u)^n } \,{\bm \pa}^{(-2)} f^{(n)} (v)
=  -\frac{\rm d}{{\rm d} t } \left(\frac{ f^{(n)}(v) }{ (v,u)^n } \right)~.
\eea
Since the line integral in (\ref{N=3action}) is a closed contour, we conclude that 
the action is  invariant under the $\cN=3 $ superconformal transformations.

\subsection{Projective multiplets  in the north chart  of two-sphere}
Without loss of generality,  we can assume  that the integration contour $\g$
in (\ref{N=3action}) does not pass through the ``{north pole}'' $v^{i}_{\rm north} \sim (0,1)$ of ${\mathbb C}P^1$.
It is then useful to introduce a complex 
({\it inhomogeneous}) coordinate $\z$ in the north chart, $\mathbb C$, 
of ${\mathbb C}P^1 ={\mathbb C} \cup \{\infty \}$: 
\bea 
 v^i = v^{1} \,(1, \z) ~,\qquad \z:=\frac{v^{2}}{v^{1}} ~,\qquad\quad 
{ i=1 ,2}
\label{Zeta}
\eea
and consider the projective multiplets in this chart.
Given a weight-$n$ projective superfield $ Q^{(n)}(z,v)$, we can associate with it 
a new object $ Q^{[n]}(z,\z )$ defined as 
\bea
Q^{(n)}(z,v)~\longrightarrow ~ Q^{[n]}(z,\z) \propto Q^{(n)}(z,v)~, \qquad 
\frac{\pa}{\pa \bar \z} Q^{[n]} =0 ~.
\eea
The explicit form of  $ Q^{[n]}(z,\z) $ depends on the multiplet under consideration.
The   $Q^{[n]}(z,\z) $ can be represented by a Laurent series
\bea
Q^{[n]}(z,\z) = \sum  \z^k Q_k (z) ~,
\label{seriess}
\eea
with $Q_k(z)$ some {ordinary $\cN=3$ superfields}. 

The covariant derivative $D^{(2)}_{\a}$ appearing in  (\ref{ana}) 
can now be represented as
\bea
D^{(2)}_{\a} = (v^1)^2 D^{[2]}_{\a}~, \qquad D^{[2]}_{\a} (\z)
&=& D^{22}_\a -2 \z D^{12}_\a +\z^2 D^{11}_\a  \non \\
&=& -\bar{\mathbb D}_\a -2 \z D^{12}_\a +\z^2 {\mathbb D}_\a~,~~~~
\eea
where we have introduced the spinor covariant derivates for $\cN=2$ superspace, 
${\mathbb D}_\a:= D^{11}_\a$ and $\bar{\mathbb D}_\a := -D^{22}_\a$, see also
subsection \ref{C1}. The analyticity conditions (\ref{ana}) 
turn into
\bea
D^{[2]}_{\a} (\z) Q^{[n]}(\z) = 0~.
\label{ancon}
\eea
These constraints have a simple and, at the same time, very important interpretation:
for all the component $\cN=3$ superfields $Q_k$ 
appearing in the series (\ref{seriess}), their dependence on $\q^\a_{12}$ 
is uniquely determined, according to (\ref{ancon}), in terms of their dependence 
on the Grassmann variables of $\cN=2$ superspace.
In other words, all information about the projective multiplet is encoded in its 
$\cN=2$ projection
\bea
Q^{[n]}(\z) |:= Q^{[n]}(\z)\big|_{\q_{12} =0} ~.
\eea

We now give two examples of superconformal projective multiplet. 
An off-shell hypermultiplet can be described in term of the so-called {\it arctic} weight-$n$
multiplet $\U^{(n)} (z, v)$ which is defined to be 
holomorphic  in the north chart
of ${\mathbb C}P^1$,
\bea
\U^{(n)} (z, v) &=&  (v^{1})^n\, \U^{[n]} (z, \z) ~, \qquad 
\U^{ [n] } (z, \z) = \sum_{k=0}^{\infty} \U_k (z) \z^k 
~, 
\label{arctic1}
\eea
and  its smile-conjugate {\it antarctic} multiplet $\breve{\U}^{(n)} (z,v) $,
 \bea
\breve{\U}^{(n)} (z,v) &=& (v^{1} \,\z \big)^{n}\, \breve{\U}^{[n]}(z,\z) ~, \qquad
\breve{\U}^{[n]}( z,\z) = \sum_{k=0}^{\infty}  {\bar \U}_k (z)\,
\frac{(-1)^k}{\z^k}~.~~~
\label{antarctic1}
\eea
The pair $\U^{[n]} ( \z)$ and $\breve{\U}^{[n]}(\z) $ constitute the so-called polar weight-$n$ multiplet.
The analyticity constraints (\ref{ancon}) imply that $\U_0$ is $\cN=2$ chiral and $\S_1$ is
$\cN=2$ complex linear, 
\bea
\bar{\mathbb D}_\a \U_0 =0~, \qquad \bar{\mathbb D}^2 \U_1 =0~,
\eea
while the other  components $\U_2,\U_3, \dots$, are unconstrained complex $\cN=2$ superfields.

Our second example is  the so-called real {\it tropical} multiplet $U^{(2n)} (z,v) $ of weight $n$ defined by 
\bea
U^{(2n)} (z,v) &=&\big({\rm i}\, v^{1} v^{2}\big)^n U^{[2n]}(z,\z) =
\big(v^{1}\big)^{2n} \big({\rm i}\, \z\big)^n U^{[2n]}(z,\z)~,  \non \\
U^{[2n]}(z,\z) &=& 
\sum_{k=-\infty}^{\infty} U_k  (z)\z^k~,
\qquad  {\bar U}_k = (-1)^k U_{-k} ~.
\label{2n-tropica1}
\eea
An example of such a multiplet with  $n=2$ is the Lagrangian $\cL^{(2)}$ in  (\ref{N=3action}).
The case $n=0$ is used to describe a vector multiplet \cite{LR-projective2}.

In the $\cN=3$ supersymmetric action (\ref{N=3action}),  the Lagrangian $\cL^{(2)}$ is a projective multiplet,
and therefore it is fully determined by its  $\cN=2$ projection $\cL^{(2)}\big|_{\q_{12}=0}$. 
The action (\ref{N=3action}) can be expressed in terms of this projection.
We recall that the integration contour $\g$
in (\ref{N=3action}) is chosen to lie outside the ``{north pole}'' $v^{i}_{\rm north} \sim (0,1)$ of ${\mathbb C}P^1$,
which allows us to use the inhomogeneous complex coordinate, $ \z$, 
 defined by $ v^i = v^{1} \,(1, \z)$.
Since the action is independent of $u_i$, the latter can be chosen to be   $ u_i =(1,0)$, 
such that $(v,u) = v^{1}\neq 0$. 
We  represent the Lagrangian in the form: 
\bea
\cL^{(2)}(z,v)={\rm i} \,v^{1}v^{2}\cL(z,\z)
= {\rm i} (v^{1})^2 \,\z\,{ \cL(z,\z)~, 
\qquad \breve{\cL} =\cL}~. 
\eea
Using the analyticity conditions $D^{[2]}_{\a} (\z) \cL(z,\z)=0$
allows us to rewrite (\ref{N=3action}) in the form:
\bea
S 
&=& \frac{1 }{ 2\pi\ri }
 \oint_\g 
 \frac{\rd\z }{  \z}
\int\rd^3 x \,{\rm d}^2 \q {\rm d}^2 {\bar \q}  \,
\cL(z,\z)\Big|_{\q_{12} =0}~.
\eea
Here the integration is carried out over the $\cN=2$ superspace.
The action is now formulated entirely in terms of $\cN=2$ superfields. 
At the same time, by construction, it is off-shell $\cN=3$ supersymmetric.

\subsection{N = 3 superconformal sigma-models}

We consider a system of interacting weight-one
 arctic  multiplets, 
$\U^{(1) I} (z,v) $, and their smile-conjugates,
$ \breve{\U}^{(1)\bar I }(z,v)$, described by a Lagrangian\footnote{The action generated by 
the Lagrangian (\ref{conformal-sm}) is real due to (\ref{smile-iso2}).}
of the form \cite{K-hyper}:
\bea
\cL^{(2)}  (\U^{(1)}, \breve{\U}^{(1)})= {\rm i} \, K (\U^{(1)I}, \breve{\U}^{(1) \bar J})~.
\label{conformal-sm}
\eea
Here $K(\F^I, {\bar \F}^{\bar J}) $ is a real function
of $n$ complex variables $\F^I$, with $I=1,\dots, n$, 
under the homogeneity condition
\bea
\F^I \frac{\pa}{\pa \F^I} K(\F, \bar \F) =  K( \F,   \bar \F)~.
\label{Kkahler22}
\eea
The function  $K(\F^I, {\bar \F}^{\bar J}) $ can be interpreted as the K\"ahler potential 
of a {\it K\"ahlerian cone} $\cM$ written in special complex coordinates in which 
the homothetic conformal Killing vector field $\c^I (\F)$ has the form $\c^I (\F)=\F^I$, 
see subsection \ref{N2 superconformal sigma-models}. 

By construction, the action generated by the Lagrangian (\ref{conformal-sm}) is invariant under
the $\cN=3$ superconformal transformation 
\be
\d \U^{(1)I} = - \Big(  \x -  \L^{(2)}  {\bm \pa}^{(-2)} \Big) \, \U^{(1)I} 
-\S \, \U^{(1)I} ~.
\ee
Keeping in mind the $\cN=2$ superconformal transformation law, eq. (\ref{genN=2sctr}), 
one could be tempted to put forward a different transformation law of the form:
\bea
\hat{\d} \U^{(1)I} = - \Big(  \x -  \L^{(2)}  {\bm \pa}^{(-2)} \Big) \, \U^{(1)I} -\S \, \c^I( \U^{(1)} )~.
\non
\eea
However, such a transformation is inconsistent with the arctic multiplet structure on two grounds:
(i) the second term in $\hat{\d} \U^{(1)I} $ is not a homogeneous function of $v^i$ of first degree 
unless  $\c^I (\F)$ is also homogenous of  first degree, which is true if $\c^I (\F) \propto \F^I$; 
(ii) the variation  $\hat{\d} \U^{(1)I}$ depends explicitly on the isotwistor $u_i$
unless   $\c^I \propto \U^{(1)I}$.
Off-shell $\cN=3$ supersymmetry requires special complex coordinates for K\"ahlerian cones.

There exists a more geometric formulation of the theory (\ref{conformal-sm}) described in detail in \cite{KLvU}.
It is realised in terms of a single weight-one arctic multiplet $\U^{(1)}$
and $n-1$ weight-zero arctic multiplets $\X^{ i} $. The corresponding Lagrangian is
\bea
K (\U^{(1)I}, \breve{\U}^{(1) \bar J}) = \U^{(1)} \breve{\U}^{(1)} \,\exp \Big\{ \cK (\X^{ i}, \breve{\X}^{\bar j}) \Big\}~,
\eea
where the original variables $\U^{(1)I}$ are related to the new ones by a holomorphic reparametrisation.
The arctic variables $\U^{(1)}$ and $\X^{ i} $ parametrise a holomorphic line bundle over a K\"ahler-Hodge 
manifold with K\"ahler potential $\cK (\vf^i, {\bar \vf}^{\bar j})$, see  \cite{KLvU} for more details.
We will not use this formulation in the present paper.

Once reformulated in $\cN=2$ superspace, 
the $\cN=3$ superconformal sigma-model action takes the from:
\bea
S &=& \frac{1}{ 2\pi\ri  } \oint_\g 
 \frac{\rd\z }{   \z}
\int \rd^3 x  \,{\rm d}^2 \q {\rm d}^2 {\bar \q}  \, 
 K (\U^I , \breve{\U}{}^{\bar J} ) ~.
 \label{N=3scsma}
\eea
Here the weight-one arctic multiplets  $ \U^I( \z) \equiv \U^{[1]I} (\z)$ and their smile-conjugates
$\breve{\U}^{\bar I} (\z) \equiv \breve{\U}^{[1] \bar I} (\z) $
embrace an infinite number of ordinary $\cN=2$ superfields.
\begin{subequations}
\bea
\U^I ( \z) &=& \sum_{k=0}^{\infty}\z^k  \U^I_k  =\F^I + \z \, \S^I  + O(\z^2) ~, 
\qquad
\bar{\mathbb D}_{ \a} \F^I =0~, 
\quad \bar{\mathbb D}^2 \S^I =0~ ,~~~~\\
\breve{\U}^{\bar J} (\z) &=& \sum_{n=0}^{\infty}  \,  (-\z)^{-n}\,
{\bar \U}_n^{\bar J}~.
\eea
\end{subequations}
We recall that the components $\U_2, \U_3, \dots$, are complex unconstrained 
$\cN=2$ superfields.

Our off-shell $\cN=3$ superconformal sigma-model (\ref{N=3scsma}) 
is determined by a real function of $n$ complex variables
 $K(\F^I, {\bar \F}^{\bar J}) $ 
which is arbitrary modulo the homogeneity condition (\ref{Kkahler22}).
We thus have a powerful scheme to generate $\cN=3$ superconformal sigma-models, 
and therefore hyperk\"ahler cones.

One can consider a more general sigma-model action, than  the superconformal theory (\ref{N=3scsma}),
of the form \cite{LR-projective}:
\bea
S &=& \frac{1}{ 2\pi\ri  } \oint_\g 
 \frac{\rd\z }{   \z}
\int \rd^3 x  \,{\rm d}^2 \q {\rm d}^2 {\bar \q}  \, 
L (\U^I , \breve{\U}{}^{\bar J}, \z ) ~.
\label{N=3scsma-SSUSY}
\eea
Here the Lagrangian is nor longer  required to obey any homogeneity condition and, moreover, 
can explicitly depend on $\z$. The action is no longer superconformal, but it is off-shell $\cN=3$ supersymmetric.
It is believed that (\ref{N=3scsma-SSUSY}) describes the most general $\cN=3$ supersymmetric sigma-model, 
see \cite{LR2010} for more details.

\subsection{Reformulation in terms of N = 2 chiral supefields}
The  $\cN=3$ superconformal sigma-model (\ref{N=3scsma}) involves the infinite set of auxiliary 
$\cN=2$ superfields  $\U^I_2, \U^I_3, \dots$, which are necessary to realise off-shell supersymmetry.
In order to describe the theory in terms of 
the physical superfields $\F^I$ and $\S^I$ only, 
all the auxiliary superfields have to be eliminated  with the aid of the 
corresponding algebraic equations of motion
\bea
\oint \frac{{\rm d} \z}{\z} \,\z^n \, \frac{\pa K(\U, \breve{\U} ) }{\pa \U^I} 
~ = ~ \oint \frac{{\rm d} \z}{\z} \,\z^{-n} \, 
\frac{\pa K(\U, \breve{\U} ) } {\pa \breve{\U}^{\bar J} } 
~ = ~ 0 ~, \qquad n \geq 2 ~ .               
\label{asfem}
\eea
Upon elimination of the auxiliaries, 
which is a difficult problem,
the superconformal symmetry becomes model-dependent 
and on-shell. The determination of its explicit form is a nontrivial technical problem. 
Fortunately, similar problems have been analysed in the case of 4D $\cN=2$ superconformal 
sigma-models in \cite{K-duality} (building on earlier works \cite{GK,AKL2}).
Here we can simply recycle the results of \cite{K-duality}.

Upon elimination of the auxiliary superfields, the action  (\ref{N=3scsma})  turns into
\bea
S [\F,  \S]   &=& \int 
\rd^3 x  \,{\rm d}^2 \q {\rm d}^2 {\bar \q} 
\, \Big\{
K \big( \F, \bar{\F} \big)+  
\cL \big(\F, \bar \F, \S , \bar \S \big)\Big\}
~,\non \\
\cL  \big(\F, \bar \F, \S , \bar \S \big)
&=&
\sum_{n=1}^{\infty}  \cL_{I_1 \cdots I_n {\bar J}_1 \cdots {\bar 
J}_n }  \big( \F, \bar{\F} \big) \S^{I_1} \dots \S^{I_n} 
{\bar \S}^{ {\bar J}_1 } \dots {\bar \S}^{ {\bar J}_n }
~,
\label{act-tab}
\eea
where $\cL_{I {\bar J} }=  - g_{I \bar{J}} \big( \F, \bar{\F}  \big) $ 
and the coefficients $\cL_{I_1 \cdots I_n {\bar J}_1 \cdots {\bar 
J}_n }$, for  $n>1$, 
are tensor functions of the K\"ahler metric
$g_{I \bar{J}} \big( \F, \bar{\F}  \big) 
= \pa_I 
\pa_ {\bar J}K ( \F , \bar{\F} )$,  the Riemann curvature $R_{I {\bar 
J} K {\bar L}} \big( \F, \bar{\F} \big) $ and its covariant 
derivatives.  
The function $\cL$ is characterised by the property 
\bea
{\S}^{I} \frac{\pa \cL}{\pa {\S}^{ I} }  &=&
{\bar \S}^{\bar I}
\frac{\pa \cL}{\pa {\bar \S}^{\bar I} }
\eea  
since the original model (\ref{N=3scsma}) 
is invariant under rigid $\sU(1)$  transformations
\be
\U(\zeta) ~~ \mapsto ~~ \U({\rm e}^{{\rm i} \a} \zeta) 
\quad \Longleftrightarrow \quad 
\U_n(z) ~~ \mapsto ~~ {\rm e}^{{\rm i} n \a} \U_n(z) ~.
\label{rfiber}
\ee
The Lagrangian $\cL \big(\F, \bar \F, \S , \bar \S \big)$ in  (\ref{act-tab})
obeys the homogeneity condition 
\bea
\Big( \F^I \frac{\pa}{\pa \F^I}  + \S^I \frac{\pa}{\pa \S^I} \Big) \cL \big(\F, \bar \F, \S , \bar \S \big)
= \cL \big(\F, \bar \F, \S , \bar \S \big)~.
\label{homo1}
\eea

Even though  the action (\ref{act-tab})  is formulated in terms of the physical superfields only, 
its Lagrangian $K \big( \F, \bar{\F} \big)+ L(\F , \bar \F, \S , \bar \S)$
is not a hyperk\"ahler potential, since  the dynamical variable $\S$ is {complex linear}. 
As discussed in section \ref{N3N4} the Lagrangian   coincides with 
the  hyperk\"ahler potential of the target space provided the theory is formulated 
in terms of $\cN=2$ chiral superfields and their conjugates only.
The theory has to be re-formulated in terms of chiral superfields by performing 
a duality transformation known as 
the generalised 
Legendre transform construction \cite{LR-projective}.
To construct a dual formulation of the theory (\ref{act-tab}), 
we consider the first-order action
\bea
S_{\rm first-order}=   \int \rd^3 x  \,{\rm d}^2 \q {\rm d}^2 {\bar \q}  \, 
\Big\{\,
K \big( \F, \bar{\F} \big)+  \cL\big(\F, \bar \F, \S , \bar \S \big)
+\J_I \,\S^I + {\bar \J}_{\bar I} {\bar \S}^{\bar I} 
\Big\}~.
\label{f-o}
\eea
Here the tangent vector $\S^I$ is complex unconstrained, 
while the one-form $\Psi_I$ is chiral, 
${\bar D}_{\dt \a} \J_I =0$.
Eliminating
$\S$'s and their conjugates, 
by using their equations of motion
\bea
\frac{\pa  }{\pa \S^I}  \cL\big(\F, \bar \F, \S , \bar \S \big)+ \J_I =0~,
\eea 
leads to the dual action
\bea
S [\F,  \J]  
&=&   \int \rd^3 x  \,{\rm d}^2 \q {\rm d}^2 {\bar \q}  \, 
{\mathbb K} (\F, \bar \F, \J, \bar \J )~, 
\label{act-ctb}
\eea
where the Lagrangian has the form
\bea
{\mathbb K} 
(\F, \bar \F, \J, \bar \J )&=&
K \big( \F, \bar{\F} \big)+    
\cH \big(\F, \bar \F, \J , \bar \J \big)~, \non \\
\cH 
\big(\F, \bar \F, \J , \bar \J \big)
&=& 
\sum_{n=1}^{\infty} \cH^{I_1 \cdots I_n {\bar J}_1 \cdots {\bar 
J}_n }  \big( \F, \bar{\F} \big) \J_{I_1} \dots \J_{I_n} 
{\bar \J}_{ {\bar J}_1 } \dots {\bar \J}_{ {\bar J}_n } 
\label{h}
\eea
and $\cH^{I {\bar J}} \big( \F, \bar{\F} \big)  =  g^{I {\bar J}} \big( \F, \bar{\F} \big) $.
The function $\cH$ is characterised by the following homogeneity properties: 
\begin{subequations}
\bea
{\J}_{I} \frac{\pa \cH}{\pa {\J}_{ I} }  &=&
{\bar \J}_{\bar I}
\frac{\pa \cH}{\pa {\bar \J}_{\bar I} }  ~, \\
\Big( \F^I \frac{\pa}{\pa \F^I}  &+&  { \J}_{ I}   \frac{\pa }{\pa { \J}_{ I} }
 \Big) \cH 
= \cH 
~.
\label{homo5}
\eea
\end{subequations}
The derivation of the above results is similar to the 4D $\cN=2$ case \cite{K-duality}. 

The Lagrangian ${\mathbb K} (\F, \bar \F, \J, \bar \J )$ in (\ref{act-ctb})
is the K\"ahler potential of a hyperk\"ahler cone. The action is still $\cN=3$ superconformal, however
the superconformal transformations form a closed algebra only on the mass shell.
To describe the symmetries of the  model, 
we introduce the condensed notation 
\be
\f^a := (\F^I\,, \J_I) ~, \qquad {\bar \f}^{\,\bar a} = ({\bar \F}^{\bar I}\,, {\bar \J}_{\bar I}), 
\label{nott}
\ee
as well as the standard symplectic matrix ${\mathbb J} =({\mathbb J}^{a b} )$, its inverse
${\mathbb J}^{-1} =(-{\mathbb J}_{a b} )$ and their complex conjugates,
\bea
{\mathbb J}^{a b} = {\mathbb J}^{\bar a \bar b} = 
\left(
\begin{array}{rc}
0 ~ &  {\mathbbm 1} \\
-{\mathbbm 1} ~ & 0  
\end{array}
\right)~,  \qquad 
{\mathbb J}_{a b} = {\mathbb J}_{\bar a \bar b} = 
\left( 
\begin{array}{rc}
0 ~ &  {\mathbbm 1} \\
-{\mathbbm 1} ~ & 0  
\end{array}
\right)~. 
\eea
As shown in Appendix C, subsection \ref{C1}, 
an arbitrary  $\cN=3$ superconformal transformation  
decomposes into two transformations  in $\cN=2$ superspace: (i) an
$\cN=2$ superconformal transformation generated by an $\cN=2$ superconformal 
Killing vector $\bm \x$; (ii) an extended superconformal transformation generated by 
a  {\it real} spinor superfield  $\r^\a $ obeying the constraints (\ref{C5}).
The action  (\ref{act-ctb})
is invariant under the $\cN=2$ superconformal transformation
\bea
\d \f^a =- {\bm \x} \f^a - \hf {\bm \s} \f^a~. 
\eea
It is also invariant under the extended superconformal transformation
\bea 
\d \f^a &=&\hf
{\bar {\mathbb D}}^2 \Big\{ \bar{\bm \r} \,  {\mathbb J}^{ab} \,\frac{\pa \mathbb K}{\pa \f^b}
\Big\} ~,
\label{7.50}
\eea
where the complex scalar ${\bm \r}$  and its complex conjugate $\bar {\bm \r}$
are defined as follows:
\be
\r_\a = {\mathbb D}_\a {\bm \r} = \bar{\mathbb D}_\a \bar {\bm \r} ~.
\ee
The transformation law (\ref{7.50}) can be derived by applying the four-dimensional construction 
of \cite{K-duality}.\footnote{Strictly speaking, 
in order to use the 4D $\cN=2$ construction of \cite{K-duality} for deriving eq. (\ref{7.50}, 
we need $\cN=4 $ supersymmetry  in three dimensions. However, it is shown in section \ref{N3N4}
that $\cN=3$ supersymmetry implies $\cN=4$.}

\section{Off-shell N = 4 superconformal sigma-models} 
\setcounter{equation}{0}

$\cN=3$ supersymmetry in three dimensions is similar to and intimately related to $\cN=4$. 
We will show in the next section
that any $\cN=3$ supersymmetric sigma-model possesses a hidden $\cN=4$ supersymmetry.
The crucial difference between the $\cN=3$ and $\cN=4$ projective superspace approaches is that 
any $\cN=3$ projective multiplet has two $\cN=4$ cousins: left and right ones. 
This can be seen from the fact that  
the $\cN=3$ projective superspace (\ref{N=3ps})
turns into
\bea
{\mathbb M}^{3|8} \times \Big\{ {\mathbb C}P^1_{\rm L}  \bigcup {\mathbb C}P^1_{\rm R} \Big\}
= \Big(  {\mathbb M}^{3|8} \times  {\mathbb C}P^1_{\rm L}  \Big) \bigcup 
\Big( {\mathbb M}^{3|8} \times  {\mathbb C}P^1_{\rm R} \Big)~,
\label{N=4ps}
\eea
in the $\cN=4$ case, see the discussion in subsection 4.3. 
However, one could also consider larger supermultiplets defined on the bi-projective superspace
${\mathbb M}^{3|8} \times  {\mathbb C}P^1 \times  {\mathbb C}P^1$.

The mirror projective spaces ${\mathbb C}P^1_{\rm L}  $ and ${\mathbb C}P^1_{\rm R}  $ 
can be parametrised
by homogeneous complex coordinates, 
or isotwistors, 
$v_{\rm L} = (v^i)$ and $v_{\rm R} = (v^{\bar k})$, 
respectively. They can be used to define two maximal subsets of 
strictly anticommuting derivatives,  $D^{(1){\bar k}}_\a:=v_{i}D_\a^{i{\bar k}}$ and 
$D^{(1){ i}}_\a:=v_{\bar k}D_\a^{i{\bar k}}$, such that 
\bsubeq
\bea
& \{D_\a^{(1){\bar k}},D_\b^{(1){\bar l}}\}=0~;\\
& \{D_\a^{(1){ i}},D_\b^{(1){j}}\}=0~.
\eea
\esubeq
These relations immediately imply that we can introduce
two types of projective multiplets, left $Q_{\rm L}^{(n)}(v_{\rm L}) $
and right  $Q_{\rm R}^{(n)}(v_{\rm R}) $ ones, which depend on the different isotwistors 
and obey the covariant constraints 
\begin{subequations}
\bea
D_\a^{(1){\bar{k}}}Q_{\rm L}^{(n)} (v_{\rm L}) &=&0~; 
\label{8.3a}\\
D_\a^{(1)i}Q_{\rm R}^{(n)} (v_{\rm R} )&=&0~.
\label{8.3b}
\eea
\end{subequations}
To define  $\cN=4$ superconformal projective multiplets, left or right, 
we should again make use of the relations  (\ref{weight}) and  (\ref{harmult1})
in each of the two sectors, left and right.
In particular, the $\cN=4$ superconformal transformation laws 
of the left and right projective multiplets are respectively
\bsubeq
\bea
\d Q^{(n)}_{\rm L} &=& - \Big(  \x -  \L_{\rm L}^{(2)}  {\bm \pa}_{\rm L}^{(-2)} \Big) \, Q_{\rm L}^{(n)} 
-n \,\S_{\rm L} \, Q^{(n)}_{\rm L} ~;~~~
\label{8.4a}
\\
\d Q^{(n)}_{\rm R} &=& - \Big(  \x -  \L_{\rm R}^{(2)}  {\bm \pa}_{\rm R}^{(-2)} \Big) \, Q_{\rm R}^{(n)} 
-n \,\S_{\rm R} \, Q^{(n)}_{\rm R}~.
\label{8.4b}
\eea
\esubeq
The objects appearing in these transformation laws have been defined in 
subsection \ref{N=4superKillings}.
In particular, associated with  the left isotwistor $v_{\rm L} = (v^i)$  is a linearly independent one
$u_{\rm L} = (u_i)$ such that $(v_{\rm L} \cdot u_{\rm L} ):= v^i u_i \neq 0$, and similarly in the right sector.

Next, we can introduce a $\cN=4$ superconformal action principle
in complete analogy with the $\cN=3$ case.
Consider a left  real weight-2 projective multiplet $\cL_{\rm L}^{(2)}(z,v_{\rm L})$
and a right real weight-2 projective multiplet $\cL_{\rm R}^{(2)}(z,v_{\rm R})$. 
Then,   the following functional 
\bea
S&=&\phantom{-} \frac{1}{2\p} \oint_{\g_{\rm L}}  { v_i {\rm d} v^i }
\int {\rm d}^3x \, D_{\rm L}^{(-4)} \cL_{\rm L}^{(2)} (z,v_{\rm L}) \Big|_{\q =0}
\non\\
&&
+\frac{1}{2\p} \oint_{\g_{\rm R}}  { v_{\bar k} {\rm d} v^{\bar k} }
\int {\rm d}^3x \, D_{\rm R}^{(-4)} \cL_{\rm R}^{(2)} (z,v_{\rm R}) \Big|_{\q =0}
\label{N=4action}
\eea
is invariant under the $\cN=4$ superconformal transformations.
Here we have defined
\bsubeq
\bea
D_{\rm L}^{(-4)} &:=&
\frac{1}{48}D^{(-2){\bar k}{\bar l}}D^{(-2)}_{{\bar k}{\bar l}}~,~~~
D^{(-2)}_{{\bar k}{\bar l}}:=D^{(-1)\g}_{{\bar k}}D_{\g{\bar l}}^{(-1)}~,
\\
D_{\rm R}^{(-4)} &:=&
\frac{1}{48}D^{(-2) ij}D^{(-2)}_{ij}~,~~~
D^{(-2)}_{ij}:=D^{(-1)\g}_{i}D_{\g j}^{(-1)}
~,
\eea
\esubeq
see subsection \ref{N=4superKillings} for the definition of 
spinor covariant derivatives $D^{(-1)i}_\a$ and $D^{(-1)\bar k}_\a$.

Supersymmetric matter can be  described by weight-$n$ arctic multiplets,
left $\U^{(n)}_{\rm L} (v_{\rm L}) $ and right   $\U^{(n)}_{\rm R} (v_{\rm R}) $, 
and their conjugate  antarctic multiplets.
In order to get  a better understanding of the difference between the left and the right arctic multiplets, 
it suffices  to study the superconformal transformation properties
of their component superfields\footnote{For each of the mirror spaces ${\mathbb C}P^1_{\rm L} $ and 
${\mathbb C}P^1_{\rm R}$,  we choose its north chart to define the expansions   
(\ref{8.7a}) and (\ref{8.7b}).}
$ \U_{{\rm L},k }$ and  $ \U_{{\rm R},k }$ defined by 
\begin{subequations}
\bea
\U^{(n)}_{\rm L} ( v_{\rm L}) &=&  (v^{1})^n\, \U^{[n]}_{\rm L} ( \z_{\rm L}) ~, \qquad 
\U^{ [n] }_{\rm L} ( \z_{\rm L}) 
= \sum_{k=0}^{\infty} \U_{{\rm L},k }\, \z_{\rm L}^k ~, \qquad   \z_{\rm L}:= \frac{v^2}{v^1}
\label{8.7a}\\
\U^{(n)}_{\rm R} ( v_{\rm R} ) &=&  (v^{\bar 1})^n\, \U^{[n]}_{\rm R} ( \z_{\rm R}) ~, \qquad 
\U^{ [n] }_{\rm R} ( \z_{\rm R}) = \sum_{k=0}^{\infty} \U_{{\rm R}, k}\,  \z_{\rm R}^k ~,\qquad
 \z_{\rm R}:=  \frac{v^{\bar 2}}{v^{\bar 1}}
 \label{8.7b}
\eea
\end{subequations}
considered as $\cN=2$ superfields.
Then,   
the analyticity constraints (\ref{8.3a}) imply that
the left components
\bea
\F_{\rm L} := \U_{{\rm L},0}~, \qquad \S_{\rm L} := \U_{{\rm L},1}
\label{8.8}
\eea
are $\cN=2$ chiral and complex linear superfields, respectively,  and similarly for the right $\cN=2$ superfields
\bea
\F_{\rm R} := \U_{{\rm R},0}~, \qquad \S_{\rm R} := \U_{{\rm R},1}~.
\label{8.9}
\eea
The other component $\cN=2$ superfields are complex unconstrained.

We know from  Appendix C, subsection \ref{C2}, that 
any arbitrary  $\cN=4$ superconformal transformation  
decomposes into three transformations  in $\cN=2$ superspace: (i) an
$\cN=2$ superconformal transformation generated by an $\cN=2$ superconformal 
Killing vector ${\bm \x}$; (ii) an extended superconformal transformation generated by 
a  {\it complex} spinor superfield  $\r^\a $ obeying the constraints
(\ref{C9}); (iii) a shadow $\sU (1)$ rotation. Using the $\cN=4$ superconformal 
transformation laws (\ref{8.4a}) and (\ref{8.4b}), it is a simple exercise to work out the transformations
of the $\cN=2$ component superfields.

The left and the right arctic multiplets turn out to have identical $\cN=2$ superconformal transformations, 
so we will omit the label `L' and `R.'
Specifically, one can show  that the component superfields $\U_k$ transform as 
\bea
\d_{\bm \x} \U_{k} &=& - \Big(  {\bm \x}
+\frac{n-k}{2}{\bm\s}
+\frac{k}{2}\bar{\bm\s}
 \Big)  \U_{k}~.
 \label{tl-36}
 \eea
As already mentioned,  the leading components 
$\F:= \U_0$ and $\S:= \U_1$ are chiral and complex linear respectively.
The transformation law (\ref{tl-36})  implies 
\begin{subequations}
\bea
\d_{\bm \x} \F &=& - \Big(  {\bm \x}
+\frac{n}{2}{\bm\s}
 \Big)  \F \quad \longrightarrow \quad \bar{\mathbb D}_\a \d_{\bm \x} \F =0\\
\d_{\bm \x} \S &=& - \Big(  {\bm \x}
+\frac{n-1}{2}{\bm\s}
+\frac{1}{2}\bar{\bm\s}
 \Big)  \S \quad \longrightarrow \quad \bar{\mathbb D}^2 \d_{\bm \x} \S =0~.
\eea
\end{subequations}
In the case of a weight-one multiplet, $n=1$, the $\bm \s$-term in the variation $\d_{\bm \x} \S$ 
drops out.

The left and the right arctic multiplets transform almost identically under the shadow $\sU (1)$ 
rotation:
\bea
\d_\a \U_{{\rm L},k} &=&  - \frac{(n-2k)}{2} \ri \,\a\U_{{\rm L},k}~, \qquad
\d_\a \U_{{\rm R},k} =   \frac{(n-2k)}{2} \ri\, \a\U_{{\rm R},k}~.
\eea

A real difference  between 
the left and the right arctic multiplets occurs only in the sector
of extended superconformal transformations. 
The left weight-$n$ arctic multiplet can be shown to transform as follows:
\bsubeq
\bea
\d \U_{{\rm L},0} &=& 
  \Big( \r^\a \mDB_\a
 -\frac{1}{2}(\mDB_{\a}{\r}^{\a})
 \Big)  \U_{{\rm L},1}
 ~,
\label{left-39a}\\
\d \U_{{\rm L},k} &=& 
  \Big(  {\bar \r}^\a \mD_\a
+ \frac{k-n-1}{2}(\mD_{\a}{\bar\r}^{\a})
 \Big)  \U_{{\rm L},k-1} \non \\
&&\qquad   + \Big( \r^\a \mDB_\a
 -\frac{k+1}{2}(\mDB_{\a}{\r}^{\a})
 \Big)  \U_{{\rm L},k+1}~, \qquad k>0~.
\label{left-39b}
\eea
\esubeq
For the right weight-$n$ arctic multiplet we get
\bsubeq
\bea
\d \U_{{\rm R},0} &=&  - \Big( {\bar\r}^\a \mDB_\a
 -\frac{1}{2}(\mDB_{\a}{\bar \r}^{\a})
 \Big)  \U_{{\rm R},1} ~,
\label{right-40a}\\
\d \U_{{\rm R},k} &=& 
 - \Big(
 { \r}^\a \mD_\a
+ \frac{k-n-1}{2}(\mD_{\a}{\r}^{\a})
 \Big)  \U_{{\rm R},k-1} \non \\
&& \qquad  - \Big( {\bar\r}^\a \mDB_\a
 -\frac{k+1}{2}(\mDB_{\a}{\bar \r}^{\a})
 \Big)  \U_{{\rm R},k+1} ~, \qquad k >0~.
\label{right-40b}
\eea
\esubeq
We see that the transformation law of the right superfields, $\U_{{\rm R},k} $, can be obtained 
from that of the left ones, $\U_{{\rm L},k} $, by applying the replacement: 
$\U_{{\rm L},k}  \to \U_{{\rm R},k} $ and $\r_\a \to -{\bar \r}_\a$.

The leading left $\cN=2$ superfields (\ref{8.8}) 
are chiral and complex linear, respectively, and similarly for the right $\cN=2$ superfields 
(\ref{8.9}).
It is important to check that their variations under the extended superconformal transformation
obey the same constraints. To make manifest the chirality of $ \d \F_{\rm L} $ and $\d \F_{\rm R}$, 
we point out the following. 
It follows from the constraints 
(\ref{C9}) that the superconformal parameters $\r_\a$ and ${\bar \r}_\a$ can be represented in the form:
\begin{subequations}
\bea
\r_\a &=& \bar{\mathbb D}_\a \bar{\bm \r}_{\rm L} ~, \qquad {\bar \r}_\a = {\mathbb D}_\a {\bm \r}_{\rm L} ~; \\
\r_\a &=& {\mathbb D}_\a {\bm \r}_{\rm R} ~, \qquad {\bar \r}_\a = \bar{\mathbb D}_\a \bar{\bm \r}_{\rm R} ~,
\eea
\end{subequations}
for some complex scalars $ {\bm \r}_{\rm L}$ and $ {\bm \r}_{\rm R}$.
Now, the variations (\ref{left-39a}) and (\ref{right-40a}) can be rewritten in the form:
\bsubeq
\bea
\d \F_{\rm L} &=& -\frac{1}{2}\mDB^2 (\bar{\bm \r}_{\rm L} \S_{\rm L})~, \\
\d \F_{\rm R} &=& \phantom{-}\frac{1}{2}\mDB^2  (\bar{\bm \r}_{\rm R}\S_{{\rm R}})~,
\eea
\esubeq
with $\bar{\mathbb D}^2 := \mDB_{\a}\mDB^{\a}$. These expressions show that $\d \F_{\rm L} $ and 
$\d \F_{\rm R} $ are indeed chiral.
Let us consider eq. (\ref{left-39b}) with $k=1$ in order to check that $\d \S_{{\rm L}}$ 
satisfies the condition
$\mDB^2 \d \S_{{\rm L}}=0$. We have
\bea
\d \S_{{\rm L}} &=& 
  \Big(  {\bar \r}^\a \mD_\a
- \frac{n}{2}(\mD_{\a}{\bar\r}^{\a})
 \Big)  \F_{{\rm L}} 
+ \Big( \r^\a \mDB_\a
 - (\mDB_{\a}{\r}^{\a})
 \Big)  \U_{{\rm L},2} \non \\
&=&  \Big(  {\bar \r}^\a \mD_\a
- \frac{n}{2}(\mD_{\a}{\bar\r}^{\a})
 \Big)  \F_{{\rm L}} 
 - \mDB_{\a} \Big( {\r}^{\a}
 \U_{{\rm L},2}  \Big)~.
\eea
Here the second term in the second line is clearly complex linear. To prove that the first term is also 
complex linear,  it suffices to use the constraints (\ref{C9}) as well as the chirality of $\F_{\rm L}$.

In conclusion, we write down a general $\cN=4$ superconformal sigma-model 
described by left and right weight-one arctic multiplets and their conjugates
\bea
S &=& \frac{1}{ 2\pi\ri  } \oint_{\g_{\rm L}} 
 \frac{\rd\z_{\rm L} }{   \z_{\rm L}}
\int \rd^3 x  \,{\rm d}^2 \q {\rm d}^2 {\bar \q}  \, 
K_{\rm L} (\U_{\rm L} , \breve{\U}_{\rm L} ) \non \\
&+&  \frac{1}{ 2\pi\ri  } \oint_{\g_{\rm R} } 
 \frac{\rd\z_{\rm R}  }{   \z_{\rm R}}
\int \rd^3 x  \,{\rm d}^2 \q {\rm d}^2 {\bar \q}  \, 
K_{\rm R} (\U_{\rm R} , \breve{\U}_{\rm R} )~.
\label{N=4scsma}
\eea
Here the K\"ahler potentials $K_{\rm L} $ and $K_{\rm R} $ obey homogeneity conditions of the type
(\ref{Kkahler22}).

\section{N = 3 SUSY implies N = 4 SUSY} 
\setcounter{equation}{0}
\label{N3N4}
It is the accepted lore that $\cN=3$ supersymmetry in three space-time dimensions
implies $\cN=4$ supersymmetry.\footnote{The standard argument (see, e.g., \cite{Bagger}) 
is as follows: $\cN$-extended supersymmetry requires $\cN-1$ anti-commuting complex structures.
In the case $\cN=3$, the target space has two such structures, $I$ and $J$. Their product $K:=I \,J$
is a third complex structure which anticommutes with $I$ and $J$, and therefore
the sigma-model is $\cN=4$ supersymmetric. This argument tells us nothing about off-shell supersymmetry.}
Here we provide two proofs of this claim by considering nonlinear sigma-models 
that possess different amounts of manifestly realised supersymmetry: (i) $\cN=2$; and (ii) $\cN=3$.

\subsection{Analysis in N = 2 superspace}
We start from a general 3D $\cN=2$ supersymmetric nonlinear $\s$-model
\cite{Zumino} 
\bea
S&=& \int {\rm d}^3 x \,{\rm d}^2 \q {\rm d}^2 {\bar \q}  \, K\big(\f^a, {\bar \f}^{\overline{b}}\big)~, 
\qquad {\mathbb \DB}_{ \b} \f^a =0~, 
\label{N=2sigma-model}
\eea
with $K(\f ,\bar \f ) $  the K\"ahler potential of a K\"ahler manifold $\cM$, 
and look for those restrictions on the target space geometry which 
guarantee the existence of a hidden  supersymmetry.
We emphasise that the sigma-model under consideration is not required to be superconformal.

The first argument why ${\cal N} =3$ implies ${\cal N}=4$ is based on an explicit
calculation. Building on the four-dimensional analysis \cite{LR,HKLR},
a general ansatz for an additional supersymmetry is of the form
\bea
\delta \phi^a = \frac12 \mDB^2 \left(\bar{\rho} \,\bar{\Omega}^a\right)~, \qquad
\d {\bar \f}^{\bar a} = \hf {\mathbb D}^2  \left({\rho} \,{\Omega}^{\bar a}\right)~,
\label{LR-ansatz0}
\eea
where $\Omega^a (\f , \bar \f)$ is a function  
associated with the  K\"ahler manifold $\cM$, 
and $\bar{\rho}$ is a constant
antichiral {\em superfield} satisfying
\bea
\mD_\alpha \bar{\rho} = \partial_{\alpha\beta} \bar{\rho} = \mDB^2\bar{\rho} = 0~.
\eea
In three dimensions it is furthermore possible to choose the
spinor component of $\rho$ to be real up to an arbitrary constant phase $\lambda$
\bea
{\rm e}^{-{\rm i }\lambda} \,\mD_\alpha \rho\big| 
= {\rm e}^{{\rm i}\lambda} \,\mDB_{\alpha} \bar{\rho}\big|~,
\eea
with $ \l \in {\mathbb R}$ some fixed parameter for  the transformation under consideration.
A real spinor parameter corresponds to one additional (third) supersymmetry. 
Thus $\bar \r$ has the following explicit form:
\bea
{\bar \r}  = \bar \t + {\rm e}^{ -{\rm i} \l }\, { \e}_\a {\bar \q}^\a~, \qquad \bar \t = {\rm const}~, 
\qquad {\e}_\a = {\bar \e}_\a ={\rm const}~.
\label{par}
\eea
In fact, the third supersymmetry transformation is obtained by setting $\bar \t =0$.
However, commuting the manifestly realised  first and second supersymmetry transformations with the third one
results in a {\it central charge transformation} which corresponds to the choice  
${\bar \r} = \bar \t \in {\mathbb C}$ in (\ref{LR-ansatz0}). For our proof below, 
it is not necessary to assume $\bar \t$  to be complex.

We would like to show  that there is no {\em special} choice of $\l$ in (\ref{par})
for which the supersymmetry
variation cancels between the $\rho$ terms and the $\bar\rho$ terms.
Varying the action (\ref{N=2sigma-model}) gives
\bea
\delta_{\rho,\bar{\rho}} S &=& \frac12 \int {\rm d}^3 x \,{\rm d}^2 \q {\rm d}^2 {\bar \q}  \,
K_a \mDB^2(\bar{\rho}\bar{\Omega}^a) 
+ {\rm c.c.} \non \\
&=& -  \frac12 \int {\rm d}^3 x \,{\rm d}^2 \q {\rm d}^2 {\bar \q}  \,
g_{a\bar b} (\mDB_\a {\bar \f}^{\bar b} ) \mDB^\a(\bar{\rho}\bar{\Omega}^a) 
+{\rm c.c.}~,
\label{var-r}
\eea
with $g_{a \bar{b} } =K_{a\bar b}:= \pa_a \pa_{\bar b} K$ the K\"ahler metric.
Choosing here ${\bar \r} = \bar \t = {\rm const}$, the variation of the action reduces to
\bea
\delta_{\t,\bar{\t}} S &=&  -  \frac12\bar \t \int {\rm d}^3 x \,{\rm d}^2 \q {\rm d}^2 {\bar \q}  \,
{\bar \o}_{ \bar{b}  \bar{c} } \, (\mDB_\a {\bar \f}^{\bar b} ) \mDB^\a {\bar \f}^{\bar c}  
+{\rm c.c.}~,
\label{var-t}
\eea
where we have denoted
\bea
{\bar \o}_{ \bar{b}  \bar{c} } :=  g_{ \bar{b} a}\, {\bar \O}^a{}_{,\bar{c} }
~, \qquad
  {\bar \O}^a{}_{,\bar{c} }:= \pa_{\bar c}  {\bar \O}^a~.
\label{two-form1}  
\eea
The two terms in (\ref{var-t}) can be seen to have different functional form. 
The condition $\delta_{\t,\bar{\t}} S =0$ is therefore equivalent to the requirement that each term on the right 
of (\ref{var-t}) vanishes separately, and thus
\bea
\int {\rm d}^3 x \,{\rm d}^2 \q {\rm d}^2 {\bar \q}  \,
{\bar \o}_{ \bar{b}  \bar{c} } \, (\mDB_\a {\bar \f}^{\bar b} ) \mDB^\a {\bar \f}^{\bar c}   =0~.
\eea
This holds if 
\bea
{\bar \o}_{ \bar{b}  \bar{c} }  =-  {\bar \o}_{ \overline{c}  \bar{b} }~,
\label{two-form2}
\eea
and hence  the target space is endowed with the {\it  two-form}\footnote{On the mass shell, 
the supersymmetry transformation  (\ref{LR-ansatz0})
takes the form:
$
\d \f^a = {\rm e}^{ -{\rm i} \l }\,{\e}_{ \a}  \,{\bar \O}^a{}_{,\bar b} \,{\bar {\mathbb D}}^{ \a} {\bar \f}^{\bar b}.
$
Since $\d \f^a$ should be a vector field on $\cM$, we conclude that ${\bar \O}^a{}_{,\bar b} $
is a tensor field on $\cM$, and therefore $\o_{ab} $ is a two-form. }
 $\o_{bc} =  g_{b \bar{a} }\, { \O}^{\bar a}{}_{,{c} }$ and its conjugate
${\bar \o}_{ \overline{b}  \overline{c} } $. We conclude that the action is invariant under 
the central charge transformation when $\omega$ is antisymmetric.
Thus there is no way that the $\tau$ and $\bar\tau$ variations could possibly
cancel each other without the variations being separately zero.

Now, the variation of the action, 
eq. (\ref{var-r}), turns into
\bea
\delta_{\rho,\bar{\rho}} S &=&
 -  \hf  {\rm e}^{ -{\rm i} \l }\, { \e}_\a 
 \int {\rm d}^3 x \,{\rm d}^2 \q {\rm d}^2 {\bar \q}  \,
g_{a\bar b}\,\bar{\Omega}^a 
 \mDB^\a {\bar \f}^{\bar b}  
+{\rm c.c.}~
\label{var-r2}
\eea
We see that the two terms in $\delta_{\rho,\bar{\rho}} S $ have different functional form.
Imposing the condition $\delta_{\rho,\bar{\rho}} S =0$ is equivalent to the fact that each term 
should vanish separately. Therefore, we have to require
\bea
\delta_{\bar \rho} S &=&
   \hf  {\rm e}^{ -{\rm i} \l }\, { \e}^\a 
 \int {\rm d}^3 x \,{\rm d}^2 \q {\rm d}^2 {\bar \q}  \,
g_{a\bar b}\,\bar{\Omega}^a 
\mDB_\a {\bar \f}^{\bar b} \non \\
&=&    \hf  {\rm e}^{ -{\rm i} \l }\, { \e}^\a 
 \int {\rm d}^3 x \,{\rm d}^2 \q {\rm d}^2 {\bar \q}  \,
\bar{\Omega}^a \mDB_\a K_a
 \equiv 0
\label{var-r3}
\eea
for the action to be invariant under the third supersymmetry.  

Before starting to analyse the  condition (\ref{var-r3}), we wish to make an important observation. 
Let us forget for a moment that ${\bar \o}_{ \bar{b}  \bar{c} }$ defined by (\ref{two-form1})
is antisymmetric, eq.  (\ref{two-form2}).  Then, we can derive several identities:
\bea\label{expns}
B_{\bar b\bar c\bar d} &:=& K_{a\bar b\bar c}\bar{\Omega}^{a}{}_{, \bar d}
+K_{a\bar d}\bar{\Omega}^{a}{}_{, \bar b\bar c}
\nonumber\\
&=&\nabla_{\bar c} {\bar \omega}_{\bar d\bar b} +
\Gamma^{\bar e}_{\bar b\bar c}( {\bar \omega}_{\bar d\bar e} +
{\bar \omega}_{\bar e\bar d}) \non\\
 &=&\frac12 \Big(\partial_{\bar b}( {\bar \omega}_{\bar c\bar d}+ {\bar \omega}_{\bar d\bar c})
 +\partial_{\bar c}( {\bar \omega}_{\bar b\bar d}+ {\bar \omega}_{\bar d\bar b})
 -\partial_{\bar d}( {\bar \omega}_{\bar b\bar c}+ {\bar \omega}_{\bar c\bar b})\Big)~.
\label{1313}
\eea
If eq.  (\ref{two-form2}) holds, then the relation (\ref{1313}) leads to 
the important result:
\bea
\nabla_{\bar a}  {\bar \o}_{ \overline{b}  \overline{c}  }&=&0 ~.
\label{se2b}
\eea

A non-trivial piece of information can be extracted from the condition  (\ref{var-r3})
without doing hard calculations (compare with \cite{K-duality}). 
Eq.  (\ref{var-r3}) tells us that the functional in the second line
must vanish identically. Let us vary this functional with respect to the antichiral 
superfields $\bar \f$s while keeping the chiral ones $\f$s fixed.  
Since ${\bar \o}_{ \bar{b}  \bar{c} }$ is antisymmetric, we obtain
 \bea
 \d_{\bar \f} 
 \int {\rm d}^3 x \,{\rm d}^2 \q {\rm d}^2 {\bar \q}  \,
\bar{\Omega}^a \mDB_\a K_a
=-2  \int {\rm d}^3 x \,{\rm d}^2 \q {\rm d}^2 {\bar \q}  \,
{\bar \o}_{ \bar{b}  \bar{c} } \,\d {\bar \f}^{\bar b} \mDB_\a  {\bar \f}^{\bar c}~.
\eea
To make this variation vanish, one has to impose
the condition
\bea
\nabla_a  {\bar \o}_{ \overline{b}  \overline{c}  }
&=& \pa_a {\bar \o}_{ \overline{b}  \overline{c}  }=0 ~, 
\label{se2}
\eea
which means that $  {\bar \o}_{ \overline{b}  \overline{c}  }$ is anti-holomorphic,
$  {\bar \o}_{ \overline{b}  \overline{c}  } =   {\bar \o}_{ \overline{b}  \overline{c}  } (\bar \f)$.
Indeed, since $\bar \f$ and $\d \bar \f$ are antichiral, we then get 
\bea
 \int {\rm d}^3 x \,{\rm d}^2 \q {\rm d}^2 {\bar \q}  \,
{\bar \o}_{ \bar{b}  \bar{c} } \,\d {\bar \f}^{\bar b} \mDB_\a  {\bar \f}^{\bar c}
= - \frac{1}{4}  \int {\rm d}^3 x \, {\rm d}^2 {\bar \q}  \,
{\bar \o}_{ \bar{b}  \bar{c} } \,\d {\bar \f}^{\bar b}{\mathbb D}^2 \mDB_\a  {\bar \f}^{\bar c}=0~.
\non
\eea

The conditions (\ref{two-form2}), (\ref{se2b}) and (\ref{se2}) prove to be sufficient 
for demonstrate that eq. (\ref{var-r3}) indeed holds. 
It is in fact instructive to compute $\delta_{\bar \rho} S$ directly without 
making use of (\ref{two-form2}), (\ref{se2b}) and (\ref{se2}).

Tedious calculations lead to 
\bea
\delta_{\bar \rho} S 
&=&    \frac{1}{16}  {\rm e}^{ -{\rm i} \l }\, { \e}_\a 
\int {\rm d}^3 x \,
{\mathbb D}^2  \Big\{ \mDB^2 K_a\mDB^\alpha\bar{\Omega}^a \Big\} \non \\
&=&
\frac{ {\rm e}^{ -{\rm i} \l }\,\epsilon^\alpha}{16} \int d^3x\, \Bigg\{ 
8 {\bar \o}_{\bar b\bar c}\Big( \Box\bar{\phi}^{\bar b}\mDB_\alpha\bar{\phi}^{\bar c}
+ \Box\bar{\phi}^{\bar c}\mDB_\alpha\bar{\phi}^{\bar b}\Big)
+16B_{\bar b\bar c\bar d}
\partial^{\beta\gamma}\bar{\phi}^{\bar b}\mDB_\gamma\bar{\phi}^{\bar d}
\partial_{\alpha\beta}\bar{\phi}^{\bar c}
\nonumber\\
&&+ \nabla_a {\bar \omega}_{\bar b\bar c} \;\Big(16\partial^{\beta\gamma}\phi^a
\mDB_\gamma\bar{\phi}^{\bar b}\partial_{\alpha\beta}\bar{\phi}^{\bar c}
-\mD^2\phi^a\mDB^2\bar{\phi}^{\bar b}\mDB_\alpha\bar{\phi}^{\bar c}
\nonumber\\
&&+8{\rm i}\, \mD^\beta\phi^a\partial_{\beta\gamma}\mDB^\gamma\bar{\phi}^{\bar b}
\mDB_\alpha\bar{\phi}^{\bar c} + 4{\rm i}\, \mD^\beta\phi^a\mDB^2\bar{\phi}^{\bar b}
\partial_{\alpha\beta}\bar{\phi}^{\bar c} \Big)
- \partial_{a}\nabla_{b} {\bar \omega}_{\bar c\bar d}\;
\mD^\beta\phi^a \mD_\beta\phi^b\mDB^2\bar{\phi}^{\bar c}\mDB_\alpha
\bar{\phi}^{\bar d}
\nonumber\\
&&+\partial_a(\Gamma^{\bar e}_{\bar b\bar c}{\bar \omega}_{\bar e\bar d})
\Big(-\mD^2\phi^a \mDB_\beta\bar{\phi}^{\bar b }\mDB^\beta\bar{\phi}^{\bar c}
\mDB_\alpha\bar{\phi}^{\bar d}
\nonumber\\
&&+4{\rm i}\, \mD_\gamma\phi^a\partial^{\gamma\beta}\bar{\phi}^{\bar b}
\mDB_\beta\bar{\phi}^{\bar c}\mDB_\alpha\bar{\phi}^{\bar d}
+4{\rm i}\, \mD_\gamma\phi^a\partial^{\gamma\beta}\bar{\phi}^{\bar c}
\mDB_\beta\bar{\phi}^{\bar b}\mDB_\alpha\bar{\phi}^{\bar d}
+4{\rm i}\,\mD^\gamma\phi^a\mDB_\beta\bar{\phi}^{\bar b}
\mDB^\beta\bar{\phi}^{\bar c}\partial_{\gamma\alpha}\bar{\phi}^{\bar d}\Big)
\nonumber\\
&&+ \partial_a\partial_b(\Gamma^{\bar f}_{\bar c\bar d}{\bar \omega}_{\bar f\bar e})
\mD^\gamma\phi^a \mD_\gamma\phi^b \mDB^\beta\bar{\phi}^{\bar c}
\mDB_\beta\bar{\phi}^{\bar d}\mDB_\alpha \bar{\phi}^{\bar e} \Bigg\}~.
\label{1717}
\eea
We see that the contributions in the first, second and third lines vanish due to
the  conditions (\ref{two-form2}), (\ref{se2b}) and (\ref{se2}). 
The remaining contributions in (\ref{1717}) are proportional either to 
\bea
\partial_a(\Gamma^{\bar e}_{\bar b\bar c}{\bar \omega}_{\bar e\bar d})
=(\partial_a\Gamma^{\bar e}_{\bar b\bar c})\, {\bar \omega}_{\bar e\bar d} =
R^{\bar e}{}_{\bar b a \bar c}\, {\bar \omega}_{\bar e\bar d}
\eea
or to  its derivative, 
$\partial_a\partial_b(\Gamma^{\bar f}_{\bar c\bar d}{\bar \omega}_{\bar f\bar e})$.
Now the fact that ${\bar \omega}_{\bar b\bar c} $
is covariantly constant, 
implies that
$R^{\bar e}{}_{\bar b a \bar c}\,{\bar \omega}_{\bar e\bar d}$ is symmetric in all barred
indices which is enough for all the remaining terms to vanish.

We have demonstrated that the conditions (\ref{two-form2}), (\ref{se2b}) and (\ref{se2}) 
guarantee that  the sigma-model action (\ref{N=2sigma-model}) 
 is invariant under  the additional {\rm third} supersymmetry  transformation (\ref{LR-ansatz0}), 
 with 
 $\bar \r$ given by eq. (\ref{par}). The important point is that  $\bar \r$ in  (\ref{LR-ansatz0})
 depends on the phase factor ${\rm e}^{-{\rm i}\l}$, where $\l$  is a fixed parameter characterising 
 the third supersymmetry transformation.  However, since  $\l$ 
does not show up in  the conditions (\ref{two-form2}), (\ref{se2b})   and (\ref{se2}), 
the action   (\ref{N=2sigma-model}) 
 is invariant under  the additional {\rm third} supersymmetry  transformation (\ref{LR-ansatz0})
 and  (\ref{par}) in which $\l$ is {\it completely arbitrary}.  
This means that the action  (\ref{N=2sigma-model}) 
 is invariant under  hidden supersymmetry transformations
of the form:
\bea
\d \f^a &=& \hf {\bar {\mathbb D}}^2 \Big( {\bar \r} (\bar \q) \, {\bar \O}^a  (\f, \bar \f  )\Big)~, 
\qquad  {\bar \r} (\bar \q) = \bar \t + {\bar \e}_\a {\bar \q}^\a~, 
\label{LR-ansatz2}
\eea 
with arbitrary {\it complex} constant parameters $\bar \t $ and ${\bar \e}_\a$.
Therefore  $\cN=3$ supersymmetry implies $\cN=4$ supersymmetry.

We now turn to describing the second argument why $\cN=3$ supersymmetry implies $\cN=4$.
For this we rewrite the third supersymmetry transformation (\ref{LR-ansatz0}) as 
\bea
\d_\l \f^a &=& \hf {\bar {\mathbb D}}^2 \Big( {\bar \r}_\l (\bar \q) \, {\bar \O}^a  (\f, \bar \f  )\Big)~. 
\label{LR-ansatz}
\eea 
The sigma-model action (\ref{N=2sigma-model}) is clearly $R$-invariant.
It does not change under $\sU(1)$ transformations
\bea
\f^a (\q) \to \f'^a (\q) = \f^a ({\rm e}^{{\rm i}\j} \q)~, \qquad 
{\bar \f}^{\bar b} (\bar  \q) \to {\bar \f} '^{\bar b} ( \bar \q) = {\bar \f}^{\bar b} ({\rm e}^{ -{\rm i}\j} \bar \q)~, 
\qquad \j \in {\mathbb R}~.
\eea
Commuting such an  infinitesimal transformation with the supersymmetry one, eq.  (\ref{LR-ansatz}), 
results in a new supersymmetry transformation of the form:
\bea
\d_{\l +\d \l} \f^a &=& \hf {\bar {\mathbb D}}^2 \Big( {\bar \r}_{\l +\d \l}(\bar \q) \, {\bar \O}^a  (\f, \bar \f  )\Big)~, 
\eea 
with $\d \l \neq0$.
Therefore, if the action  (\ref{N=2sigma-model}) is invariant under the third supersymmetry 
(\ref{LR-ansatz}) and (\ref{par}), for some fixed $\l$, 
it is in fact invariant under a one-parameter family of supersymmetry transformations 
corresponding 
to all possible values for $\l \in \mathbb R$ in   (\ref{LR-ansatz}).
This means that $\cN=3$ supersymmetry implies $\cN=4$ supersymmetry.
Instead of the ansatz (\ref{LR-ansatz}), we can now look for a hidden supersymmetry 
transformation
of the form (\ref{LR-ansatz2}).
Therefore,  we can recycle, word for word,  the four-dimensional derivation given in \cite{K-duality} 
of the  results of \cite{HKLR}
devoted to the formulation of general 4D $\cN=2$ supersymmetric nonlinear sigma-models
in terms of $\cN=1$ chiral superfields.

On the mass shell, 
\bea
 {\bar {\mathbb D}}^2 K_a=0 ~,~
\label{mass-shell}
\eea
 and the first and the second supersymmetry transformations
generate the $\cN=2$ super-Poincar\'e algebra {\it without} central charge
provided
\bea
{\bar \O}^a{}_{, \bar c}  \, { \O}^{\bar c}{}_{,  b}
=- \d^a{}_b~.
\label{se3}
\eea
In fact, the closure of the supersymmetry algebra
requires two more conditions
\bea
 {\bar {\mathbb D}}^2 {\bar \O}^a&=&0 ~,\\
{\bar \O}^d{}_{, \bar b}  \nabla_d {\bar \O}^a{}_{, \bar c}
- {\bar \O}^d{}_{, \bar c}  \nabla_d {\bar \O}^a{}_{, \bar b} &=&0 ~.
\eea
They hold due to (\ref{se2}) -- (\ref{se3}).
The detailed derivation of the above results can be found in \cite{K-duality}.

Let $J \equiv J_3 $ be the complex structure chosen on the target space $\cM$,
\bea
J_3 = \left(
\begin{array}{cc}
{\rm i} \, \d^a{}_b  ~ & ~ 0 \\
0 ~ &   -{\rm i} \, \d^{\bar a}{}_{\bar b}  
\end{array}
\right)~.
\label{J3}
\eea
The above consideration shows that there are two more complex structures defined as
\bea
J_1 = \left(
\begin{array}{cc}
0  ~ & ~ {\bar \O}^a{}_{, \bar b} \\
{\O}^{\bar a}{}_{,  b} ~ &   0
\end{array}
\right)~, \qquad 
J_2 = \left(
\begin{array}{cc}
0  ~ &  {\rm i}\,  {\bar \O}^a{}_{, \bar b} \\
-{\rm i}\, {\O}^{\bar a}{}_{,  b} ~ &   0
\end{array}
\right)~
\label{J1J2}
\eea
such that $\cM$ is K\"ahler with respect to all of them, and 
the operators $J_A = (J_1,J_2,J_3) $ form the quaternionic algebra:
\be
J_A \,J_B = -\d_{AB} \, {\mathbbm 1} + \ve_{ABC}J_C~.
\ee
As a result, it has been demonstrated that the target space $\cM$ is a hyperk\"ahler 
manifold.

As is seen from (\ref{J1J2}), the complex structures are given in terms of 
the tensor fields ${\bar \O}^a{}_{, \bar b}$ and ${\O}^{\bar a}{}_{,  b}$, while the supersymmetry 
transformation (\ref{LR-ansatz}) involves ${\bar \O}^a$ and ${\O}^{\bar a}$. 
The latter can be constructed using the K\"ahler potential \cite{HKLR}:
\be
{\bar \O}^a = \o^{ab}  \big(\f \big) K_b \big(\f , \bar \f \big)~.
\label{barOmega}
\ee
Under the  K\"ahler transformations 
\be
K \big(\f , \bar \f \big)  \quad \longrightarrow \quad K\big(\f , \bar \f \big) 
+  \L \big(\f  \big) + {\bar \L} \big( \bar \f \big)   ~,
\ee
${\bar \O}^a $ changes  as follows:
$ \o^{ab}  K_b \to  \o^{ab}  K_b +  \o^{ab}  \L_b$.
However, the supersymmetry variation
$\d \f^a = \hf {\bar {\mathbb D}}^2 \big({\bar \r}( {\bar \q}) \,{\bar \O}^a \big)$ in (\ref{LR-ansatz}) 
is invariant under the K\"ahler transformations.

\subsection{Off-shell N = 3 SUSY implies off-shell N = 4 SUSY} 
Consider the off-shell $\cN=3$ superconformal sigma-model generated by the Lagrangian
(\ref{conformal-sm}) and (\ref{Kkahler22}). Upon reduction to $\cN=2$ superspace, its action 
takes the form (\ref{N=3scsma}). Comparing this action 
with that of the off-shell $\cN=4$
superconformal sigma-model formulated in $\cN=2$ superspace, eq. (\ref{N=4scsma}), 
we see that the former is identical in its form with either the left or the right sector of the latter.
This means that the theory (\ref{N=3scsma}) can be lifted to an off-shell $\cN=4$ superconformal 
sigma-model realised only in terms of left weight-one arctic multiplets and their conjugates, 
or only in terms of their right mirrors.
If the K\"ahlerian cone $\cM$, for which $K(\F , \bar \F) $ is the K\"ahler potential, 
factorises, $\cM = \cM_1 \times \cM_2$, then we can use both left and right arctic multiplets
and their conjugates in order to realise $\cN=4$ extensions of the two sectors of the $\cN=3$ sigma-model.

\section{General N = 3, 4 superconformal sigma-models} 
\setcounter{equation}{0}
\label{off-shell-N3N4}
We now have all prerequisites available 
to develop a chiral formulation in $\cN=2$ superspace for the most general 
$\cN=3$ and $\cN=4$ superconformal nonlinear $\s$-models. Given a hyperk\"ahler cone $\cM$, 
we pick one of its complex structures, say $J_3$, and introduce complex 
coordinates $\f^a$ compatible with it. In these coordinates, $J_3$ has the form 
(\ref{J3}).  Two other complex structures, $J_1$ and $J_2$, become
\bea
J_1 = \left(
\begin{array}{cc}
0  ~ & ~ {g}^{a \bar c} {\bar \o}_{\bar c \bar b} \\
{g}^{ \bar a c } { \o}_{ c  b}
 ~ &   0
\end{array}
\right)~, \qquad 
J_2 = \left(
\begin{array}{cc}
0  ~ &  {\rm i}\,   {g}^{a \bar c} {\bar \o}_{\bar c \bar b}  \\
-{\rm i}\,   {g}^{ \bar a c } { \o}_{ c  b}~ &   0
\end{array}
\right)~,
\label{J1J2-hkc}
\eea
where ${g}_{a \bar b} $ be the hyperk\"ahler metric, and 
$\o_{ab}$ the holomorphic symplectic two-form. 
Let  $\c =\Big(\c^a (\f), \bar\c^{\bar b}(\bar \f )  \Big) $ be the homothetic conformal Killing vector field
associated with $\cM$, 
\begin{subequations}
\bea
\nabla_a \c^b &=& \d_a^b~, \qquad {\bar \nabla}_{\bar a} \c^b ={\bar \pa}_{\bar a} \c^b =0
 \label{HCKVa} \\
 \c_a := g_{a\bar b}  {\bar \c}^{\bar b}&=& \pa_a K~, \qquad g_{a \bar b} 
=\pa_a {\bar \pa}_{\bar b} K~, 
\label{HCKVb}
\eea
\end{subequations}
with the hyperk\"ahler potential $K$ given by 
\bea
 K  =g_{a \bar b}  \c^a  {\bar \c}^{\bar b} =K_a \c^a~.
 \label{HCKVc}
 \eea
With this hyperk\"ahler potential chosen, our goal is  to prove that 
the $\cN=4$ supersymmetric $\s$-model 
\bea
S&=& \int {\rm d}^3 x \,{\rm d}^2 \q {\rm d}^2 {\bar \q}  \, K\big(\f^a, {\bar \f}^{\overline{b}}\big)~, 
\qquad {\mathbb \DB}_{ \b} \f^a =0 
\label{N=2SCSM}
\eea
is $\cN=4$ superconformal. 
In accordance with the analysis given in subsection \ref{N2 superconformal sigma-models}, 
the action (\ref{N=2SCSM})  is invariant under the $\cN=2$ superconformal transformation
\bea
\d \f^a =- {\bm \x} \f^a - \hf {\bm \s} \c^a (\f)~, 
\label{9.5}
\eea
with $\bm \x$ an arbitrary $\cN=2$ superconformal Killing vector.

\subsection{N = 3 superconformal invariance}
To start with, we prove that the action  (\ref{N=2SCSM}) is $\cN=3$ superconformal.
As shown in Appendix C, subsection \ref{C1}, 
an arbitrary  $\cN=3$ superconformal transformation  
decomposes into two transformations  in $\cN=2$ superspace: (i) an
$\cN=2$ superconformal transformation generated by an $\cN=2$ superconformal 
Killing vector $\bm \x$; (ii) an extended superconformal transformation generated by 
a  {\it real} spinor superfield  $\r^\a (x,\q , \bar \q)$ obeying the constraints
\bea
{\mathbb D}_{(\a}\r_{\b)}={\mathbb \DB}_{(\a}\r_{\b)}=0~,
\label{SCK-N3-->N2}
\eea
and therefore
\bea
\pa_{(\a\b}\, \r_{\g)}
={\mathbb D}^2\r_{\a}={\mathbb \DB}^2\r_{\a}=0~.
\eea
The general solution of the constraints (\ref{SCK-N3-->N2}) is 
\bea
\r^\a (x, \q , \bar \q) =\e^\a-\ri\l \q^\a+\ri\bar{\l} {\bar \q}^\a
+{\eta}_\b x^{\a\b}
+\ri\eta^\a\q^\b {\bar \q}_\b ~.
\eea
Here the real spinor  $\e^\a$ generates the third $Q$-supersymmetry, 
the complex scalar $\l$ the off-diagonal $\sSU(2)$ transformation
and the real spinor ${\eta}^\a$ the third $S$-supersymmetry transformation.
Associated with $\r^\a$ is a complex scalar ${\bm \r}$  and its complex conjugate $\bar {\bm \r}$
defined as follows:
\be
\r_\a = {\mathbb D}_\a {\bm \r} ~, \qquad \r_\a = \bar{\mathbb D}_\a \bar {\bm \r} ~.
\ee
The scalar $\bar{\bm \r}$ is defined modulo arbitrary shifts of the form:
\be
\bar{\bm \r} ~ \to ~\bar{\bm \r} + \vf ~, \qquad  \bar{\mathbb D}_\a \vf =0~.
\ee
This freedom is not strong enough to make ${\bm \r}$ and $\bar{\bm \r}$ 
coincide,\footnote{A simple way to justify this statement is as follows.
According to the analysis given in appendix C.1,
$\L^{11}|=\frac{\ri}{2}\mD_{\a}{\r}^{\a}$
and its complex conjugate is $(\L^{11})^*=\L^{22}|=\frac{\ri}{2}\mDB_{\a}\r^{\a}$.
These relations can be rewritten in terms of $\bm \rho$ and $\bar{\bm \rho}$
as $\L^{11}|=\frac{\ri}{2}\mD_{\a}\mDB^{\a}{\bar{\bm \r}}$
and
$\L^{22}|=\frac{\ri}{2}\mDB_{\a}\mD^{\a}{\bm \r}$, and thus
$\bm \rho$ has to be complex, for  $\L^{11}|$ is complex in general.}
although 
 ${\bm \r}$ and $\bar{\bm \r}$ obey the reality condition:
\be
{\mathbb D}_\a {\bm \r} = \bar{\mathbb D}_\a \bar {\bm \r} ~.
\label{9.11}
\ee

We have seen that the sigma-model   (\ref{N=2SCSM}) is invariant 
under the $\cN=2$ superconformal transformations (\ref{9.5}).
Let us now  define the extended superconformal transformation:
\bea 
\d \f^a &=&\hf
{\bar {\mathbb D}}^2 \Big\{ \bar{\bm \r} \,  { \o}^{ab} \c_b \Big\} ~.
\label{hkc-esc}
\eea
Our goal is to prove that the action    (\ref{N=2SCSM})  is  invariant under (\ref{hkc-esc}). 
The variation of the action is 
\bea
\d S&=& -\hf    \int \rd^3 x\, {\rm d}^2 \q {\rm d}^2 {\bar \q} \, 
\big({\bar {\mathbb D}}_{ \a} \c_a \big)
\big(  {\bar {\mathbb D}}^{ \a}\bar {\bm \r} \big) \, \o^{ab}\c_b +{\rm c.c.} \non \\
&=& 
-\hf    \int \rd^3 x\,{\rm d}^2 \q {\rm d}^2 {\bar \q} \, {\bar \r}_{ \a} \big({\bar {\mathbb D}}^{ \a} {\bar \f}^{\bar c} \big)
g_{\bar c\, a}  \, \o^{ab}\c_b +{\rm c.c.}  \non \\
&=&  -\hf    \int \rd^3 x\,{\rm d}^2\q {\rm d}^2{\bar \q}
\, {\bar \r}_{ \a} \big({\bar {\mathbb D}}^{ \a} {\bar \f}^{\bar a} \big)
{\bar \o}_{\bar a \bar b} \,{\bar \c}^{\bar b} +{\rm c.c.} ~
\eea
Since the tensor fields ${\bar \o}_{\bar a \bar b} $ and ${\bar \c}^{\bar b}$ are anti-holomorphic, 
the combination $  {\bar \o}_{\bar a \bar b} \,{\bar \c}^{\bar b}$
appearing in the integrand is antichiral. 
As a result, doing the Grassmann integral $\int {\rm d}^2\q$ gives
\bea
\d S&=& 
\frac{1}{4}  \int \rd^3 x\,{\rm d}^2{\bar \q}\,
({\mathbb D}^\a {\bar \r}_{ \a})\, {\bar \o}_{\bar a \bar b} \,{\bar \c}^{\bar b}\,{\mathbb D}^\b {\bar {\mathbb D}}_{\b} {\bar \f}^{\bar a} 
+\frac{1}{8}  \int \rd^3 x\,{\rm d}^2{\bar \q}\,
 {\bar \r}_{ \a} {\bar \o}_{\bar a \bar b} \,{\bar \c}^{\bar b}\,{\mathbb D}^2 {\bar {\mathbb D}}^{ \a} {\bar \f}^{\bar a} 
 +{\rm c.c.} 
 ~~~~~
\eea
In deriving the first term, we have used the identity ${\mathbb D}_{(\a}\r_{\b)}=0$.
The above variation vanishes, 
for  ${\mathbb D}^\b {\bar {\mathbb D}}_{\b} {\bar \f} \equiv 0$ and 
${\mathbb D}^2 {\bar {\mathbb D}}_{ \b} {\bar \f} \equiv 0$ for any antichiral superfield $\bar \f$.
 
\subsection{N = 4 superconformal invariance}
As shown in Appendix C, subsection \ref{C2}, 
an arbitrary  $\cN=4$ superconformal transformation  
decomposes into three transformations  in $\cN=2$ superspace: (i) an
$\cN=2$ superconformal transformation generated by an $\cN=2$ superconformal 
Killing vector $\x$; (ii) an extended superconformal transformation generated by 
a  {\it complex} spinor superfield  $\r^\a (x,\q , \bar \q)$ obeying the constraints
(\ref{C9}); (iii) a shadow $\sU (1)$ rotation.
In  regard to transformation (ii), the only difference from the $\cN=3$ case considered earlier is that 
the parameter $\r^\a$ is now complex.
Since $\r^\a$ is complex, the general solution of the constraints 
(\ref{C9}) is 
\bea
\r^\a=\e^\a+\l_{\rm R}\q^\a
-\bar{\l}_{\rm L}\qb^\a
+\bar{\eta}_\b x^{\a\b}
+\ri\bar{\eta}^\a\q^\b\qb_\b 
~.
\eea
Here the complex parameter $\e^\a$ generates the third and fourth $Q$-supersymmetries;
the complex parameters $\l_{\rm L}$ and $\l_{\rm R}$ generate 
off-diagonal $\sSU(2)_{\rm L}$ and $\sSU(2)_{\rm R}$ transformations 
associated with $\L_{\rm L}^{11}$ and $\L_{\rm R}^{11}$;
finally,  the complex parameter ${\eta}^\a$ generates the third and fourth
$S$-supersymmetry transformations.

The constraints 
(\ref{C9}) imply that we can represent
\begin{subequations}
\bea
\r_\a &=& \bar{\mathbb D}_\a \bar{\bm \r}_{\rm L} ~, \qquad {\bar \r}_\a = {\mathbb D}_\a {\bm \r}_{\rm L} ~; \\
\r_\a &=& {\mathbb D}_\a {\bm \r}_{\rm R} ~, \qquad {\bar \r}_\a = \bar{\mathbb D}_\a \bar{\bm \r}_{\rm R} ~,
\eea
\end{subequations}
for some complex scalars $ {\bm \r}_{\rm L}$ and $ {\bm \r}_{\rm R}$.
Unlike the $\cN=3$ case, the mutually conjugate  parameters $ {\bm \r}_{\rm L,R}$ and $\bar{\bm \r}_{\rm L,R}$
do not obey any additional reality condition like (\ref{9.11}).

In the $\cN=4$ case, it is possible to define two types of extended superconformal transformations: 
\begin{subequations}
\bea 
\d_{\rm L} \f^a &=&\hf
{\bar {\mathbb D}}^2 \Big\{ \bar{\bm \r}_{\rm L} \,  { \o}^{ab} \c_b \Big\} ~, \\
\d_{\rm R} \f^a &=&\hf
{\bar {\mathbb D}}^2 \Big\{ \bar{\bm \r}_{\rm R} \,  { \o}^{ab} \c_b \Big\} ~.
\eea
\end{subequations}
Such transformations leave the action    (\ref{N=2SCSM})  invariant, 
for the proof given at the end of the previous subsection  carries over without any change.

${}$Finally, we define the infinitesimal shadow $\sU (1)$ transformation of $\f^a$: 
\bea
\d \f^a = -\frac{\rm i}{2} \a \, \c^a (\f) ~, \qquad \bar \a = \a~.
\eea
Because of the identity (\ref{HCKVc}), this 
transformation leaves  the action invariant.
It should be remarked that the shadow chiral rotation is generated 
by the Killing vector 
\bea
\u = {\rm i} \,\c^a (\f) \frac{\pa}{\pa \f^a} -{\rm i} \, {\bar \c}^{\bar a} (\bar \f)  \frac{\pa}{\pa {\bar \f}^{\bar a}} ~.
\eea

\section{Conclusion}
In this paper we have elaborated on various aspect of three-dimensional $\cN\leq4$ 
superconformal sigma-models from the superspace point of view.
The original motivation for the research presented in this paper was the desire to explore
the additional opportunities offered by superspace as compared with the component analysis
given in \cite{BCSS}. 

We have not studied gauged superconformal sigma-models. 
The procedures of gauging the target-space isometries 
for sigma-models formulated in superspace are well-elaborated, see in particular
\cite{HitchinKLR,HKLR}, and can be naturally used in three dimensions.
\\

\noindent
{\bf Acknowledgements:}\\
GT-M acknowledges the hospitality and support of the University of Western Australia during the final stage 
of this project. 
The work  of SMK is supported in part by the Australian Research Council under contract No. DP1096372.
The work of JHP is supported by the National Research Foundation of Korea (NRF) grant funded 
by the Korea government (MEST) through the Center for Quantum Spacetime (CQUeST) of 
Sogang University with grant number 2005-0049409.
The work of GT-M is supported by the European 
Commission, Marie Curie Intra-European Fellowships under contract No.
PIEF-GA-2009-236454.  The research of R.v.U. was supported by the Czech ministry of education 
under contract No. MSM0021622409.

\appendix

\section{3D notation and conventions}
\setcounter{equation}{0}
Our spinor conventions in three space-time dimensions (3D) are compatible with
the 4D two-component spinor 
formalism used by Wess and Bagger \cite{WB} and also adopted in \cite{BK}.
Specifically, we start from the 4D sigma-matrices 
\begin{subequations}
\bea
(\s_{\mun } )_{\a \dt \b}&:=& ({\mathbbm 1}, \phantom{-}\vec{\s} ) ~, \qquad {\mun }=0,1,2,3 \\
(\tilde{\s}_{\mun } )^{{\dt \a}  \b}&:=& ({\mathbbm 1}, - \vec{\s} ) ~, \qquad {\mun }=0,1,2,3
\eea
\end{subequations}
and delete the matrices with space index $\mun =2$. This leads to the 3D gamma-matrices 
\begin{subequations}
\bea
(\s_{\mun } )_{\a \dt \b}\quad & \longrightarrow & \quad (\g_m )_{\a  \b} = (\g_m)_{\b\a} ~
=({\mathbbm 1}, \s_1, \s_3) ~,\\
(\tilde{\s}_{\mun } )^{\dt \a  \b}\quad & \longrightarrow & \quad (\g_m )^{\a  \b} = (\g_m)^{\b\a} 
=\ve^{\a \g} \ve^{\b \d} (\g_m)_{\g \d} ~,
\eea
\end{subequations}
where the spinor indices are  raised and lowered using
the $\sSL(2,{\mathbb R})$ invariant tensors
\bea
\ve_{\a\b}=\left(\begin{array}{cc}0~&-1\\1~&0\end{array}\right)~,\qquad
\ve^{\a\b}=\left(\begin{array}{cc}0~&1\\-1~&0\end{array}\right)~,\qquad
\ve^{\a\g}\ve_{\g\b}=\d^\a_\b
\eea
using the standard rule:
\bea
\psi^{\a}=\ve^{\a\b}\psi_\b~, \qquad \psi_{\a}=\ve_{\a\b}\psi^\b~.
\eea
By construction, the matrices $ (\g_m )_{\a  \b} $ and $ (\g_m )^{\a  \b} $ are real and symmetric.
Using the properties of the 4D sigma-matrices, we can immediately read off the properties 
of the 3D gamma-matrices. In particular, for the
matrices 
\be
\g_m:=(\g_m)_\a{}^{\b}=\ve^{\b\g}(\g_m)_{\a\g}
\ee
one readily  obtains the relations
\bsubeq
\bea
&\{\g_m,\g_n\}=2\eta_{mn}{\mathbbm 1}~,
\\
&\g_m\g_n=\eta_{mn}{\mathbbm 1}+\ve_{mnp}\g^p~,
\eea
\esubeq
where the 3D Minkowski metric is $\eta_{mn}=\eta^{mn}={\rm diag}(-1,1,1)$
and the Levi-Civita tensor is normalised as $\ve_{012}=-\ve^{012}=-1$.
Another useful relation is the following
\bea
(\g^m)_{\a \b} (\g_m)_{\g \d} = 2\ve_{\a(\g}  \ve_{\d)\b} ~.
\eea

To comply with the tradition, we will label the 3D vector indices by values $0,\,1,\,2$.
Given a three-vector $V_m$, it can be equivalently described by a symmetric bi-spinor $V_{\a\b}$
defined as 
\bea
V_{\a\b}:=(\g^m)_{\a\b}V_m=V_{\b\a}~,\qquad
V_m=-\hf(\g_m)^{\a\b}V_{\a\b}~.
\eea

\section{The super-Poincar\'e group} 
\setcounter{equation}{0}

\allowdisplaybreaks
The $\cN$-extended super-Poincar\'e group in three space-time dimensions, 
$ {\mathfrak P}(3|\cN)$, can naturally be realised as a subgroup of the superconformal group 
$\sOSp(\cN|2, {\mathbb R})$. Any element $g \in {\mathfrak P}(3|\cN)$ can uniquely be represented 
in the form:
\begin{subequations} 
\bea
g &=& s(a, \e) \, h( M) ~, \\
s(a, \e) &=& \exp \left(
\begin{array}{c | c ||c}
  0  ~& ~ 0~ &~0   \\
\hline 
-{a}^{\a \b}  ~& ~  0~&~ -\sqrt{2}\e^\a{}_J
\\
\hline
\hline
{\rm i}\sqrt{2}\, \e_I{}^\b ~& ~0~&~0
\end{array}
\right) \non\\
&=& 
\left(
\begin{array}{c | c ||c}
  \d_\a{}^\b  ~& ~ 0~ &~0   \\
\hline 
-{a}^{\a \b} +\frac{\rm i}{2}\ve^{\a\b} \e^2 ~& ~  \d^\a{}_\b~&~ -\sqrt{2}\e^\a{}_J
\\
\hline
\hline
{\rm i}\sqrt{2}\, \e_I{}^\b ~& ~0~&~\d_{IJ}
\end{array}
\right) 
~, 
\label{B.1b}
\\
h(M) &=& \exp \left(
\begin{array}{c | c ||c}
  M  ~& ~ 0~ &~0   \\
\hline 
0  ~& ~  (M^{-1})^{\rm T}~&~ 0
\\
\hline
\hline
0 ~& ~0~&~{\mathbbm 1}_\cN
\end{array}
\right) ~, \qquad
M \in \sSL(2, {\mathbb R})~.
\label{B.1c}
\eea
\end{subequations}
In eq. (\ref{B.1b}), the bosonic  $a^{\a \b}=a^{\b \a} = a^m (\g_m)^{\a\b}$ and fermionic 
$\e_I{}^\a = \e^\a{}_I \equiv \e^\a_I$ parameters are real.

{$\cN$-extended Minkowski superspace} is  the homogeneous space
\bea
{\mathbb M}^{3|2\cN} 
={\mathfrak P}(3|\cN)\big/ {\sSL}(2,{\mathbb R}) 
~,
\eea
where ${\sSL}(2,{\mathbb R})$ is identified with the set of all matrices 
$h(M)$ defined in  (\ref{B.1c}).
The points of ${\mathbb M}^{3|2\cN}$
can be parametrised by the variables 
\bea
 z^M = (x^m, \q^\a_I)
\eea
which correspond to the following coset representative:
\bea
s(z):= s(x, \q) 
= \left(
\begin{array}{c | c ||c}
  \d_\a{}^\b  ~& ~ 0~ &~0   \\
\hline 
-{x}^{\a \b} +\frac{\rm i}{2}\ve^{\a\b} \q^2 ~& ~  \d^\a{}_\b~&~ -\sqrt{2}\q^\a{}_J
\\
\hline
\hline
{\rm i}\sqrt{2}\, \q_I{}^\b ~& ~0~&~\d_{IJ}
\end{array}
\right) ~, \qquad x^{\a \b} = x^m (\g_m)^{\a\b}~.
\eea
The supersymmetry transformation $s(0, \e) $ acts on the superspace according to 
the law $s(x, \q)  \to s(x', \q ') = s(0,\e) s(x, \q) $, and thus
\bea
x'^{\a\b} = x^{\a \b} + {\rm i} ( \e^\a_I  \q^\b_I  +   \e^\b_I \q^\a_I)~, \qquad
\q'^\a_I = \q^\a_I + \e^\a_I~.
\eea
These results can be rewritten in terms of $z^A =(x^a , \q^\a_I)$ as 
\bea
z'^A =z^A -{\rm i} \, \e^\b_J Q^J_\b \,z^A~, 
\eea
where we have introduced
the supersymmetry generators
\bea
Q^I_\a ={\rm i}\, \frac{\pa}{\pa \q^\a_I} + (\g^m)_{\a\b}\, \q^\b_I \pa_m
= {\rm i}\, \frac{\pa}{\pa \q^\a_I} +   \q^\b_I \pa_{\b\a}~.
\eea
From here we immediately  read off the spinor covariant derivatives
\bea
D^I_\a =\frac{\pa}{\pa \q^\a_I} + {\rm i}  (\g^m)_{\a\b}\, \q^\b_I \pa_m
=  \frac{\pa}{\pa \q^\a_I} + {\rm i}  \q^\b_I \pa_{\b\a}~,
\eea
which obey  the anti-commutation relations 
\bea
\big\{ D^I_\a , D^J_\b \big\} = 2{\rm i}\, \d^{IJ}  (\g^m)_{\a\b}\,\pa_m~.
\eea
As compared with the supersymmetry in four dimensions, 
the spinor covariant derivatives possess unusual conjugation properties.
Specifically, given an arbitrary superfield $F$ 
and $\bar{F}:={(F)}^*$ its complex conjugate, the following relations holds
\bea
{(D_\a^IF)}^*=-(-1)^{\e(F)}D_\a^I\bar{F} ~,
\eea
where $\e(F)$ denotes the  Grassmann parity of $F$.


\section{N = 2 reduction for N = 3 and N = 4 superconformal Killing vector fields}
\setcounter{equation}{0}
The $\cN$-extended superconformal Killing vector fields were studied in
section \ref{Superconformal Killing vectors}. 
Here we describe the reduction of 
the $\cN=3$ and $\cN=4$ superconformal Killing vectors to $\cN=2$ superspace. 
The results of this appendix are used in the
sections \ref{N=3conf}, \ref{N3N4} and \ref{off-shell-N3N4}.
The analysis and the results given below  are 
analogous to those described 
in the the 4D  $\cN=2$ case
in \cite{K-hyper,K-duality}.

We recall that $\cN=3$ superspace is parametrised by the coordinates
$z^A = (x^a , \q^\a_{ij})$, while its $\cN=4$ cousin by $z^A=(x^a, \q^\a_{i \bar j})$.
Embedded into these superspaces is the $\cN=2$ superspace
parametrised by variables $z^{\frak A}=(x^a,\q^\a,\bar{\q}^\a)$, where the odd complex coordinate $\q^\a$ 
is defined as  $\q^\a=\q^\a_{11}=\q^\a_{1{\bar 1}}$, and for its conjugate $\qb^\a=(\q^\a)^*$
we have $\qb^\a=\q^\a_{22}=\q^\a_{2{\bar 2}}$.

Let  $\F(z^A)$ be an arbitrary  $\cN=3$  or $\cN=4$ superfield. 
Its $\cN=2$ projection is
\bea
\F|:=\F(z^A)\big|_{\q_{\perp}=0}
~,
\eea
where $\q_{\perp}$  
stands for  $\q^\a_{12}$ in the $\cN=3$ case, and 
$(\q^\a_{1\ot},\q^\a_{2\oo})$ in the $\cN=4$ case.
Given a vector field $V=V^A (z) D_A$ on the  $\cN=3$  or $\cN=4$ superspace, its
$\cN=2$ projection is defined as
\be
V|=V^A | \,D_A~.
\ee
The important point is that the covariant derivatives  $D_\a^{11}=D_\a^{1{\bar 1}}$ and 
$D_\a^{22}=D_\a^{2\ot}$ depend only on $\q^\a$ and $\qb^\a$. The explicit representation 
of the $\cN=2$ covariant derivatives is
\bsubeq
\bea
\mD_\a&=& \frac{\pa}{\pa\q^\a}+\ri \qb^\b \pa_{\a\b}=
D_\a^{11}=D_\a^{1\oo}~,
\\
\mDB_\a&=& -\frac{\pa}{\pa\qb^\a}-\ri \q^\b \pa_{\a\b} 
= -D_\a^{22}=-D_\a^{2\ot}~.
\eea
\esubeq
They satisfy the anticommutation relations (\ref{N=2acd}).

\subsection{N = 3  superconformal Killing vector fields}
\label{C1}
Let $\bm \x$ be a $\cN=3$ superconformal Killing vector. 
Consider its $\cN=2$ projection
\bea
&&\x|:=\x^A|D_A= {\bm \x} 
+2\ri\r^\a D_\a^{12}~, \qquad {\bm \x} = {\bm \x}^a\pa_a+{\bm \x}^\a \mD_\a-{\bar {\bm \x}}^\a \mDB_\a
~.
\label{C.3}
\eea
Making use of eq. (\ref{master-5.33}) gives 
\bea
\big[ \x ,  D^{11}_\a\big] \big| =
{[}\x|,\mD_\a{]}=
{\bm\o}_{\a}{}^{\b}\mD_\b
+\frac{1}{4}\big({\bm\s}-3\bar{\bm\s}\big) \mD_\a
-\L^{11}|D_{\a}^{12}
~.
\label{C.4}
\eea
The parameters in (\ref{C.3}) and  (\ref{C.4}) 
are related to those in  (\ref{master-5.33}) as follows:
\bsubeq
\bea
&&
{\bm \x}^a:=\x^a|~,\qquad
{\bm \x}^\a:=\x^\a_{11}|~,\qquad
{\bar {\bm \x}}^\a:=\x^\a_{22}|~,\qquad
\r^\a:=-\ri\x^\a_{12}|=\overline{\r^\a}~,
\\
&&
\s|=\frac{1}{3}\pa^a{\bm \x}_a
=\frac{1}{2}\big(
\mD_\a{\bm \x}^\a
-\mDB_\a\bar{\bm\x}^\a
\big)
~,\qquad
\o_{\a\b}|={\bm \o}_{\a\b}=
-\mD_{(\a}{\bm \x}_{\b)}
=\mDB_{(\a}\bar{\bm \x}_{\b)}~,
\\
&&
\L^{11}|=
\frac{\ri}{2}\mD_{\a}{\r}^{\a}
~,~~~
\L^{12}|=
-\frac{1}{8}\big(
\mD_{\a}{\bm \x}^{\a}
+\mDB_{\a}\bar{\bm \x}^{\a}
\big)
~,\\
&&{\bm\s}=\frac{1}{4}\Big(\mD_\a{\bm\x}^\a-3\mDB_\a\bar{\bm\x}^\a\Big)
=\Big(2\L^{12}|+\s|\Big) ~,\qquad
\mDB_\a{\bm\s}=0~.
\eea
\esubeq
The $\cN=3$ superconformal transformation generated by $\x$ induces two different transformations 
of $\cN=2$ superfields. They are:\\
${}\quad$ {\bf 1.} A $\cN=2$ superconformal transformation. It is generated by $\bm \x$   which, 
due to (\ref{C.4}), is a $\cN=2$ superconformal vector field.
Its components ${\bm \x}^a, {\bm \x}^\a,\bar{\bm \x}^\a$ as well as the 
descendants ${\bm\o}_{\a\b},\s|$ and $\L^{12}|$, which are introduced above, 
correspond to 
the $\cN=2$ parameters ${\x}^a, {\x}^\a,\bar{\x}^\a,\o_{\a\b},\s$ and $({\ri}/{2})\L$ 
of subsection (\ref{N2superKilling}).\\
${}\quad$ {\bf 2.} An extended superconformal transformation.
It is generated by the real spinor parameter $\r^\a$ under the constraints
\bea
\mD_{(\a}\r_{\b)}=\mDB_{(\a}\r_{\b)}=0~~~
\Longrightarrow~~~
\pa_{(\a\b}\r_{\g)}
=\mD^2\r_{\a}=\mDB^2\r_{\a}=0
~.
\label{C5}
\eea
The general solution to these constraints is
\bea
\r^\a=\e^\a
-\ri\l\q^\a+\ri\bar{\l}\qb^\a
+{\eta}_\b x^{\a\b}
+\ri\eta^\a\q^\b\qb_\b
~,
\eea
with $\e^\a,\l$ and $\eta^\a$ constant parameters.
The {\it real} parameter $\e^\a$ generates the third $Q$-supersymmetry; 
the {\it complex} parameter  $\l =\L^{11}|_{\q=0}$ generates an off-diagonal $\sSU(2)$ transformation;
finally,  the {\it real} parameter ${\eta}^\a$ generates the third $S$-supersymmetry transformation.

\subsection{N = 4 superconformal Killing vector fields}
\label{C2}
In the $\cN=4$ case, the analysis is similar to that given in the previous subsection.
The $\cN=2$ projection of the $\cN=4$ superconformal Killing vector field is
\bsubeq
\bea
&&\x|:=\x^A|D_A= {\bm \x}
+\r^\a D_\a^{1\ot}
-{\bar \r}^\a D_\a^{2\oo}
~, \qquad {\bm \x} = {\bm \x}^a\pa_a+{\bm \x}^\a \mD_\a-{\bar {\bm \x}}^\a \mDB_\a
\\
&&{[}\x|,\mD_\a{]}=
{\bm\o}_{\a}{}^{\b}\mD_\b
+\frac{1}{4}\big({\bm\s}-3\bar{\bm\s}\big) \mD_\a
-(\L_{\rm L})^{11}|D_{\a}^{2\oo}
-(\L_{\rm R})^{\oo\oo}|D_{\a}^{1 \ot}~,
\eea
\esubeq
                where
\bsubeq
\bea
&&{\bm \x}^a:=\x^a|~,\qquad
{\bm \x}^\a:=\x^\a_{1\oo}|~,\qquad
\r^\a:=\x^\a_{1\ot}|~,
\\
&&\s|=\frac{1}{3}\pa^a{\bm \x}_a
=\frac{1}{2}\big(
\mD_\a{\bm \x}^\a
-\mDB_\a\bar{\bm\x}^\a
\big) ~,\qquad
{\bm \o}_{\a\b}=\o_{\a\b}|=
-\mD_{(\a}{\bm \x}_{\b)}
=\mDB_{(\a}\bar{\bm \x}_{\b)}
~,~~~~~
\\
&&\L_{\rm L}^{11}|=
-\frac{1}{2}\mD_{\a}{\bar\r}^{\a}~,\qquad
\L_{\rm R}^{\oo\oo}|=
\frac{1}{2}\mD_{\a}\r^{\a}
~,
\\
&& \L_{\rm L}^{12}|+\L_{\rm R}^{{\bar 1}{\bar 2}}| =
-\frac{1}{4}\big(
\mD_{\a}{\bm \x}^{\a}
+\mDB_{\a}\bar{\bm \x}^{\a}
\big)
~,\\
&& \L_{\rm L}^{12}|-\L_{\rm R}^{{\bar 1}{\bar 2}}|=
\frac{1}{4}\big(
D_{\a}^{2\oo}\x_{2\oo}^{\a}|
-D_{\a}^{1\ot}\x_{1\ot}^{\a}|
\big)
~,~~~~~~~~~
\\
&&
{\bm\s}=\frac{1}{4}\Big(\mD_\a{\bm\x}^\a-3\mDB_\a\bar{\bm\x}^\a\Big)
=
 \L_{\rm L}^{12}|+\L_{\rm R}^{\oo\ot}|+\s|
 ~,\qquad
\mDB_\a{\bm\s}=0~.
\eea
\esubeq
Associated with $\x$ are three types 
of $\cN=2$ superfields, specifically:\\
${}\quad$ {\bf 1.} 
A $\cN=2$ superconformal transformation. It is generated by
the  $\cN=2$ superconformal vector field $\bm \x$.
Its components
${\bm \x}^a, {\bm \x}^\a,\bar{\bm \x}^\a$
and the descendants ${\bm\o}_{\a\b},\s|$ and $(\L_{\rm L}^{12}|+\L_{\rm R}^{12}|)$
 should be respectively identified
with the $\cN=2$ superconformal parameters ${\x}^a, {\x}^\a,\bar{\x}^\a,\o_{\a\b},\s$ and $({\ri}/{2})\L$ 
introduced in  subsection (\ref{N2superKilling}).  \\
${}\quad$ {\bf 2.} An extended superconformal transformation.
It is generated by 
the complex parameter $\r^\a$ satisfying  the constraints
\bea
\mD_{(\a}\r_{\b)}=\mDB_{(\a}\r_{\b)}=0~~~
\Longrightarrow~~~
\pa_{(\a\b}\r_{\g)}
=\mD^2\r_{\a}=\mDB^2\r_{\a}=0 ~.
\label{C9}
\eea
The constraints are solved by
\bea
\r^\a=\e^\a+\l_{\rm R}\q^\a
-\bar{\l}_{\rm L}\qb^\a
+\bar{\eta}_\b x^{\a\b}
+\ri\bar{\eta}^\a\q^\b\qb_\b
~.
\eea
Here the complex parameter $\e^\a$ generates the third and fourth $Q$-supersymmetries;
the complex  parameters  $\l_{\rm L} = \L_{\rm L}^{11}|_{\q=0}$ 
and $\l_{\rm R}=\L_{\rm R}^{11}|_{\q=0}$ generate
off-diagonal $\sSU(2)_{\rm L}$ and $\sSU(2)_{\rm R}$ transformations; 
finally, the complex parameter ${\eta}^\a$ generates the third and fourth
$S$-supersymmetry transformations.\\
${}\quad$ {\bf 3.} A shadow $\sU(1)$ rotation is generated by 
\bea
\a:= {\rm i}( \L_{\rm R}^{\oo\ot}| - \L_{\rm L}^{12}|) ={\rm const}~.
\eea
In $\cN=4$ superspace, it describes  a $\sU(1)$ phase transformation of $\q^\a_{1\ot},\q^\a_{2\oo}$ only,
with $\q^\a_{1\oo},\q^\a_{2\ot}$ kept unchanged.


\footnotesize{

}

\end{document}